\documentclass[runningheads]{llncs}

\usepackage{amsmath,amssymb}
\usepackage{algorithm}
\usepackage[noend]{algpseudocode}
\usepackage[inline]{enumitem}
\usepackage{graphicx}
\usepackage{subcaption}
\usepackage{wrapfig}
\usepackage{listings}
\usepackage{mathtools}
\usepackage{textcomp}
\usepackage{tabularx}
\usepackage{xcolor}
\definecolor{vred}{rgb}{0,0,0}
\definecolor{vblue}{rgb}{0,0,0}
\usepackage{xspace}
\usepackage{url}
\usepackage{sansmath}
\usepackage{tikz}
\usetikzlibrary{positioning,fit,arrows.meta}
\usetikzlibrary{positioning}
\usetikzlibrary{trees, shapes.geometric}
\usepackage[prefixflatinterpret,prefixflatbracketmodalinterpret,prefixflatsetinterpret,sidenotecalculus,longseqcontext,silentconst]{logic}
\usepackage[prefixflatinterpret,prefixflatbracketmodalinterpret,differentialdL,simplenames,silentconst]{dL}
\usepackage{hyperref}

\usepackage{prettyref}
\newcommand{\rref}[2][]{\prettyref{#2}}
\newrefformat{sec}{Section\,\ref{#1}}
\newrefformat{subsec}{Section\,\ref{#1}}
\newrefformat{def}{Def.\,\ref{#1}}
\newrefformat{thm}{Theorem\,\ref{#1}}
\newrefformat{prop}{Proposition\,\ref{#1}}
\newrefformat{lem}{Lemma\,\ref{#1}}
\newrefformat{cor}{Corollary\,\ref{#1}}
\newrefformat{ex}{Example\,\ref{#1}}
\newrefformat{tab}{Table\,\ref{#1}}
\newrefformat{fig}{Fig.\,\ref{#1}}
\newrefformat{app}{Appendix\,\ref{#1}}
\newrefformat{alg}{Algorithm\,\ref{#1}}
\newrefformat{model}{Model\,\ref{#1}}
\newrefformat{line}{Line\,\ref{#1}}

\newrefformat{eq}{Eq.\,\eqref{#1}}
\newrefformat{case}{Case\,\eqref{#1}}
\newrefformat{appendix}{Appendix\,\ref{#1}}

\newcommand{\finalNode}{\kwd{end}\xspace}
\newcommand{\minabsorbed}{\ensuremath{a_{min}}}
\newcommand{\maxdischarged}{\ensuremath{d_{max}}}
\newcommand{\absorbed}{\texttt{absbd}\xspace}
\newcommand{\tempDiff}{\texttt{temp\_diff}\xspace}
\newcommand{\discharged}{\texttt{disch}\xspace}
\newcommand{\generated}{\ensuremath{g}}
\newcommand{\timer}{\texttt{deadline}\xspace}
\newcommand{\flow}{\texttt{flow}\xspace}
\newcommand{\maxflow}{\ensuremath{F}}
\newcommand{\latency}{\ensuremath{T}\xspace}
\newcommand{\clock}{\texttt{t}\xspace}
\newcommand{\preci}[1]{\phi_{#1}}

\newcommand{\examplewinningc}{\ensuremath{\discharged \leq \maxdischarged \land (\timer \leq 0 \limply \absorbed \geq \minabsorbed)}\xspace}

\newcommand{\predator}{\texttt{y}\xspace}
\newcommand{\prey}{\texttt{x}\xspace}
\newcommand{\predatormin}{\ensuremath{y_{\textrm{mn}}}}
\newcommand{\preymin}{\ensuremath{x_{\textrm{mn}}}\xspace}
\newcommand{\lotkaa}{\ensuremath{a}}
\newcommand{\lotkab}{\ensuremath{b}}
\newcommand{\lotkag}{\ensuremath{g}}
\newcommand{\lotkad}{\ensuremath{d}}
\newcommand{\predatoradd}{\texttt{y\_add}\xspace}
\newcommand{\preyadd}{\texttt{x\_add}\xspace}

\newcommand{\trainpos}{\texttt{trainPos}\xspace}
\newcommand{\vel}{\texttt{vel}\xspace}
\newcommand{\trainacc}{\texttt{trainAcc}\xspace}
\newcommand{\brakeslope}{\texttt{brakeSlope}\xspace}
\newcommand{\airbrake}{\texttt{airBrake}\xspace}
\newcommand{\epos}{\ensuremath{e}\xspace}
\newcommand{\aone}{\ensuremath{a_1}\xspace}
\newcommand{\atwo}{\ensuremath{a_2}\xspace}
\newcommand{\azero}{\ensuremath{a_0}\xspace}
\newcommand{\bzero}{\ensuremath{b_0}\xspace}
\newcommand{\apb}{\ensuremath{A_{pb}}\xspace}
\newcommand{\pressurechangerate}{\ensuremath{p}\xspace}

\newcommand{\vdpxone}{\texttt{x1}\xspace}
\newcommand{\vdpxtwo}{\texttt{x2}\xspace}
\newcommand{\vdpyone}{\texttt{y1}\xspace}
\newcommand{\vdpytwo}{\texttt{y2}\xspace}
\newcommand{\vdpxmax}{\ensuremath{x_{max}}\xspace}
\newcommand{\vdpm}{\ensuremath{m}\xspace}
\newcommand{\vdpb}{\ensuremath{b}\xspace}

\newcommand{\chemA}{\texttt{A}\xspace}
\newcommand{\chemB}{\texttt{B}\xspace}
\newcommand{\chemC}{\texttt{C}\xspace}
\newcommand{\chemTemp}{\texttt{Temp}\xspace}
\newcommand{\chemTmax}{\ensuremath{T_{max}}\xspace}
\newcommand{\chemkA}{\ensuremath{k_A}\xspace}
\newcommand{\chemkB}{\ensuremath{k_B}\xspace}
\newcommand{\chemkC}{\ensuremath{k_C}\xspace}
\newcommand{\chemkT}{\ensuremath{k_T}\xspace}
\newcommand{\chemisOn}{\texttt{isOn}\xspace}

\usepackage{float}
\floatstyle{ruled}
\newfloat{model}{tbp}{model}
\floatname{model}{Model}
\newcounter{modelline}
\makeatletter
\newcommand{\mline}[1]{{\refstepcounter{modelline}\ltx@label{#1}}~\text{\scriptsize{\themodelline}}\quad}
\makeatother

\newcommand{\kwd}[1]{{\normalfont \textsf{#1}}\xspace} 

\newcommand{\safe}{\kwd{contract}}

\newcommand{\assumptions}{\kwd{assume}}

\newcommand{\plant}{\kwd{plant}}
\newcommand{\ctrl}{\kwd{ctrl}}

\newcommand{\invariant}{\kwd{inv}}

\newcommand{\equ}{\kwd{equ}}
\newcommand{\pre}{\kwd{pre}}
\newcommand{\const}{\kwd{const}}
\newcommand{\preypost}{\ensuremath{\prey\ast}}
\newcommand{\predatorpost}{\ensuremath{\predator\ast}}

\newcommand{\init}{\kwd{init}}

\newcommand{\seq}{\,;\,}
\newcommand{\D}[1]{\ensuremath{#1'}}

\begin{document}

\title{LLM-Powered Automatic Theorem Proving and Synthesis for Hybrid Systems and Games}

\author{
  Aditi Kabra\inst{1} \and
  Jonathan Laurent\inst{2} \and
  Ruben Martins\inst{1} \and
  Stefan Mitsch\inst{3} \and
  Andr\'e Platzer\inst{2}
}

\institute{
  Carnegie Mellon University, Pittsburgh, USA\\
  \email{akabra@cs.cmu.edu, rmatins@cs.cmu.edu}
  \and
  Karlsruhe Institute of Technology, Karlsruhe, Germany\\
  \email{jonathan.laurent@kit.edu, platzer@kit.edu}
  \and
  DePaul University, Chicago, USA\\
  \email{smitsh@depaul.edu}
}

\maketitle

\begin{abstract}
  Hybrid games model cyber-physical systems (CPS), like cars, trains, and airplanes, where discrete control decisions interact with continuous physical dynamics.
  We use Large Language Models (LLMs) to scale formal verification and synthesis for hybrid systems and games for a high-level hybrid games symbolic logic, differential game logic (\dGL).
  This combination of a logic with the right expressivity and automation of the interactive theorem proving process using LLMs brings within reach a challenging class of CPS verification/synthesis problems, that were previously well out of range of automatic theorem proving.
  We demonstrate it on five challenging case studies, all beyond the reach of existing automatic techniques.
  Verification succeeds for all five, and the synthesis of control solutions succeeds for four of the five.
\end{abstract}

\section{Introduction}
\label{sec:introduction}

Systems like cars, trains, and airplanes, where discrete control software interacts with continuous physical dynamics, are modeled by \emph{hybrid systems} and \emph{hybrid games}.
Their verification and formally justified synthesis is important because they are often safety-critical, but mathematically complex, and thus difficult to get exactly right.
As a consequence, extensive research has gone into developing formal methods for hybrid systems and games, with different approaches making different tradeoffs.
Some tools gain impressive scalability by limiting the systems that they handle to mathematically simpler dynamics such as piecewise constant \cite{10.5555/1517424.1517433} or affine \cite{10.1145/3567425,DBLP:conf/hvc/RayGDBBG15}.
Other tools very effectively handle nonlinear dynamics over short time horizons \cite{DBLP:conf/arch/Althoff21,DBLP:conf/hybrid/BogomolovFFPS19} using numerical techniques, but lose precision of prediction over longer time horizons, struggling when time is unbounded.
The general trend is that scalability is gained at the cost of expressiveness and precision.
On the most expressive end are deductive, symbolic techniques in high-level logics \cite{DBLP:conf/lics/Platzer12a,DBLP:conf/cade/FultonMQVP15,DBLP:conf/hybrid/ChaochenJR95,DBLP:conf/icfem/WangZZ15}.
They accept nonlinear dynamics interacting with discrete control in rich, program-like structures, permit reasoning in terms of \emph{symbolic parameters} that make theorems adaptable and re-usable, and cleanly handle infinite state space as well as unbounded domains including unbounded time horizons, all with precise reasoning free of discretization or approximation.
However, they have one big disadvantage: despite efforts towards automation \cite{DBLP:journals/fmsd/SogokonMTCP22,DBLP:conf/sas/GhorbalSP14}, they generally require effort-intensive interactive theorem proving by an expert, which becomes a bottleneck for the complexity of systems handled and broader adoption.

\begin{wrapfigure}{r}{0.39\textwidth}
  \centering
  \includegraphics[width=0.39\textwidth]{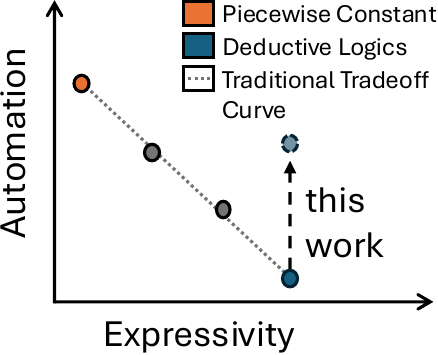}
  \caption{Schematic representation of tradeoffs in formal methods applied to Hybrid Systems.}
  \label{fig:tradeoffs}
\end{wrapfigure}

LLMs have recently shown promise in automating (and therefore scaling) interactive theorem proving for mathematics and (purely discrete) software \cite{Polu2020GenerativeLM,10.1145/3713082.3730382,hilbert}.
This paper explores the same premise for the data-scarce, mathematically complex domain of hybrid systems and games.
We identify five challenging case studies in verification and control synthesis and create LLM-assisted pipelines to prove and synthesize for them.
The implication is a change in the tradeoff landscape of formal methods for hybrid systems, as the application of LLMs brings unprecedented automation to symbolic, deductive verification and synthesis, illustrated schematically in \rref{fig:tradeoffs}.

We create a pipeline that lets the LLM interact with the interactive theorem prover \KeYmaeraX over multiple rounds, analyzing the problem, and summarizing and fixing its own mistakes.
With this verification engine available, we turn to \emph{synthesizing} the space of correct control.
We synthesize \emph{control envelopes}, or families of controllers that all verifiably meet the control objectives.
A \emph{symbolic} control envelope is parametric in symbols in its input, e.g., a train control solution in terms of symbolic train weight \(w\) that can be any real number.
We use LLMs strategically within a recent control envelope synthesis framework \cite{arxiv-general,DBLP:conf/tacas/KabraLMP24} to automatically synthesize verified \emph{symbolic control envelopes}.
The resulting tool verifies all and synthesizes four of the five case studies, which are all not solved by existing automatic tools.
Our key contributions are an \emph{automatic theorem proving system} and a \emph{synthesis pipeline} that scale symbolic logic based verification and synthesis for hybrid systems and games, with the pipeline design justified by ablation experiments on challenging case studies.
Our implementation is available online at \url{https://figshare.com/s/7f39a6756a7efa2e460e}.

\section{Related Work}
\label{sec:related-work}

Hybrid games, and in particular the tactic language of \KeYmaeraX, are significantly more data-scarce than \emph{LLM-based verification} in mathematics \cite{Polu2020GenerativeLM,hilbert} and software \cite{10.1145/3713082.3730382,verus,dafny}, with few proof and synthesis examples available online.
To our knowledge, this is the first work to explore LLM-based verification and synthesis for hybrid systems and games using a deductive, high-level symbolic logic like differential game logic \cite{DBLP:journals/tocl/Platzer15}.
\emph{Theorem provers for hybrid systems} traditionally focus on a mix of interactive proof techniques and proof heuristics for automation \cite{ARCH25:ARCH_COMP25_Category_Report_HSTP}. 
HHL Prover \cite{DBLP:conf/icfem/WangZZ15} and HHLPy \cite{DBLP:conf/fm/ShengBZ23} use proof scripts with annotated invariants for loops and differential equations \cite{ARCH25:ARCH_COMP25_Category_Report_HSTP}.
IsaVODEs \cite{DBLP:journals/jar/MuniveFGSLH24} uses mostly scripted proofs with automation provided by Isabelle \cite{DBLP:conf/lpar/PaulsonB10}.
Plaidypvs \cite{DBLP:journals/corr/abs-2404-15214} embeds differential dynamic logic in PVS.
\KeYmaeraX \cite{DBLP:conf/cade/FultonMQVP15} selects proof tactics based on the syntactic shape of proof obligations; proof heuristics provide a manually selected priority between tactic alternatives, attempted in order with timeouts. 
These proof heuristics achieve automation in hybrid systems, nonlinear continuous systems, and hybrid games, outperforming even proof scripts of other hybrid systems theorem provers \cite{ARCH23:ARCH_COMP23_Category_Report_HSTP,ARCH25:ARCH_COMP25_Category_Report_HSTP}.
This paper introduces a reasoning pipeline to push automation beyond the reach of \KeYmaeraX heuristics.

\section{Automated Theorem Proving}
\label{sec:atp}

We design an automated theorem proving (ATP) framework that uses LLMs to prove hybrid game/systems theorems in \KeYmaeraX.
A strength of our framework is its generality: we design the pipeline and prompts to focus on generic information about all parts of \dGL, and the pipeline works across all five case studies without any case-study-specific engineering.

\paragraph{Background.} Hybrid \emph{systems} extend \emph{discrete} software by also modeling \emph{continuous} dynamics which allows them to model control systems governed by discrete code in a continuous physical environment.
Hybrid \emph{games} extend systems by adding nondeterministic choice points resolved \emph{adversarially} by players.
They can be used to specify the \emph{space} of control within which a controller player must find a strategy to achieve its control objectives despite potentially adversarial environmental conditions, e.g., a car that must reach the destination while remaining safe despite other cars driving aggressively.
In this paper, hybrid games are represented in differential game logic (\dGL) \cite{DBLP:journals/tocl/Platzer15}.
Differential dynamic logic (\dL) \cite{Platzer18} is the single-player subset of \dGL that models hybrid \emph{systems}.
\dGL can be read similar to an imperative programming language.
This paper does not require a detailed understanding, but an introduction to \dGL \cite{Platzer18,DBLP:journals/tocl/Platzer15} can be found in the literature, with the central details recalled in \rref{app:appendix-dgl}.

\subsection{Pipeline Description}
\label{sec:atp-pipeline}

\begin{figure}[t]
  \centering
  \includegraphics[width=\textwidth]{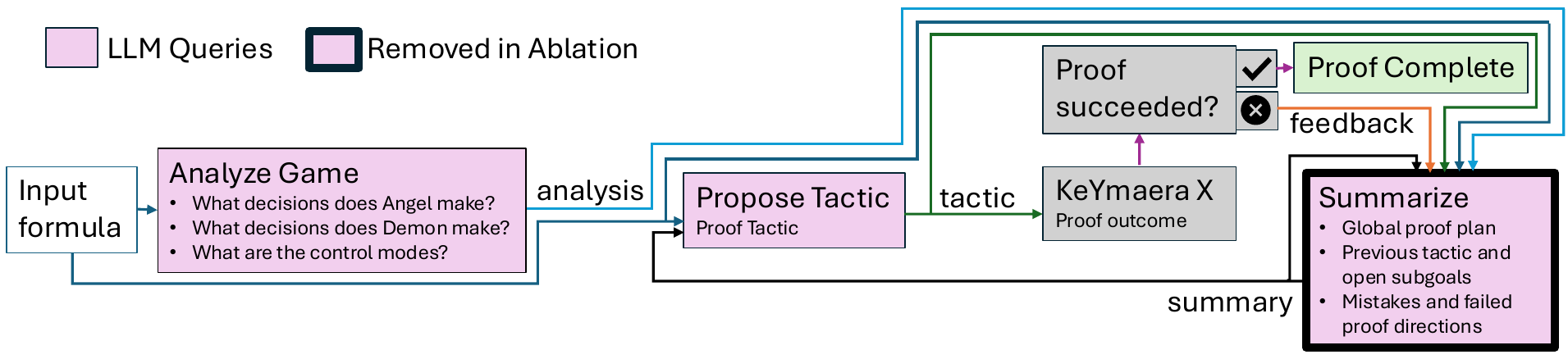}
  \caption{Automated Theorem Proving Pipeline. \rref{app:verification-prompts} shows prompts.}
  \label{fig:atp-pipeline}
\end{figure}

Our automated theorem proving system follows the pipeline in \rref{fig:atp-pipeline}.
A natural design is a conversational agent that iteratively proposes tactics that \KeYmaeraX checks.
With access to the full conversation history, it corrects mistakes based on KeYmaera X output.
This system fails to find proofs for three of our case studies.
Therefore, to condense long contexts with limited signal, we add a summarization-and-reflection step.
To make the LLM aware of the high-level meaning of the theorem, we add an initial problem analysis step.
The analysis appears in subsequent prompts, having a small positive effect on robustness.

In the \emph{Analyze Game} step, given information about the syntax of \dGL, the LLM produces an analysis of the input formula that contains four parts: what actions are controlled by which player, what modes (regions of different qualitative behavior) the game has, and comments about the overall gameplay pattern.
Then the proving loop begins.
This consists of three steps repeated until the proof is complete.
First, the LLM is prompted to propose a tactic. It is given a guide on the Bellerophon tactic language \cite{DBLP:conf/itp/FultonMBP17} that \KeYmaeraX supports, the formula to prove, and a summary of past proof attempts, if available.
The second step is to run the proposed tactic in \KeYmaeraX.
If the tactic completes the proof, the pipeline ends successfully.
Otherwise, \KeYmaeraX produces some output indicating where the proof failed.
This output proceeds to the third step, \emph{Summarize}.
In this step, the LLM uses the summary from the previous round (if available), and the proposed tactic and outcome of the present round to come up with a summary of the new proof state, currently known information, and suggestions/speculations about what to do next.
This summary feeds back into the next tactic proposal step, continuing the loop.
\rref{app:verification-prompts} lists the detailed prompts for each query.
The pipeline is implemented using the Delphyne framework \cite{oracular-programming-2025} for oracular programming with LLMs.

\subsection{Case Studies}
\label{sec:other-case-studies-overview}

Five case studies are carefully developed to pose \emph{challenging verification and synthesis problems}.
The first is about maintaining safe populations under the \emph{non-linear} predator-prey \emph{Lotka-Volterra} model given the ability to introduce additional predators and prey at unpredictable intervals of \emph{unbounded duration}.
The second case study is about verifying the ability of a \emph{train} control system to maintain safety over train dynamics with \emph{Davis resistance (nonlinear) and time-dependent airbrakes}. This case study poses structural complexity with a \emph{nested inner loop} over \emph{different modes} of air brake operation.
The third is maintaining safe temperatures during a \emph{chemical reaction}. This system again requires reasoning about complex dynamics and different, arbitrarily alternating control modes over unbounded iterations. The solution to the dynamics lies \emph{outside the decidable (polynomial) fragment of real arithmetic}, requiring non-trivial \emph{deductive reasoning about differential equations via invariants}.
The fourth is about the ability of a \emph{coolant} system to meet heat absorption quotas in a nuclear power plant while limiting coolant discharge, with a \emph{complex, timed, reach-avoid objective}, different discrete modes, and dynamics that behave very differently depending on initial conditions of variables.
Anecdotally, it took a human expert three days to manually prove this case study.
The fifth is about choosing the right initial conditions for a \emph{Van der Pol} oscillator to maintain a safety property, which again requires reasoning about \emph{nonlinear dynamics} over unbounded iterations, and therefore, effectively, \emph{unbounded time}.
The \emph{chaotic} nature of the Van der Pol oscillator dynamics makes the \emph{choice of initial conditions} delicate.
\emph{\rref{app:other-case-studies} presents these case studies in detail.}

\subsection{Outcomes}
\label{sec:outcomes}

\rref{tab:evaluation-results-verification} summarizes the outcomes of running the tool on the case studies.
A successful outcome means that the tool was able to find a proof within eight runs and a total LLM call budget of twenty four dollars.
\emph{\rref{app:verification-details} reports details.}
None of the case studies can be solved by \KeYmaeraX's existing automation alone \cite{DBLP:journals/fmsd/SogokonMTCP22}.
An ablation experiment demonstrate the significance of the summarization step.
Instead of an LLM-generated summary, the tactic proposal step is given a list of all previously proposed tactics and their outcomes, solving only two of the five case studies. 
With the summarization step, all five case studies are solved.
All experiments in this paper use OpenAI GPT-5 for LLM calls.

\begin{table}[b]
  \centering
  \begin{tabular}{c|c|c}
    \hline
    Traditional Proof Automation & Agent with Full History & Agent with Summarization \\
    \hline
    0/5 & 2/5 & 5/5 \\
    \hline
  \end{tabular}
  \caption{Verification: Adding a summarization step improves success rate compared to providing the agent the full history of tactic attempts and outcomes.}
  \label{tab:evaluation-results-verification}
\end{table}

\subsection{Verification Impact by Example}
\label{sec:verification-impact-by-example}

This section illustrates how the combination of a symbolic logic with the right level of expressivity and LLMs to automate reasoning resolves challenges beyond the reach of existing automatic techniques.
For simplicity of explanation, we focus on one case study in particular, but the style of challenges and their resolutions generalize to all case studies and most \dGL theorems.

\paragraph{Case Study: Lotka-Volterra.}
We introduce the first case study as a running example. A forest department must maintain predator population \(\predator\) and prey population \(\prey\) above minimum thresholds \(\predatormin\) and \(\preymin\) respectively, for arbitrarily long.
Because of unpredictable funding cycles, they can introduce additional prey and predators at unpredictable intervals, that could last arbitrarily long.
This case study is inspired by an ARCH-COMP benchmark \cite{ARCH25:ARCH_COMP25_Category_Report}, but is modified by the addition of discrete control choices and adversarial game play.

\paragraph{\rref{model:lotka-volterra} Explanation.}
\rref{model:lotka-volterra} shows the \dGL formula modeling this problem via a \emph{game}, with two players, canonically called Angel (representing the forest department) and Demon (representing uncertainties). 
Angel chooses the number of animals to introduce (\rref{line:lotka-control} and \ref{line:lotka-addition})\footnote{%
  In the game \(\preyadd := \ast\), Angel freely chooses any real number to assign to \(\preyadd\). Then, \(\ptest{\preyadd \geq 0}\) asserts that \(\preyadd\) must be non-negative.
}, while Demon adversarially chooses the length of the interval till the next opportunity to introduce animals (\rref{line:lotka-plant}), as well as how many iterations the game runs for (\rref{line:lotka-winning-condition}), forcing Angel to maintain safe populations for arbitrarily long\footnote{%
  In the game \(\{x'=f(x)\}^d\), the state evolves per the flow of the differential equation \(x'=f(x)\) for a duration chosen by Demon.
  In the game \((\alpha)^\times\), Demon chooses how many times to repeat the game \(\alpha\).
}.
Angel wins if in the end populations remain above the minimum thresholds (\rref{line:lotka-winning-condition}).
The overall modal formula of form \(\langle\alpha\rangle\phi\) is true in states where Angel has a strategy to play in \(\alpha\) so that regardless of how Demon plays in \(\alpha\), she will win (i.e., postcondition \(\phi\) holds).
\rref{model:lotka-volterra} asks us to prove that, under the assumptions of Lines \ref{line:lotka-assumptions-1} and \ref{line:lotka-assumptions-2}, Angel can always win.

\begin{model}
  \setcounter{modelline}{0}
  \caption{Lotka-Volterra Population Control. \textcolor{gray}{Gray text} is part of the verification problem but is excluded during synthesis, where it is synthesized automatically.}
  \label{model:lotka-volterra}
  \vspace{-0.5em}
  \begin{spreadlines}{0.3ex}
  \begin{align*}
    \text{\assumptions} &\,\left| 
      \begin{aligned}
        \mline{line:lotka-assumptions-1} &\textrm{\textcolor{gray}{\(\prey \leq \frac{\generated}{\lotkad} \land \predator \leq \frac{\lotkaa}{\lotkab} \land \frac{\generated}{\lotkad} \geq \preymin \land \frac{\lotkaa}{\lotkab} \geq \predatormin \land \)}} \prey > 0 \land \predator > 0 \\
        \mline{line:lotka-assumptions-2} & \quad \land \lotkaa > 0 \land \lotkab > 0 \land \lotkad > 0 \land \generated > 0 \land \prey \geq \preymin \land \predator \geq \predatormin \limply\\
      \end{aligned}
    \right.\\
    \text{\ctrl} &\,\left| 
      \begin{aligned}
        \mline{line:lotka-control} &\langle ( \ \ \preyadd := *; \ptest{\preyadd \geq 0} \seq \predatoradd := *; \ptest{\predatoradd \geq 0} \seq \\
        \mline{line:lotka-addition} &\phantom{\langle ( }\ \  \prey := \prey + \preyadd \seq \predator := \predator + \predatoradd \seq
      \end{aligned}
    \right. \\
    \text{\plant} &\,\left| \mline{line:lotka-plant} \phantom{\langle ( }\ \  \{ \prey' = \lotkaa \cdot \prey - \lotkab \cdot \prey \cdot \predator, \ \predator' = \lotkad \cdot \prey \cdot \predator - \generated \cdot \predator \}^d
        \right.\\
    \text{\safe} &\,\left| \mline{line:lotka-winning-condition} \phantom{\langle })^\times \rangle (\prey \geq \preymin \land \predator \geq \predatormin)
      \right.
  \end{align*}
  \end{spreadlines}
  \vspace{-\baselineskip}
  \vspace{-0.5em}
\end{model}

This case study is symbolic, with different behavior depending on the values of the parameters \lotkaa, \lotkab, \lotkad, \lotkag, and variables \prey and \predator.
Our reasoning must account for all uncountably infinite possible values that these symbols can take.
To verify this formula, we will necessarily have to reason about the \emph{nonlinear dynamics of the Lotka-Volterra equation over unbounded time horizons} (\rref{line:lotka-plant}).
While only a few automatic solvers can reason about nonlinear dynamics, even these cannot reason about reachability over unbounded time horizons\footnote{For instance, at ARCH-COMP 2025, where benchmark 5 involved Lotka-Volterra dynamics, the aim was only to solve for 3.64 time units.}.
Then we must consider the discrete and adversarial dynamics: Angel chooses \emph{some} positive real number of animals to introduce (\rref{line:lotka-control}) and we must verify whether \emph{any} of her uncountably infinite choices will work over the nonlinear, unbounded-time dynamics.
Additionally, we must consider what happens if this entire process repeats \emph{arbitrarily many times} in a loop (\rref{line:lotka-winning-condition}).
This explosion in possible behaviors is hard for traditional automatic techniques to deal with.

Yet a symbolic proof is possible by translating a high-level physical understanding of the problem to a clever choice of deductive proof rules.
Looking closely at the qualitative behavior of the Lotka-Volterra dynamics reveals a way for Angel to win.
The dynamics have an \emph{equilibrium point} at \((\prey,\predator) = (\lotkag/\lotkad, \lotkaa/\lotkab)\), which is above the minimum populations threshold per assumptions \(\lotkaa/\lotkab \geq \predatormin\) and \(\lotkag/\lotkad \geq \preymin\) (\rref{model:lotka-volterra}, \rref{line:lotka-assumptions-1}).
If Angel manages to set the populations to this equilibrium point at the start of each iteration, then she will satisfy her winning condition at the end of the iteration, regardless of how long Demon runs the dynamics.
Since Angel can introduce any non-negative number of animals, all she needs at the start of each iteration is that predator and prey populations are no greater than the equilibrium point.
Finally, many iterations of this manoeuver string together correctly over loop iterations because at the end of each iteration, populations are at the equilibrium point, satisfying the requirement that populations are no greater than the equilibrium point.

To translate this high-level physical reasoning to a formal proof, the right choice of \dGL proof rules\footnote{
\rref{fig:lotka-volterra-rules} in \rref{app:appendix-dgl} lists all the rules referenced in this discussion; the intent is not to dwell on their individual details, but to highlight that the proof requires a sequence of complex rule applications that resist straightforward (traditional) automation.}
is critical, and traditionally provided by a human expert who understands what proof rule is applicable when.
To show the first step in our reasoning above, that \((\prey,\predator) = (\lotkag/\lotkad, \lotkaa/\lotkab)\) is an equilibrium point of the dynamics,
the rule most commonly used to prove invariants of differential equations (the \emph{differential invariant} rule) does not immediately apply.
Instead, a quick proof is possible using the more rarely used \emph{differential radical invariant} \cite{DBLP:conf/tacas/GhorbalP14} rule (\rref{eq:dri-rule}).
It allows us to conclude that polynomial equation \(h_j=0\) is an invariant of the dynamics if we can show that its Lie derivatives \(\mathcal{L}_d(h_j), \mathcal{L}^2_d(h_j), \ldots \mathcal{L}_d^N(h_j)\) are zero, where \(N\) is the finite derivative order at which the chain of ideals \(\langle h \rangle \subseteq \langle h, \mathcal{L}_p(h) \rangle \subseteq \cdots\) reaches a fixed point.
\begin{equation}
  \label{eq:dri-rule}
\textrm{
\begin{calculus}
  \dinferenceRule[dRI|dRI]{differential radical invariant rule}
  {\linferenceRule
    { \lsequent{\Gamma, Q}{\land_i^N \land_j \mathcal{L}_p^i (h_j) =0}
    }
    {\lsequent{\Gamma}{\dbox{x'=p\ \&\ Q}{\land_j h_j=0}} }
  }{}
\end{calculus}
}
\end{equation}
\rref{eq:lotka-dri-proof} shows how this applies to our case. Read from bottom to top, the differential radical invariant rule (dRI) applies to produce a sequent that is propositional in the decidable fragment of real arithmetic, that an automatic prover like Z3 \cite{DBLP:journals/corr/DeMouraM08} can discharge, establish an equilibrium point of these nonlinear dynamics.
\begin{equation}
  \label{eq:lotka-dri-proof}
\textrm{
\begin{calculus}
  \scalebox{0.8}{\(
  \linfer
  {\lclose}
  {\linfer[dRI]
    { \lsequent{\prey - \lotkag/\lotkad = 0 \land \predator - \lotkaa/\lotkab = 0}{\lotkad(\lotkaa \prey - \lotkab \prey \predator) = 0  \land  \lotkab(\lotkad \prey \predator - \lotkag \predator) = 0}
    }
    {\lsequent{\prey-\lotkag/\lotkad = 0 \land \predator - \lotkaa/\lotkab = 0}{\dbox{\prey' = \lotkaa \prey - \lotkab \prey \predator, \ \predator' = \lotkad \prey \predator - \lotkag \predator}{(\prey-\lotkag/\lotkad = 0 \land \predator - \lotkaa/\lotkab = 0)}} }
  }
  \)}
\end{calculus}
}
\end{equation}
This reasoning step must be carefully combined with the other reasoning steps to complete the proof.
In this example we also need to apply rules such as 
the \emph{loop} rule with the correct invariant \(J\equiv\prey \geq \preymin \land \predator \geq \predatormin \land \prey \leq \lotkag/\lotkad \land \predator \leq \lotkaa/\lotkab \land \lotkag/\lotkad \geq \preymin \land \lotkaa/\lotkab \geq \predatormin \land \lotkaa > 0 \land \lotkab > 0 \land \lotkad > 0 \land \lotkag > 0\) to demonstrate the inductiveness of Angel's strategy over many iterations, and the \emph{exists right} rule with a correct expression for Angel's choice of \preyadd and \predatoradd.
Identifying and applying these rules requires both an understanding of the high-level solution strategy of the problem as well as \dGL's proof rules, leading to difficulties in automation using traditional techniques which lack the former and need careful, case-by-case encoding of the latter.
%
%

Proofs can be expressed succinctly as \emph{tactic} scripts, which are sequences of proof rules with their relevant parameters that when applied in order, complete the proof.
Our pipeline was able to automatically generate a 49 line tactic script that completed the proof for this case study, listed in \rref{app:lv-tactic}.

\section{Synthesis}
\label{sec:synthesis-start}

Verifying \rref{model:lotka-volterra} shows \emph{there exists} some way for Angel to play to win the game.
Synthesis is instead about \emph{constructing} a concrete policy that the user desires, which is useful for developing controllers and runtime monitors/supervisors that ensure control remains compliant with a verified nondeterministic policy.
The user provides the game and describes their desired solution informally, and the tool computes the corresponding formal nondeterministic policy, verified to be correct (never lose the game or get stuck).

\subsection{Subvalue Maps}
\label{sec:subvalue-map}

In this paper, we represent nondeterministic policies as \emph{subvalue maps}~\cite{arxiv-general}, a symbolic equivalent of value functions in reinforcement learning or value tables in dynamic programming.
\dGL games consist of \emph{subgames}, that compose together with operators like sequential composition (\(\seq\)).
An Angelic subvalue map delineates a nondeterministic policy for Angel by mapping each subgame to a \emph{formula}, and constraining Angel to only make choices that result in her entering the next subgame in a state where its mapped formula is true.
\rref{fig:example-envelope} shows a subvalue map for the running example the induces the policy of setting populations to the equilibrium point.
Here, the subgame \(\ptest{\preyadd \geq 0}\), that follows Angel's choice of \preyadd, is mapped to formula \(\phi_\prey \equiv \cdots\ \land\ \lotkad \cdot (\prey + \preyadd) = \lotkag\ \land\ \preyadd\geq 0 \), constraining Angel in the preceding subgame \(\preyadd:=\ast\) to only choose nonnegative \preyadd values set the dynamics to their equilibrium point.

%
\newcommand{\astSpacing}{0.5cm}%
\newcommand{\precAstAnnot}[2]{\textcolor{gray}{\scriptsize $\preci{#2}$} \ $#1$}%
\newcommand{\AlgGuess}{\textbf{Guess }}%
\newcommand{\AlgCheck}{\textbf{Check }}%
\newcommand{\AlgNoCheck}{\textbf{\st{Check} }}%
\newcommand{\AlgDeduce}{\textbf{Deduce }}%
\newcommand{\ExRelaxedLoop}{n \!:=\! * \seq\! ?n \!\ge\! 0 \seq\! v \!:=\! v\!+\!an}
\newcommand{\AlgDef}{\equiv}%
\newcommand{\AlgSubApprox}{\Leftarrow}%
\newcommand{\AlgEqv}{\Leftrightarrow}%
\begin{figure}[t]


    
    \scriptsize

    
      
   
   \begin{minipage}{0.28\textwidth}
   \noindent \textbf{1. Input}: \rref{model:lotka-volterra}\\
   Angel freely chooses \(\preyadd\) and \(\predatoradd\).

   \medskip
   \hrule
   \medskip

    \noindent \textbf{3. Control policy derived from subvalue map}\\
    \smallskip
    In state \(\sigma\):
    \vspace{-1em}
    \begin{enumerate}
      \item At subgame \(\preyadd:=\ast\), \(\prey:=e\) permitted iff \(\sigma(\prey\mapsto e) \models \phi_\prey.\)
      \item At subgame \(\predatoradd:=\ast\), \(\predator:=e\) permitted iff \(\sigma(\predator\mapsto e) \models \phi_\predator.\)
    \end{enumerate}
    where \(\sigma(\varphi \mapsto e)\) is the state obtained by updating \(\sigma\) to map variable \(\varphi\) to value \(e\).
    \medskip
   \end{minipage}
   \hfill
   \vrule
   \hfill
   \begin{minipage}{0.7\textwidth}
    \noindent \textbf{2. Synthesized subvalue map (some subgames unexpanded)}\\
     \scriptsize
     \vspace{-5pt}
    \begin{tikzpicture}
      [
        baseline=(current bounding box.north),
        level distance={\astSpacing},
        every node/.style={draw,rectangle,align=center},
        sibling distance=2cm,
        grow=down,
        scale=0.9,
        transform shape
      ]
      \node (root) {\precAstAnnot{\times}{\textrm{\invariant}}}
        child {
          node {\precAstAnnot{\seq}{\prey}}
          child {
            node {\precAstAnnot{\seq}{\prey}}
            [sibling distance=2.2cm]
            child {
              node {\precAstAnnot{\seq}{\prey}}
              [sibling distance=4.5cm]
              child {
                node {\precAstAnnot{\seq}{\prey}}
                [sibling distance=2.2cm]
                child {node {\precAstAnnot{\preyadd := *}{\prey}}}
                child {node {\precAstAnnot{?\preyadd \geq 0}{\preypost}}}
              }
              child {
                node {\precAstAnnot{\seq}{\preypost}}
                [sibling distance=2.2cm]
                child {node {\precAstAnnot{\predatoradd := *}{\predator}}}
                child {node {\precAstAnnot{?\predatoradd \geq 0}{\predatorpost}}}
              }
            }
            child {node {\precAstAnnot{\ctrl \textrm{ line } \ref{line:lotka-addition}}{\equ+}}}
          }
          child {
            node {\precAstAnnot{\plant}{\equ}}
          }
        };
      \node[right=0.5cm of root] {\precAstAnnot{\finalNode}{\safe}};
    \end{tikzpicture}

    \par\medskip
    where \(\phi_\const \equiv \lotkaa>0 \land \lotkab>0 \land \lotkad>0 \land \generated>0\),
    \(\phi_\pre \equiv \preymin \lotkad \leq \lotkag \land \predatormin \lotkab \leq \lotkaa\),
    \\
    \(\phi_\safe \equiv \prey \geq \preymin \land \predator \geq \predatormin\), \hfill
    \(\phi_\equ \equiv \phi_\const \land \phi_\pre \land \lotkad \prey = \lotkag \land \lotkab \predator = \lotkaa\),\\
    \(\phi_\invariant \equiv \phi_\const \land \phi_\safe \land \phi_\pre \land \lotkad \prey \leq \lotkag \land \lotkab \predator \leq \lotkaa \land \prey > 0 \land \predator > 0 \),\\
    \(\phi_{\equ+} \equiv \phi_\const \land \phi_\pre \land \lotkad \cdot (\prey + \preyadd) = \lotkag \land \lotkab \cdot (\predator + \predatoradd) = \lotkaa \),\\
    \(\phi_\predatorpost \equiv \phi_{\equ+} \land \predatoradd \geq 0\), \hfill
    \(\phi_\predator \equiv \phi_\const \land \phi_\pre \land \lotkad \cdot (\prey + \preyadd)= \lotkag \land \lotkab \predator \leq \lotkaa\),\\
    \(\phi_\preypost \equiv \phi_\predator \land \preyadd \geq 0 \), \hfill
    \(\phi_\prey \equiv \phi_\const \land \phi_\pre \land \lotkad \prey \leq \lotkag \land \lotkab \predator \leq \lotkaa\).
   \end{minipage}

    \hrule

    \smallskip
    \(\ctrl\) line \ref{line:lotka-addition} and \(\plant\) refer to the corresponding parts of \rref{model:lotka-volterra}, and \(\finalNode\) is the special node representing the end of the game.
    \(\preci{\safe}\) is the same correct control condition from \rref{model:running-example-verification}, that is, \(\examplewinningc\).
    The other formulas are synthesized by the pipeline.
  \caption{A \emph{subvalue map} for the running example.}
  \label{fig:example-envelope}
\end{figure}%

Once the LLM guesses a policy (subvalue map), we can verify if it is correct (non-losing, never stuck) by checking some \dGL formulas identified by existing frameworks \cite{arxiv-general,DBLP:conf/tacas/KabraLMP24}.
For example, in the subvalue map of \rref{fig:example-envelope}, to check that the subvalue of the loop, \(\phi_\invariant\), is correct, we make sure
\(\phi_\invariant \limply \phi_\safe \land \phi_\prey\) holds, that is, Angel won't lose regardless of whether Demon continues the loop or not. If he continues the loop, Angel will win into \(\phi_\safe\), and if he repeats the loop, Angel will still win because \(\phi_\prey\) holds, which inductively tells us that she has a winning strategy from the start of the loop body \(\prey := \ast\).
While this check was simple (no modalities), for other subgames, checks can involve modal \dGL formulas, for which we use the verification tool of \rref{sec:atp}.

For synthesis, the same five case studies are used but with some initial assumptions missing (it will be the task of the synthesis pipeline to come up with all conditions necessary to make a strategy succeed).
Additionally, each is associated with an informal natural language description of which of the many possible control solutions to synthesize (listed in \rref{app:other-case-studies}).
Given the running example and a request to synthesize the solution that sets the system to an equilibrium point as soon as possible,
our tool automatically identifies the corresponding subvalue map.
It computes that Angel must pick \(\preyadd\) and \(\predatoradd\) such that \(\prey + \preyadd = \lotkag/\lotkad\) and \(\predator + \predatoradd = \lotkaa/\lotkab\), and that this policy only works when, initially, \(\prey \leq \lotkag/\lotkad\), \(\predator \leq \lotkaa/\lotkab\), \(\lotkag/\lotkad\geq\preymin\), and \(\lotkaa/\lotkab\geq\predatormin\).

\subsection{Pipeline Description}
\label{sec:synthesis-pipeline}

\begin{figure}[t]
  \centering
  \includegraphics[width=\textwidth]{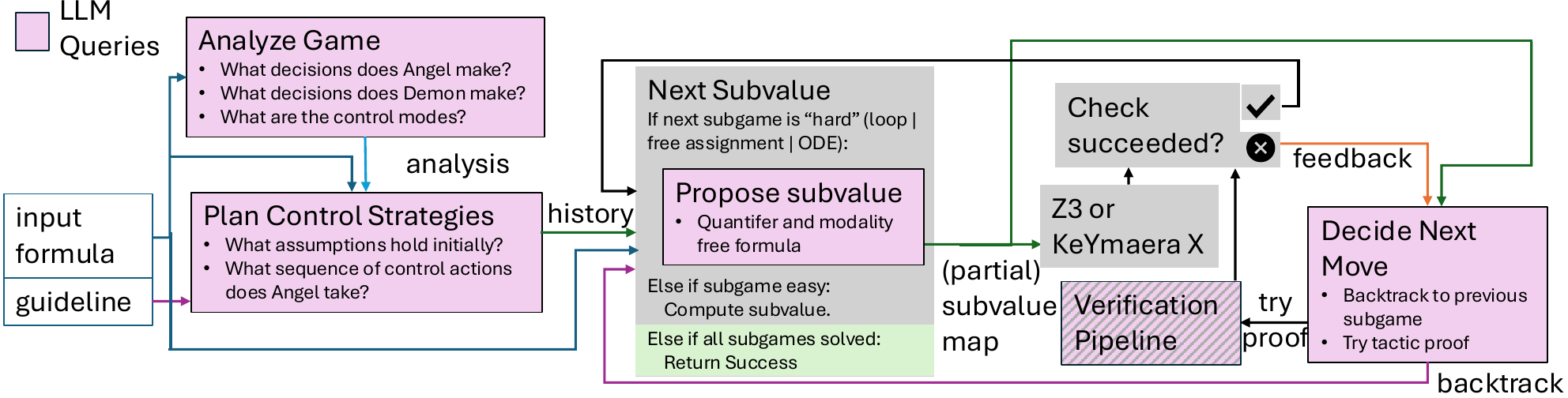}
  \caption{Synthesis Pipeline. \rref{app:synthesis-prompts} shows prompts.}
  \label{fig:synthesis-pipeline}
\end{figure}

Weakest precondition calculus for normal programs \cite{10.1145/360933.360975} starts at the end of the game and works backwards, requiring loop invariant guesses and retrospective checks at loops.
Similarly, our pipeline executes a precondition calculus for subvalue maps \cite{arxiv-general} on the game backwards from end to start, computing subvalues as preconditions.
Some subgames, such as simple assignments, are easy to symbolically execute.
Computing the precondition of others: loops, continuous dynamics, and nondeterministic choices, is expensive or undecidable in general.
For these, the LLM is prompted to guess subvalue map entries which encode Angel's winning strategy.
A retrospective check using Z3 \cite{DBLP:journals/corr/DeMouraM08} when possible and KeYmaera X when necessary recovers soundness.
\rref{fig:synthesis-pipeline} shows the synthesis pipeline.

Initially, the LLM analyzes the game, identifying player actions and control modes, and comes up with a detailed strategy for Angel based on the informal guideline provided by the user.
The strategy is expressed as a sequence of steps that Angel will take,
stored to provide context in later steps.
In the main subvalue map computation phase, the \emph{Next Subvalue} step goes backwards to the next unprocessed subgame, and either symbolically computes the subvalue if it is easy, or prompts the LLM to guess it otherwise.
In the latter case, the subvalue is checked using Z3 or KeYmaera X with existing, LLM-free automation.
Guesses and their outcomes are stored to provide context for future guesses\footnote{
  Control flow is more complicated for the special case of loops, where the loop invariant is guessed first, then the subvalue map of the loop body is computed assuming the invariant holds at the end of the body, and finally the loop invariant is retrospectively checked for inductiveness given the computed subvalue map of the body. 
}.
If the check succeeds, the pipeline continues to the next unprocessed subgame.
Otherwise, there are two possible recovery strategies, that the LLM chooses between.
If the lemma was too complicated to verify using existing automation, it should try using the automatic theorem proving system from \rref{sec:atp}.
Otherwise, the subvalue may be genuinely incorrect.
The incorrectness may lie at the current subgame, or result from a bad earlier subvalue guess that makes the current subgame unsolvable (e.g., a non-inductive loop invariant causing problems in the loop body subvalues).
The LLM should choose to backtrack to the subgame where the bad guess was made, and subvalue computation resumes from there.

\subsection{Outcomes}

\label{sec:synthesis-outcomes}
\begin{table}[b]
  \centering
  \begin{tabular}{c|c|c}
    \hline
    \shortstack{Previous Work} & \shortstack{Without Recovery and LLM-Assisted Proving} & \shortstack{Full Pipeline} \\
    \hline
    0/5 & 2/5 & 4/5 \\
    \hline
  \end{tabular}
  \caption{Synthesis: ablation shows the benefit of the recovery segment of the pipeline (detailed proving and custom backtracking).}
  \label{tab:evaluation-results-synthesis}
\end{table}

\rref{tab:evaluation-results-synthesis} summarizes the results of our experiments.
A case study is solved when one of four runs of the synthesis tool produces a correct subvalue map within a total budget of ten dollars for LLM calls.
As a baseline, we compare to the tool associated with the only previous work general enough to synthesize control envelopes for our problems \cite{arxiv-general}.
It fails, although it could succeed if extended with the right custom heuristics on a case-by-case basis.
In contrast, our pipeline is very general, free of the need for case-by-case encoding of heuristics, and solves four out of five case studies.
We also perform an ablation experiment where we do not allow the LLM to use the proving system from \rref{sec:atp} or perform custom backtracking.
Instead, upon failure, we simply repeat LLM call points until the proof succeeds. 
\rref{app:synthesis-details} shows further details.



\bibliographystyle{splncs04}
\bibliography{refs}

@inproceedings{DBLP:conf/fm/ShengBZ23,
  author       = {Huanhuan Sheng and
                  Alexander Bentkamp and
                  Bohua Zhan},
  editor       = {Marsha Chechik and
                  Joost{-}Pieter Katoen and
                  Martin Leucker},
  title        = {HHLPy: Practical Verification of Hybrid Systems Using Hoare Logic},
  booktitle    = {Formal Methods - 25th International Symposium, {FM} 2023, L{\"{u}}beck,
                  Germany, March 6-10, 2023, Proceedings},
  series       = {Lecture Notes in Computer Science},
  volume       = {14000},
  pages        = {160--178},
  publisher    = {Springer},
  year         = {2023},
  url          = {https://doi.org/10.1007/978-3-031-27481-7\_11},
  doi          = {10.1007/978-3-031-27481-7\_11},
  bibsource    = {dblp computer science bibliography, https://dblp.org}
}

@article{DBLP:journals/jar/MuniveFGSLH24,
  author       = {Jonathan Juli{\'{a}}n Huerta y Munive and
                  Simon Foster and
                  Mario Gleirscher and
                  Georg Struth and
                  Christian Pardillo Laursen and
                  Thomas Hickman},
  title        = {IsaVODEs: Interactive Verification of Cyber-Physical Systems at Scale},
  journal      = {J. Autom. Reason.},
  volume       = {68},
  number       = {4},
  pages        = {21},
  year         = {2024},
  url          = {https://doi.org/10.1007/s10817-024-09709-2},
  doi          = {10.1007/S10817-024-09709-2},
  bibsource    = {dblp computer science bibliography, https://dblp.org}
}

@inproceedings{DBLP:journals/corr/abs-2404-15214,
  author       = {J. Tanner Slagel and
                  Mariano M. Moscato and
                  Lauren M. White and
                  C{\'{e}}sar A. Mu{\~{n}}oz and
                  Swee Balachandran and
                  Aaron Dutle},
  editor       = {Temur Kutsia and
                  Daniel Ventura and
                  David Monniaux and
                  Jos{\'{e}} F. Morales},
  title        = {Embedding Differential Dynamic Logic in {PVS}},
  booktitle    = {Proceedings 18th International Workshop on Logical and Semantic Frameworks,
                  with Applications and 10th Workshop on Horn Clauses for Verification
                  and Synthesis, {LSFA/HCVS} 2023, and 10th Workshop on Horn Clauses
                  for Verification and SynthesisRome, Italy {\&} Paris, France,
                  1-2 July, 2023 {\&} 23rd April 2023},
  series       = {{EPTCS}},
  volume       = {402},
  pages        = {43--62},
  year         = {2023},
  url          = {https://doi.org/10.4204/EPTCS.402.7},
  doi          = {10.4204/EPTCS.402.7},
  bibsource    = {dblp computer science bibliography, https://dblp.org}
}

@inproceedings{ARCH23:ARCH_COMP23_Category_Report_HSTP,
  author    = {Stefan Mitsch and Huanhuan Sheng and Bohua Zhan and Shuling Wang and Simon Foster and Jonathan Julian Huerta Y Munive},
  title     = {ARCH-COMP23 Category Report: Hybrid Systems Theorem Proving},
  booktitle = {Proceedings of 10th International Workshop on Applied Verification of Continuous and Hybrid Systems (ARCH23)},
  editor    = {Goran Frehse and Matthias Althoff},
  series    = {EPiC Series in Computing},
  volume    = {96},
  publisher = {EasyChair},
  bibsource = {EasyChair, https://easychair.org},
  issn      = {2398-7340},
  url       = {/publications/paper/TtcQ},
  doi       = {10.29007/57g4},
  pages     = {170-188},
  year      = {2023}}

@inproceedings{ARCH25:ARCH_COMP25_Category_Report_HSTP,
  author    = {Stefan Mitsch and Ismail Patel and Hari Hara Sudhan Kannan and Xiangyu Jin and Bohua Zhan and Shuling Wang},
  title     = {ARCH-COMP25 Category Report: Hybrid Systems Theorem Proving},
  booktitle = {Proceedings of 12th Int. Workshop on Applied Verification for Continuous and Hybrid Systems},
  editor    = {Goran Frehse and Matthias Althoff},
  series    = {EPiC Series in Computing},
  volume    = {108},
  publisher = {EasyChair},
  bibsource = {EasyChair, https://easychair.org},
  issn      = {2398-7340},
  url       = {/publications/paper/ZlZZ},
  doi       = {10.29007/pd1w},
  pages     = {152-168},
  year      = {2025}}

@book{Platzer18,
  author    = {Andr{\'e} Platzer},
  title     = {Logical Foundations of Cyber-Physical Systems},
  publisher = {Springer},
  address   = {Cham},
  year      = {2018},
  isbn      = {978-3-319-63587-3},
  e-isbn    = {978-3-319-63588-0},
  doi       = {10.1007/978-3-319-63588-0}
}

@inproceedings{DBLP:conf/lics/Platzer12a,
  author        = {Andr{\'e} Platzer},
  title         = {Logics of Dynamical Systems},
  booktitle     = {LICS},
  year          = {2012},
  pages         = {13-24},
  doi           = {10.1109/LICS.2012.13},
  longbooktitle = {Proceedings of the 27th Annual ACM/IEEE
                   Symposium on Logic in Computer Science, LICS
                   2012, Dubrovnik, Croatia, June 25???28, 2012},
  publisher     = {IEEE},
  isbn          = {978-1-4673-2263-8}
}

@article{DBLP:journals/tocl/Platzer15,
  author  = {Andr{\'e} Platzer},
  title   = {Differential Game Logic},
  journal = {{ACM} Trans. Comput. Log.},
  volume  = {17},
  number  = {1},
  year    = {2015},
  pages   = {1:1--1:51},
  doi     = {10.1145/2817824},
  issn    = {1529-3785}
}

@inproceedings{DBLP:journals/corr/DeMouraM08,
  address   = {Berlin, Heidelberg},
  author    = {de Moura, Leonardo and Bj{\o}rner, Nikolaj},
  booktitle = {Tools and Algorithms for the Construction and Analysis of Systems},
  editor    = {Ramakrishnan, C. R. and Rehof, Jakob},
  pages     = {337--340},
  publisher = {Springer},
  title     = {{Z3}: An Efficient {SMT} Solver},
  year      = {2008}
}

@article{DBLP:journals/fmsd/SogokonMTCP22,
  author      = {Andrew Sogokon and
                 Stefan Mitsch and
                 Yong Kiam Tan and
                 Katherine Cordwell and
                 Andr{\'{e}} Platzer},
  title       = {Pegasus: Sound Continuous Invariant Generation},
  journal     = {Form. Methods Syst. Des.},
  longjournal = {Formal Methods in System Design},
  year        = {2022},
  volume      = {58},
  number      = {1},
  pages       = {5-41},
  doi         = {10.1007/s10703-020-00355-z},
  issn        = {0925-9856},
  note        = {Special issue for selected papers from FM'19}
}

@inproceedings{DBLP:conf/cade/FultonMQVP15,
  author        = {Nathan Fulton and
                   Stefan Mitsch and
                   Jan-David Quesel and
                   Marcus V{\"o}lp and
                   Andr{\'{e}} Platzer},
  title         = {{KeYmaera X}: An Axiomatic Tactical Theorem Prover for Hybrid Systems},
  booktitle     = {CADE},
  longbooktitle = {International Conference on Automated Deduction, CADE'15, Berlin, Germany, Proceedings},
  year          = {2015},
  pages         = {527-538},
  doi           = {10.1007/978-3-319-21401-6\_36}
}

@inproceedings{DBLP:conf/itp/FultonMBP17,
  author        = {Nathan Fulton and
                   Stefan Mitsch and
                   Rose Bohrer and
                   Andr{\'e} Platzer},
  title         = {Bellerophon: Tactical Theorem Proving
                   for Hybrid Systems},
  booktitle     = {ITP},
  longbooktitle = {Interactive Theorem Proving,
                   International Conference, ITP 2017},
  year          = {2017},
  pages         = {207-224},
  month         = {},
  doi           = {10.1007/978-3-319-66107-0\_14},
  editor        = {Mauricio Ayala-Rinc{\'o}n and
                   C{\'e}sar A. Mu{\~n}oz},
  publisher     = {Springer},
  series        = {LNCS},
  volume        = {10499},
  isbn          = {978-3-319-66106-3}
}

@INPROCEEDINGS{DBLP:conf/tacas/KabraLMP24,
  author    = {Aditi Kabra and
               Jonathan Laurent and
               Stefan Mitsch and
               Andr{\'e} Platzer},
  title     = {{CESAR}: Control Envelope Synthesis via Angelic Refinements},
  booktitle = {Tools and Algorithms for the Construction and Analysis of Systems},
  year      = {2024},
  pages     = {144--164},
  doi       = {10.1007/978-3-031-57246-3\_9},
  editor    = {Bernd Finkbeiner and
               Laura Kov{\'{a}}cs},
  longbooktitle = {Tools and Algorithms for the Construction
               and Analysis of Systems. TACAS 2024},
  publisher = {Springer},
  series    = {LNCS},
  volume    = {14570},
  isbn      = {978-3-031-57246-3},
}

@article{train,
  author  = {Kabra, Aditi and Mitsch, Stefan and Platzer, André},
  journal = {IEEE Transactions on Computer-Aided Design of Integrated Circuits and Systems},
  title   = {Verified Train Controllers for the Federal Railroad Administration Train Kinematics Model: Balancing Competing Brake and Track Forces},
  year    = {2022},
  volume  = {41},
  number  = {11},
  pages   = {4409-4420},
  doi     = {10.1109/TCAD.2022.3197690}
}

@article{10.1145/360933.360975,
author = {Dijkstra, Edsger W.},
title = {Guarded commands, nondeterminacy and formal derivation of programs},
year = {1975},
issue_date = {Aug. 1975},
publisher = {Association for Computing Machinery},
address = {New York, NY, USA},
volume = {18},
number = {8},
issn = {0001-0782},
doi = {10.1145/360933.360975},
journal = {Commun. ACM},
month = aug,
pages = {453–457},
numpages = {5}
}

@misc{arxiv-general,
  title         = {Hybrid Game Control Envelope Synthesis},
  author        = {Aditi Kabra and Jonathan Laurent and Stefan Mitsch and André Platzer},
  year          = {2025},
  eprint        = {2508.05997},
  archiveprefix = {arXiv},
  primaryclass  = {cs.LO},
  note          = {arXiv:2508.05997},
  doi           = {10.48550/arXiv.2508.05997}
}

@inproceedings{10.1007/978-3-031-15008-1_8,
author = {Bohrer, Rose},
title = {Chemical Case Studies in KeYmaera X},
year = {2022},
isbn = {978-3-031-15007-4},
publisher = {Springer-Verlag},
address = {Berlin, Heidelberg},
doi = {10.1007/978-3-031-15008-1\_8},
booktitle = {Formal Methods for Industrial Critical Systems: 27th International Conference, FMICS 2022, Warsaw, Poland, September 14-15, 2022, Proceedings},
pages = {103-120},
numpages = {18},
location = {Warsaw, Poland}
}

@inproceedings{10.5555/1517424.1517433,
author = {Bu, Lei and Li, You and Wang, Linzhang and Li, Xuandong},
title = {BACH: Bounded reachAbility CHecker for linear hybrid automata},
year = {2008},
isbn = {9781424427352},
publisher = {IEEE Press},
booktitle = {Proceedings of the 2008 International Conference on Formal Methods in Computer-Aided Design},
articleno = {9},
numpages = {4},
location = {Portland, Oregon},
series = {FMCAD '08}
}

@article{10.1145/3567425,
author = {Kundu, Atanu and Das, Sarthak and Ray, Rajarshi},
title = {
SAT-Reach: A Bounded Model Checker for Affine Hybrid Systems},
year = {2023},
issue_date = {March 2023},
publisher = {Association for Computing Machinery},
address = {New York, NY, USA},
volume = {22},
number = {2},
issn = {1539-9087},
doi = {10.1145/3567425},
journal = {ACM Trans. Embed. Comput. Syst.},
month = jan,
articleno = {26},
numpages = {36}
}

@inproceedings{DBLP:conf/hvc/RayGDBBG15,
  author       = {Rajarshi Ray and
                  Amit Gurung and
                  Binayak Das and
                  Ezio Bartocci and
                  Sergiy Bogomolov and
                  Radu Grosu},
  editor       = {Nir Piterman},
  title        = {XSpeed: Accelerating Reachability Analysis on Multi-core Processors},
  booktitle    = {Hardware and Software: Verification and Testing - 11th International
                  Haifa Verification Conference, {HVC} 2015, Haifa, Israel, November
                  17-19, 2015, Proceedings},
  series       = {Lecture Notes in Computer Science},
  volume       = {9434},
  pages        = {3--18},
  publisher    = {Springer},
  year         = {2015},
  doi          = {10.1007/978-3-319-26287-1\_1},
  bibsource    = {dblp computer science bibliography, https://dblp.org}
}

@inproceedings{DBLP:conf/arch/Althoff21,
  author       = {Matthias Althoff},
  editor       = {Goran Frehse and
                  Matthias Althoff},
  title        = {Guaranteed State Estimation in {CORA} 2021},
  booktitle    = {8th International Workshop on Applied Verification of Continuous and
                  Hybrid Systems (ARCH21), Brussels, Belgium, July 9, 2021},
  series       = {EPiC Series in Computing},
  volume       = {80},
  pages        = {161--175},
  publisher    = {EasyChair},
  year         = {2021},
  doi          = {10.29007/7M2K},
  bibsource    = {dblp computer science bibliography, https://dblp.org}
}

@inproceedings{DBLP:conf/hybrid/BogomolovFFPS19,
  author       = {Sergiy Bogomolov and
                  Marcelo Forets and
                  Goran Frehse and
                  Kostiantyn Potomkin and
                  Christian Schilling},
  editor       = {Necmiye Ozay and
                  Pavithra Prabhakar},
  title        = {JuliaReach: a toolbox for set-based reachability},
  booktitle    = {Proceedings of the 22nd {ACM} International Conference on Hybrid Systems:
                  Computation and Control, {HSCC} 2019, Montreal, QC, Canada, April
                  16-18, 2019},
  pages        = {39--44},
  publisher    = {{ACM}},
  year         = {2019},
  doi          = {10.1145/3302504.3311804},
  bibsource    = {dblp computer science bibliography, https://dblp.org}
}

@inproceedings{DBLP:conf/hybrid/ChaochenJR95,
  author       = {Chaochen Zhou and
                  Ji Wang and
                  Anders P. Ravn},
  editor       = {Rajeev Alur and
                  Thomas A. Henzinger and
                  Eduardo D. Sontag},
  title        = {A Formal Description of Hybrid Systems},
  booktitle    = {Hybrid Systems {III:} Verification and Control, Proceedings of the
                  {DIMACS/SYCON} Workshop on Verification and Control of Hybrid Systems,
                  October 22-25, 1995, Ruttgers University, New Brunswick, NJ, {USA}},
  series       = {Lecture Notes in Computer Science},
  volume       = {1066},
  pages        = {511--530},
  publisher    = {Springer},
  year         = {1995},
  doi          = {10.1007/BFB0020972},
  bibsource    = {dblp computer science bibliography, https://dblp.org}
}

@inproceedings{DBLP:conf/icfem/WangZZ15,
  author       = {Shuling Wang and
                  Naijun Zhan and
                  Liang Zou},
  editor       = {Michael J. Butler and
                  Sylvain Conchon and
                  Fatiha Za{\"{\i}}di},
  title        = {An Improved {HHL} Prover: An Interactive Theorem Prover for Hybrid
                  Systems},
  booktitle    = {Formal Methods and Software Engineering - 17th International Conference
                  on Formal Engineering Methods, {ICFEM} 2015, Paris, France, November
                  3-5, 2015, Proceedings},
  series       = {Lecture Notes in Computer Science},
  volume       = {9407},
  pages        = {382--399},
  publisher    = {Springer},
  year         = {2015},
  doi          = {10.1007/978-3-319-25423-4\_25},
  bibsource    = {dblp computer science bibliography, https://dblp.org}
}

@inproceedings{ARCH25:ARCH_COMP25_Category_Report,
  author    = {Luca Geretti and Julien Alexandre Dit Sandretto and Matthias Althoff and Luis Benet and Pieter Collins and Marcelo Forets and Stefan Mitsch and Ismail Patel and Maximilian Perschl and Christian Schilling and Joris Tillet},
  title     = {ARCH-COMP25 Category Report: Continuous and Hybrid Systems with Nonlinear Dynamics},
  booktitle = {Proceedings of 12th Int. Workshop on Applied Verification for Continuous and Hybrid Systems},
  editor    = {Goran Frehse and Matthias Althoff},
  series    = {EPiC Series in Computing},
  volume    = {108},
  publisher = {EasyChair},
  issn      = {2398-7340},
  doi       = {10.29007/7br2},
  pages     = {39-70},
  year      = {2025}}

@INPROCEEDINGS{DBLP:conf/rssrail/MitschGBGP17,
  author    = {Stefan Mitsch and
               Marco Gario and
               Christof J. Budnik and
               Michael Golm and 
               Andr{\'{e}} Platzer},
  title     = {Formal Verification of Train Control with
               Air Pressure Brakes},
  booktitle = {RSSRail},
  longbooktitle = {RSSRail 2017: Reliability, Safety,
               and Security of Railway Systems},
  year      = {2017},
  pages     = {173-191},
  editor    = {Alessandro Fantechi and
               Thierry Lecomte and
               Alexander Romanovsky},
  volume    = {10598},
  doi       = {10.1007/978-3-319-68499-4\_12},
  series    = {LNCS},
  publisher = {Springer},
}

@INPROCEEDINGS{DBLP:conf/tacas/GhorbalP14,
  author    = {Khalil Ghorbal and
               Andr{\'{e}} Platzer},
  title     = {Characterizing Algebraic Invariants by
               Differential Radical Invariants},
  booktitle = {TACAS},
  year      = {2014},
  pages     = {279-294},
  doi       = {10.1007/978-3-642-54862-8\_19},
  editor    = {Erika {\'A}brah{\'a}m and
               Klaus Havelund},
  longbooktitle = {Tools and Algorithms for the Construction
               and Analysis of Systems - 20th International
               Conference, TACAS 2014, Held as Part of the
               European Joint Conferences on Theory and
               Practice of Software, ETAPS 2014, Grenoble,
               France, April 5-13, 2014. Proceedings},
  publisher = {Springer},
  series    = {LNCS},
  volume    = {8413},
  isbn      = {978-3-642-54861-1},
}

@INPROCEEDINGS{DBLP:conf/sas/GhorbalSP14,
  author    = {Khalil Ghorbal and
               Andrew Sogokon and
               Andr{\'{e}} Platzer},
  title     = {Invariance of Conjunctions of Polynomial
               Equalities for Algebraic Differential
               Equations},
  booktitle = {SAS},
  year      = {2014},
  pages     = {151-167},
  editor    = {Markus M{\"u}ller-Olm and
               Helmut Seidl},
  longbooktitle = {Static Analysis - 21th International
               Symposium, SAS 2014, Munich, Germany,
              September 11-13, 2014. Proceedings},
  publisher = {Springer},
  series    = {LNCS},
  volume    = {8723},
  isbn      = {978-3-319-10935-0},
  doi       = {10.1007/978-3-319-10936-7\_10}
}

@article{oracular-programming-2025,
  title={Oracular Programming: A Modular Foundation for Building LLM-Enabled Software},
  author={Laurent, Jonathan and Platzer, Andr{\'e}},
  journal={arXiv preprint arXiv:2502.05310},
  year={2025}
}

@article{Polu2020GenerativeLM,
  title={Generative Language Modeling for Automated Theorem Proving},
  author={Stanislas Polu and Ilya Sutskever},
  journal={ArXiv},
  year={2020},
  volume={abs/2009.03393},
}

@inproceedings{10.1145/3713082.3730382,
author = {Qin, Jianxing and Du, Alexander and Zhang, Danfeng and Lentz, Matthew and Zhuo, Danyang},
title = {Can Large Language Models Verify System Software? A Case Study Using FSCQ as a Benchmark},
year = {2025},
isbn = {9798400714757},
publisher = {Association for Computing Machinery},
address = {New York, NY, USA},
doi = {10.1145/3713082.3730382},
booktitle = {Proceedings of the 2025 Workshop on Hot Topics in Operating Systems},
pages = {34–41},
numpages = {8},
location = {Banff, AB, Canada},
series = {HotOS '25}
}

@inproceedings{hilbert,
title = {Hilbert: Recursively Building Formal Proofs with Informal Reasoning},
booktitle = {NeurIPS Workshop},
author = {Sumanth Varambally and Thomas Voice and Yanchao Sun and Zhifeng Chen and Rose Yu and Ke Ye},
year = {2025},
URL = {https://arxiv.org/abs/2509.22819}
}

@inproceedings{DBLP:conf/lpar/PaulsonB10,
  author       = {Lawrence C. Paulson and
                  Jasmin Christian Blanchette},
  editor       = {Geoff Sutcliffe and
                  Stephan Schulz and
                  Eugenia Ternovska},
  title        = {Three years of experience with Sledgehammer, a Practical Link Between
                  Automatic and Interactive Theorem Provers},
  booktitle    = {The 8th International Workshop on the Implementation of Logics, {IWIL}
                  2010, Yogyakarta, Indonesia, October 9, 2011},
  series       = {EPiC Series in Computing},
  volume       = {2},
  pages        = {1--11},
  publisher    = {EasyChair},
  year         = {2010},
  doi          = {10.29007/36DT},
  bibsource    = {dblp computer science bibliography, https://dblp.org}
}

@inproceedings{dafny,
  author       = {K. Rustan M. Leino},
  editor       = {Edmund M. Clarke and
                  Andrei Voronkov},
  title        = {Dafny: An Automatic Program Verifier for Functional Correctness},
  booktitle    = {Logic for Programming, Artificial Intelligence, and Reasoning - 16th
                  International Conference, LPAR-16, Dakar, Senegal, April 25-May 1,
                  2010, Revised Selected Papers},
  series       = {Lecture Notes in Computer Science},
  volume       = {6355},
  pages        = {348--370},
  publisher    = {Springer},
  year         = {2010},
  url          = {https://doi.org/10.1007/978-3-642-17511-4\_20},
  doi          = {10.1007/978-3-642-17511-4\_20},
  bibsource    = {dblp computer science bibliography, https://dblp.org}
}

@article{verus,
  author       = {Andrea Lattuada and
                  Travis Hance and
                  Chanhee Cho and
                  Matthias Brun and
                  Isitha Subasinghe and
                  Yi Zhou and
                  Jon Howell and
                  Bryan Parno and
                  Chris Hawblitzel},
  title        = {Verus: Verifying Rust Programs using Linear Ghost Types},
  journal      = {Proc. {ACM} Program. Lang.},
  volume       = {7},
  number       = {{OOPSLA1}},
  pages        = {286--315},
  year         = {2023},
  url          = {https://doi.org/10.1145/3586037},
  doi          = {10.1145/3586037},
  bibsource    = {dblp computer science bibliography, https://dblp.org}
}

\appendix

\section{Details of Verification Experiments}
\label{app:verification-details}
We report further details on the verification experiments.
All experiments consist of two sets of four parallel, independent pipeline runs (where each pipeline run consists of many iterations of the inner verification loop). These four parallel runs terminate only when any one of them succeeds, or when they collectively run out of a twelve dollar budget.
The choice of twelve dollars is because beyond this point, we find reasoning to diverge, suggesting limited benefit to further iterations of the proving loop.
Thus for each case study the verification experiment allows eight attempts with a total budget of twenty four dollars.

Summing over the eight runs, we list the number of LLM calls made and the dollar budget consumed for LLM calls in \rref{tab:verification-numcalls-results}.
Smaller numbers indicate that the proof was reached early.
The 12.00 in the Lotka Volterra case study indicates that the proof was only reached in the second set of runs.
We found that the time taken to run was dominated by LLM reasoning model response time.

\begin{table}[h]
\centering
\begin{tabular}{|l|c|c|c|}
\hline
\textbf{Model} & \textbf{Number of LLM Calls} & \textbf{Dollar Cost} & \textbf{Time (minutes)} \\
\hline
Coolant & 89 + 64 & 6.9 + 5.8 & 47.0+39.9 \\
Train & 87 + 63 & 7.0 + 5.2 & 43.9+31.4 \\
Lotka Volterra & 176 + 57 & 12.0 + 3.6 & 93.0+27.5 \\
Chemical Reaction & 31 + 49 & 2.0 + 3.7 & 16.7+31.7 \\
Van der Pol & 44 + 59 & 3.3 + 5.5 & 26.9+41.3 \\
\hline
\end{tabular}
\caption{Verification experiments: Number of LLM calls, time to run, and dollar cost for two sets of four parallel runs per case study.}
\label{tab:verification-numcalls-results}
\end{table}

In the ablation where we disable the summary step, only the Chemical Reaction and Van der Pol case studies complete successfully.
For Chemical Reaction, the number of LLM calls actually decreases, while for Van der Pol, it increases (\rref{tab:results-no-summary}).

\begin{table}
\centering
\begin{tabular}{|l|c|c|}
\hline
\textbf{Model} & \textbf{Number of LLM Calls} & \textbf{Dollar Cost} \\
\hline
Chemical Reaction & 28 + 24 & 2.7 + 2.4 \\
Van der Pol & 51 + 99 & 4.9 + 10.3 \\
\hline
\end{tabular}
\caption{Verification experiments without summary step: Number of LLM calls and dollar cost for two sets of four parallel runs per case study. The costs for all other case studies exceeds the budget.}
\label{tab:results-no-summary}
\end{table}

\section{Details of Synthesis Experiments}
\label{app:synthesis-details}
We discuss further details on the synthesis experiments.
All experiments consist of four parallel pipeline runs that terminate when any one of them succeeds, or when they collectively run out of a ten dollar budget for LLM calls.
The ten dollar budget, again, is chosen because of diminishing returns beyond this point.
In addition to the ablation discussed in the main text, we also tried an ablation with the initial analysis and control strategy planning step removed.
The result was that three out of the five case studies received a verified subvalue map, but one of these was different than the desired solution per the provided guideline.
In either case, this is lower than the four problems solved by the full pipeline, indicating that the initial analysis segment is helpful.

Like in \rref{app:verification-details}, we next report the number of LLM calls, dollar cost, and time taken for each case study by the full pipeline in \rref{tab:synthesis-numcalls-results}.

\begin{table}[h]
\centering
\begin{tabular}{|l|c|c|c|}
\hline
\textbf{Model} & \textbf{Number of LLM Calls} & \textbf{Dollar Cost} & \textbf{Time (minutes)} \\
\hline
Coolant & 115 & 10.0 & 83.8 \\
Chemical Reaction & 16 & 0.7 & 6.8 \\
Van der Pol & 26 & 0.7 & 6.3 \\
Lotka Volterra & 58 & 2.8 & 23.7 \\
Train & 89 & 5.5 & 41.4 \\
\hline
\end{tabular}
\caption{Synthesis experiments: Number of LLM calls, time to run, and dollar cost for each case study.}
\label{tab:synthesis-numcalls-results}
\end{table}

In the ablation without proving or backtracking (recovery choice), as the LLM does not guide backtracking, we need a separate \emph{search policy} to decide how to do retries.
We use depth-first search with a maximum branching factor of three.
That is, when a subvalue check fails, we first retry it twice, and if both fail, we backtrack to the second most recent subvalue guess and retry it, and so on.
The two benchmarks that succeed in this ablation are Chemical Reaction and Van der Pol, with the number of LLM calls and dollar costs listed in \rref{tab:synthesis-no-recovery-choice}.
\begin{table}
\centering
\begin{tabular}{|l|c|c|}
\hline
\textbf{Model} & \textbf{Number of LLM Calls} & \textbf{Dollar Cost} \\
\hline
Chemical Reaction & 17 & 0.9 \\
Van der Pol & 22 & 0.7 \\
\hline
\end{tabular}
\caption{Synthesis experiments without recovery choice: Number of LLM calls and dollar cost for each case study. The costs for all other case studies exceeds the budget.}
\label{tab:synthesis-no-recovery-choice}
\end{table}

\section{Tactic for \rref{model:lotka-volterra}}
\label{app:lv-tactic}
The tactic that the LLM discovers to prove \rref{model:lotka-volterra} is as follows:

\begin{verbatim}
unfold;
loop("x>=xmin & y>=ymin & x<=g/d & y<=a/b & g/d>=xmin &
      a/b>=ymin & a>0 & b>0 & d>0 & g>0", 1); <(
  /* init */
  auto; print("Init subgoal after auto"),
  /* step */
  unfold; unfold;
  print("Step: after exposing existentials/diamond");
  /* witness for xadd and discharge ?xadd>=0 */
  existsR(1, "g/d - x");
  unfold; <(
    QE("Z3"); print("Step: discharged ?xadd>=0"),
    /* continue with yadd */
    unfold; existsR(1, "a/b - y");
    unfold; <(
      QE("Z3"); print("Step: discharged ?yadd>=0"),
      /* push deterministic assignments into box over ODE */
      unfold; unfold;
      print("Step: after pushing assignments");
      /* split post x>=xmin & y>=ymin */
      boxAnd(1); andR(1); <(
        /* prove [ODE] x>=xmin */
        dC("x=g/d & y=a/b", 1); <(
          /* use: with equality in domain, finish */
          cut("1=1"); <(
            dW(1); QE("Z3"),
            QE("Z3") using "1=1"
          ),
          /* show: [{LV}](x=g/d & y=a/b) via dRI */
          dC("d*x - g = 0 & a - b*y = 0", 1); <(
            /* use: from q1=0 & q2=0 derive result */
            cut("1=1"); <(
              dW(1); QE("Z3"),
              QE("Z3") using "1=1"
            ),
            /* show: [{LV}](q1=0 & q2=0) */
            boxAnd(1); andR(1); <(
              dRI(1); QE("Z3"),
              dRI(1); QE("Z3")
            )
          )
        ),
        /* prove [ODE] y>=ymin: symmetric */
        dC("x=g/d & y=a/b", 1); <(
          cut("1=1"); <(
            dW(1); QE("Z3"),
            QE("Z3") using "1=1"
          ),
          /* same fix for the show branch here */
          dC("d*x - g = 0 & a - b*y = 0", 1); <(
            cut("1=1"); <(
              dW(1); QE("Z3"),
              QE("Z3") using "1=1"
            ),
            boxAnd(1); andR(1); <(
              dRI(1); QE("Z3"),
              dRI(1); QE("Z3")
            )
          )
        )
      )
    )
  ),
  /* post */
  auto; print("Post subgoal after auto")
)
\end{verbatim}

Here, the seemingly redundant \texttt{cut("1=1")} steps are a workaround to a bug in \KeYmaeraX that incorrectly prevents application of \texttt{dW} on certain subgoal shapes. The print statements are for the LLM to track progress and get information about the proof state in case of failure. The use of \texttt{QE("Z3")} indicates that the Z3 SMT solver is used to discharge the corresponding proof obligation. In the prompt, the LLM is encouraged to break down the proof rather than using auto so that the proof remains stable despite changes to the automation and quicker to check.

\section{Other Case Studies}
\label{app:other-case-studies}

\paragraph{Case Study: Coolant.}
Consider a coolant system in a nuclear power plant that must achieve the control objective of absorbing \emph{at least} \(\minabsorbed\) units of heat while expending \emph{at most} \(\maxdischarged\) units of coolant in a given time frame.
The coolant system controller chooses how much coolant to send through the system, and then goes to sleep for some amount of time bounded by maximum latency \(\latency\), repeatedly polling in a control loop\footnote{%
We use the typographic convention of monospaced font for \texttt{variables} that change over the course of a game and math font for each \emph{symbolic parameter} \(P\) that can represent any real number which remains constant over the course of a game.}.
\rref{model:running-example-verification} shows the formula modeling this system, that we must prove valid.
We must reason directly in terms of symbolic parameters such as required absorption \(\minabsorbed\) and maximum flow rate \(F\).
The system has different control modes, can run for arbitrarily long (because of loop iterations), and dynamical behavior qualitatively changes depending on the initial temperature, all of which our reasoning will precisely account for.

\begin{model}
  \setcounter{modelline}{0}
  \caption{Case study: Coolant System. \(\phi_\init\) is a long formula listed in \rref{fig:loop-pre}. It is not provided in the synthesis variant of the problem, and must be computed.}
  \label{model:running-example-verification}
  \begin{align*}
    \text{\assumptions} &\,\left| 
      \begin{aligned}
        &\mline{line:coolant-assumptions-1} \textrm{\textcolor{gray}{\(\phi_\init\)}} \land T > 0 \land F > 0 \land \minabsorbed > 0 \land \maxdischarged > 0 \land c_1 > 0 \land c_2 > 0 \limply\\
      \end{aligned}
    \right.\\
    \text{\ctrl} &\,\left| \mline{line:control} \langle ( \ \ (\ptest{\absorbed < \minabsorbed}^d \seq \flow := F) \cap (\absorbed \geq \minabsorbed^d \seq \flow := 0) \seq
      \right. \\
    \text{\plant} &\,\left| \mline{line:plant} \phantom{\langle\ } \quad \clock := 0; \left\{ \begin{array}{l} \absorbed' = \flow \cdot c_1 \cdot \tempDiff, \\
      \discharged' = \flow, \\ \tempDiff' = -\flow \cdot c_2 \cdot \tempDiff + \generated, \\ \clock' = 1, \\ \timer' = -1\ \&\ \clock \leq T \end{array} \right\}^d
        \right.\\
    \text{\safe} &\,\left| \mline{line:winning-condition} \phantom{\langle\ } )^\times \rangle (\examplewinningc)
      \right.
  \end{align*}
\end{model}

\paragraph{\rref{model:running-example-verification} Explanation.}
We provide an explanation of the \dGL model for this case study.
\rref{line:coolant-assumptions-1} shows initial assumptions expressing physical constraints, e.g., that the maximum flow rate \(F\) is positive.
Formula \(\phi_\init\) is a complex formula listed in \rref{fig:loop-pre} representing initial conditions that make safe control possible: the discharge limit should not already have been exceeded, and if the heat absorption quota is not already met, then there is enough time and discharge capacity left to still meet it.
At \rref{line:control}, the control loop begins.

\begin{figure}[t]
  \begin{equation}
    \label{eq:loop-pre}
    \begin{aligned}
    &\preci{\invariant} \equiv \overbracket{\discharged \leq \maxdischarged}^{\text{discharge within limit}} \land 
    \overbracket{
      \maxflow > 0
      \land c_1 > 0
      \land c_2 > 0
      \land \latency > 0}^{\text{physics constraints}} \land\ (\\
              &\quad\overbrace{\absorbed \geq \minabsorbed}^{\text{already absorbed enough}}
                \lor 
              (\overbrace{\timer \geq 0}^{\text{there's time left}}
                    \land\ (\\
                    &\begin{aligned}
                    &\text{Current} \\
                    & \tempDiff \\
                    &\text{lower} \\
                    &\text{bounds} \\
                    &\text{all future} \\
                    & \tempDiff
                    \end{aligned}
                    \left|
                    \begin{aligned}
                      &(\overbrace{\tempDiff \leq \generated/(\maxflow \cdot c_2)}^{\text{\(\tempDiff\) only increases}} \land \overbrace{c_1 \cdot \tempDiff \cdot (\maxdischarged -}^{\text{can absorb enough \(\cdots\)}}\\
                          &  \overbrace{\discharged - \maxflow \cdot \latency) \geq \minabsorbed - \absorbed}^{\text{\(\cdots\) without exceeding discharge budget}} \land  \\
                        & \overbrace{\timer \cdot c_1 \cdot \tempDiff \cdot \maxflow \geq \minabsorbed - \absorbed}^{\text{can absorb enough before deadline}} ) \lor
                    \end{aligned}
                    \right.\\
                    &\begin{aligned}
                    & \frac{\generated}{\maxflow \cdot c_2} \\
                    &\text{lower} \\
                    &\text{bounds} \\
                    & \tempDiff
                    \end{aligned}
                    \left|
                    \begin{aligned}
                      &(\overbrace{\tempDiff > \generated/(\maxflow \cdot c_2)}^{\text{\(\tempDiff\) bound}} \land \overbrace{c_1 \cdot (\generated/(\maxflow \cdot c_2))\cdot}^{\text{can absorb enough \(\cdots\)}} \\
                        & \overbrace{(\maxdischarged - \discharged - \maxflow \cdot \latency) \geq \minabsorbed - \absorbed}^{\text{ \(\cdots\) without exceeding discharge budget}} \land  \\
                      & \overbrace{\timer \cdot c_1 \cdot \generated/c_2 \geq \minabsorbed - \absorbed}^{\text{can absorb enough before deadline}}))))
                    \end{aligned}
                    \right.\\
    \end{aligned}
  \end{equation}
  \caption{Formula \(\phi_\init\) for the Coolant case study.}
  \label{fig:loop-pre}
\end{figure}

The controller assigns a coolant flow rate.
The choice \(\flow := F\) is chosen in states where the absorption quota has not yet been met (\(\absorbed < \minabsorbed\)), and the choice \(\flow := 0\) is chosen otherwise.
At \rref{line:plant}, the controller goes to sleep and the state evolves per the physical dynamics, where the heat absorbed is proportional to temperature difference between the coolant and the system, total coolant discharged is tracked by variable \(\discharged\), the system generates heat at rate \(\generated\), and the auxiliary \(\timer\) variable counts down to when the system must have met its absorption obligations.
The controller stays asleep for at most \(\latency\) time units, though it may wake up earlier, with variable \(\clock\) tracking time elapsed.
Finally, at \rref{line:winning-condition}, the control loop closes, with the syntax \((\alpha)^\times\).
The control loop repeats arbitrarily many times, and we must verify that all possible number of iterations are safe.
The modal formula \(\langle\alpha\rangle \phi\) is true when \emph{all} possible runs result in a state satisfying postcondition \(\phi\).
Therefore, the overall formula of \rref{model:running-example-verification} is valid when, given the assumptions of \rref{line:coolant-assumptions-1}, all executions lead to a state where the control objective is satisfied, that is, \(\examplewinningc\) holds.
This case study is an example of a hybrid \emph{system}, since there is no adversarial game play.

\paragraph{Case Study: Train.}
A train must stop before its \emph{end of motion authority} \(\epos\), beyond which may lie other trains.
The train can brake using two mechanisms: a normal brake that decelerates at most at rate \(\bzero\), and an air brake that ramps up deceleration gradually, as pressure leaks from brake pipes, at rate \(\pressurechangerate\) until it reaches maximum deceleration \(\apb\).
This case study is inspired by a previous case study on verification of trains with air brakes \cite{DBLP:conf/rssrail/MitschGBGP17}, but with the addition of air brakes \cite{train} and adversarial game dynamics.

\begin{model}
  \setcounter{modelline}{0}
  \caption{Case study: Train Control. \textcolor{gray}{Gray} parts are not included in the synthesis variant of the problem, and must be computed. \textcolor{brown}{Brown} text is not a part of the verification problem, but added to the synthesis problem.}
  \label{model:train}
  \begin{align*}
    \text{\assumptions} &\,\left| 
      \begin{aligned}
        \mline{line:train-assumptions-1} &\textrm{\textcolor{gray}{$\epos - \trainpos - \frac{\vel^2}{2 \cdot \bzero} > 0 \land $}} \azero > 0 \land \bzero > 0 \land \aone < 0 \land \atwo < 0 \land T > 0  \\
        \mline{line:train-assumptions-2} &\quad \land \apb < 0 \land \pressurechangerate < 0 \land \vel \geq 0 \textrm{\textcolor{brown}{$\land \brakeslope=0 \land \airbrake=0$}} \limply\\
      \end{aligned}
    \right.\\
    \text{\ctrl} &\,\left| 
      \begin{aligned}
        \mline{line:train-control-1} &\langle ( \ t := 0 \seq \\
        \mline{line:train-control-2} &\phantom{\langle ( }\ (\ (\trainacc := *; \ptest{-\bzero \leq \trainacc < \azero} \seq \\
        \mline{line:train-control-3} &\phantom{\langle ( ( }\ \ \ \brakeslope := 0 \seq \airbrake := 0)\ \cup \\
        \mline{line:train-control-4} &\phantom{\langle ( (}\ \ \ (\trainacc := -\bzero \seq \brakeslope := \pressurechangerate) \ )^d \seq
      \end{aligned}
    \right. \\
    \text{\plant} &\,\left| 
      \begin{aligned}
        \mline{line:train-plant-1} &\phantom{\langle ( } \left( \left\{ \begin{array}{l} \trainpos' = \vel, \\ \vel' = \trainacc + \aone \cdot \vel + \atwo \cdot \vel^2 + \airbrake, \\ \airbrake' = \brakeslope, \\ t' = 1 \ \&\ t \leq T \land \vel \geq 0 \land \airbrake \geq \apb \end{array} \right\}^d \right. \cap \\
        \mline{line:train-plant-2} &\phantom{\langle ((( }\left. \left\{ \begin{array}{l} \trainpos' = \vel, \\ \vel' = \trainacc + \aone \cdot \vel + \atwo \cdot \vel^2 + \airbrake, \\ \airbrake' = 0, \\ t' = 1 \ \&\ t \leq T \land \vel \geq 0 \land \airbrake \leq \apb \end{array} \right\}^d \right)^\times
      \end{aligned}
        \right.\\
    \text{\safe} &\,\left| \mline{line:train-winning-condition} \phantom{\langle })^\times \rangle (\epos - \trainpos > 0)
      \right.
  \end{align*}
\end{model}

\paragraph{\rref{model:train} Explanation.} \rref{model:train} shows the \dGL formula modeling the train control problem via a game between Angel (the train controller) and Demon (environmental uncertainties).
Angel chooses between the control modes of normal operation and air brake application (\rref{line:train-control-3}).
In normal operation, she may choose any acceleration between \(-\bzero\) and \(\azero\).
Then, Demon chooses how long the train will run before Angel gets to re-evaluate her decision, bounded by a maximum time latency \(T\) (\rref{line:train-plant-1} and \ref{line:train-plant-2}).
The physical dynamics operate in two modes, depending on whether the air brakes are ramping up (\rref{line:train-plant-1}) or have stabilized at their maximum (\rref{line:train-plant-2}).
This is modeled by ODEs for each mode within an inner loop that allows switching between them.
Demon controls the number of iterations of the outer control loop (\rref{line:train-control-1} to \ref{line:train-plant-2}), forcing Angel to maintain safety for arbitrarily long.
The overall modal formula of form \(\langle\alpha\rangle\phi\) is true in states where Angel has a strategy to play so that regardless of how Demon plays, she will win, that is, the train controller has a way to maintain safety (\rref{line:train-winning-condition}).
To reason correctly about this problem we must again reason over the nonlinear dynamics of the train, and ensure safety over unbounded time horizons.
The mode switching inner-loop structure adds a further layer of complexity.

\paragraph{Case Study: Van der Pol.}
A pair of coupled Van der Pol oscillators must be set so that the sum of their positions \vdpxone and \vdpxtwo never exceeds a given bound \vdpxmax.
The system dynamics are governed by parameters \vdpm and \vdpb.
The oscillators can run arbitrarily long, and safety must be ensured throughout.
This case study requires reasoning about nonlinear dynamics over unbounded time horizons.
\rref{model:vanderpol} shows the full control formula.
Angel sets initial conditions at \rref{line:vdp-ctrl-1} and \rref{line:vdp-ctrl-2}, then Demon chooses how long the oscillators run at \rref{line:vdp-plant}, after which safety is checked at \rref{line:vdp-winning-condition}.
Then Demon chooses how long to run the oscillators at \rref{line:vdp-plant}, and whether to repeatedly further run them in the loop at \rref{line:vdp-winning-condition}.
Angel wins in the end if the sum of the positions never exceeds \vdpxmax (\rref{line:vdp-winning-condition}).
This case study is inspired by one appearing in ARCH-COMP \cite{ARCH25:ARCH_COMP25_Category_Report}, but is modified to add adversarial game dynamics and a loop structure that makes formal reasoning more challenging.

\begin{model}
  \setcounter{modelline}{0}
  \caption{Case study: Coupled Van der Pol Oscillators. \textcolor{gray}{Gray} parts are not included in the synthesis variant of the problem, and must be computed.}
  \label{model:vanderpol}
  \begin{align*}
    \text{\assumptions} &\,\left| \mline{line:vdp-assumptions} \textrm{\textcolor{gray}{\(\vdpxmax > 0 \limply \)}}
    \right.\\
    \text{\ctrl} &\,\left| 
      \begin{aligned}
        \mline{line:vdp-ctrl-1} &\langle \vdpxone := * \seq \vdpxtwo := * \seq \vdpyone := * \seq \vdpytwo := * \seq \\
        \mline{line:vdp-ctrl-2} &\phantom{\langle} \ptest{\vdpm > 0 \land \vdpb > 0}^d \seq ( \ 
      \end{aligned}
    \right.\\
    \text{\plant} &\,\left| \mline{line:vdp-plant} \phantom{\langle (} \left\{ \begin{array}{l} \vdpxone' = \vdpyone, \\ \vdpyone' = \vdpm \cdot (1 - \vdpxone^2) \cdot \vdpyone + \vdpb \cdot (\vdpxtwo - \vdpxone) - \vdpxone, \\ \vdpxtwo' = \vdpytwo, \\ \vdpytwo' = \vdpm \cdot (1 - \vdpxtwo^2) \cdot \vdpytwo - \vdpb \cdot (\vdpxtwo - \vdpxone) - \vdpxtwo \end{array} \right\}^d
        \right.\\
    \text{\safe} &\,\left| \mline{line:vdp-winning-condition} \phantom{\langle} )^\times \rangle (\vdpxone + \vdpxtwo \leq \vdpxmax)
      \right.
  \end{align*}
\end{model}

\paragraph{Case Study: Chemical Reaction.}
An exothermic chemical reaction \(A + B \rightarrow C\) is taking place.
The reaction rates depend on the temperature of the reactor \(\chemTemp\), which increases in proportion to the reaction rate.
The reaction rate also depends on the concentrations of the reactants \(A\) and \(B\).
The safety requirement is that the temperature never exceeds a maximum safe temperature \(\chemTmax\).
A time-triggered controller can turn the reaction on and off, getting to revise its decision in at most every \(T\) time units.
However, we do not have direct access to the controller, and instead conservatively safeguard the system by not putting in too much reactant to begin with.
The verification challenge is to check whether regardless of what the controller does, safety is always ensured.
The physical reactor setup in this case study is similar to one from the literature \cite{10.1007/978-3-031-15008-1_8}, but with an important change in setup: the controller is not directly accessible, and we want to verify safety for all possible controller actions.

\rref{model:chemical-reaction} shows the full control formula.
We make some initial assumptions about the quantities of reactant and signs of the rate constants at \rref{line:chem-assumptions-1} and \rref{line:chem-assumptions-2}.
Then Demon chooses whether to turn the reaction on or off at \rref{line:chem-ctrl-1} in a time-triggered control loop, also controlled by Demon.
In the end, Angel wins if the temperature never exceeds \chemTmax (\rref{line:chem-winning-condition}).
This system again requires reasoning about complex dynamics over different, arbitrarily alternating control modes over unbounded iterations.
The solution to the dynamics lies outside the decidable (polynomial) fragment of real arithmetic, requiring non-trivial deductive reasoning about differential equations via invariants. 

\begin{model}
  \setcounter{modelline}{0}
  \caption{Case study: Chemical Reaction Control. \textcolor{gray}{Gray} parts are not included in the synthesis variant of the problem, and must be computed.}
  \label{model:chemical-reaction}
  \begin{align*}
    \text{\assumptions} &\,\left| 
      \begin{aligned}
        \mline{line:chem-assumptions-1} & \chemkA > 0 \land \chemkB > 0 \land \chemkC > 0 \land \chemkT \geq 0 \land T > 0 \land \chemA \geq 0 \land \chemB \geq 0 \textrm{\textcolor{gray}{\(\land \chemC \geq 0\)}} \\
        \mline{line:chem-assumptions-2} &\quad \textrm{\textcolor{gray}{\( \land \chemTemp \geq 0 \land (\chemkB \cdot \chemA \leq \chemkA \cdot \chemB \land \chemkA \cdot \chemTemp + \chemkT \cdot \chemA \leq \chemkA \cdot \chemTmax \)}} \\
        \mline{line:chem-assumptions-3} &\quad \quad \textrm{\textcolor{gray}{\( \lor \chemkA \cdot \chemB \leq \chemkB \cdot \chemA \land \chemkB \cdot \chemTemp + \chemkT \cdot \chemB \leq \chemkB \cdot \chemTmax) \)}} \limply\\
      \end{aligned}
    \right.\\
    \text{\ctrl} &\,\left| 
      \begin{aligned}
        \mline{line:chem-ctrl-1} &\langle ( \ ( \ \chemisOn := 0 \ \cap \ \chemisOn := 1 \ ) \seq \clock := 0 \seq
      \end{aligned}
    \right. \\
    \text{\plant} &\,\left| \mline{line:chem-plant} \phantom{\langle (}\ \left\{ \begin{array}{l} \chemA' = -\chemA \cdot \chemB \cdot \chemTemp \cdot \chemisOn \cdot \chemkA, \\ \chemB' = -\chemA \cdot \chemB \cdot \chemTemp \cdot \chemisOn \cdot \chemkB, \\ \chemC' = \chemA \cdot \chemB \cdot \chemTemp \cdot \chemisOn \cdot \chemkC, \\ \chemTemp' = \chemA \cdot \chemB \cdot \chemTemp \cdot \chemisOn \cdot \chemkT, \\ \clock' = 1 \ \&\ \clock \leq T \land \chemA \geq 0 \land \chemB \geq 0 \land \chemC \geq 0 \land \chemTemp \geq 0 \end{array} \right\}^d
        \right.\\
    \text{\safe} &\,\left| \mline{line:chem-winning-condition} \phantom{\langle} )^\times \rangle (\chemTemp \leq \chemTmax)
      \right.
  \end{align*}
\end{model}

\paragraph{Note:} In the synthesis variants of problems, assumptions are written not in the form 
\[ \textrm{assumptions}\limply \langle \alpha \rangle \textrm{postcondition},\]
but rather as
\[\langle ? \textrm{assumptions (excluding gray parts)}^d \seq \alpha \rangle \textrm{postcondition}.\]
This does not make a semantic difference, but allows subvalue map computation to account for these assumptions as it passes over the assumptions subgame, and produce a subvalue for the overall game that does not redundantly include the assumptions.

In the \emph{synthesis} variants of the problems, the following guidelines are used:
\begin{itemize}
  \item For \emph{Lotka Volterra}: Set the system to its equilibrium point as soon as possible, and assume what you must so that the equilibrium point is good enough to maintain the post condition.
  \item For \emph{train}: Find the control solution for when the train can stop by braking continuously, even conservatively ignoring rolling resistance and air brakes.
  \item For \emph{chemical reaction}: Find the control solution for the situation where the reactant runs out before temperature can become too large.
  \item For \emph{coolant}: Find the solution where initially enough heat has not been absorbed, but over time, enough is absorbed before the timer requirement or discharge capacity can be exceeded.
  \item For \emph{Van der Pol}: Keep the two oscillators in antiphase, maintaining the postcondition this way.
\end{itemize}

\section{Differential Game Logic}
\label{app:appendix-dgl}
This appendix recalls the syntax and semantics of differential game logic (\dGL) \cite{Platzer18}.
\rref{fig:dGL-calculus} summarizes the axioms and proof rules of \dGL.
A full explanation is in \cite{Platzer18}.
The dual operator $\alpha^d$ switches the role of players in $\alpha$. $(\alpha \cup \beta)^d$, for example, is the same as $\alpha^d \cap \beta^d$.

The set of \dGL~formulas is generated by the following grammar (where \m{{\sim}\in\{<,\leq,=,\geq,>\}} and $\theta_1,\theta_2$ are arithmetic expressions in~\m{+,-,\cdot,/} over the reals, \(x\) is a variable, \(\alpha\) is a hybrid game):
\[
\phi \coloneqq \theta_1 \sim \theta_2
\mid \lnot \phi
\mid \phi \land \psi
\mid \phi \lor \psi
\mid \phi \limply \psi
\mid \lforall{x}{\phi}
\mid \lexists{x}{\phi}
\mid \dbox{\alpha}{\phi}
\mid \ddiamond{\alpha}{\phi}
\]

The syntax for games is as follows.
\begin{equation}
\label{eq:dgl-grammar}
\begin{aligned}
   \alpha \equiv\  x:=e \mid \alpha;\beta \mid \ptest{Q} &\mid \alpha^\ast \mspace{4mu} \mid \alpha\cup\beta \mid x:=\ast \mspace{2.5mu} \mid \{x'=f(x) \& Q\}\phantom{^d}\\
   \mid \mspace{1.5mu} ! \mspace{1.5mu} Q &\mid \alpha^\times \mid \alpha\cap\beta \mid x:=\otimes \mid \{x'=f(x) \& Q\}^d
\end{aligned}
\end{equation}

The semantics \cite{Platzer18} of \dGL is as follows.

{\renewcommand{\D}[2][]{\ifthenelse{\equal{#1}{}}{#2'}{\frac{d#2}{d{#1}}}}%
\newcommand{\I}{\ensuremath{\phi}}

\begin{definition}[\dGL semantics] \label{def:dGL-formula-semantics}
The \emph{semantics of a \dGL formula} $\phi$ is the subset \m{\imodel{\I}{\phi}\subseteq\linterpretations{\Sigma}{V}} of states in which $\phi$ is true.
It is defined inductively as follows
\begin{enumerate}
\item \(\imodel{\I}{p(\theta_1,\dots,\theta_k)} = \{\iportray{\I} \in \linterpretations{\Sigma}{V} \with (\ivaluation{\I}{\theta_1},\dots,\ivaluation{\I}{\theta_k})\in\iget[const]{\I}(p)\}\)
\item \(\imodel{\I}{\theta_1\sim\theta_2} = \{\iportray{\I} \in \linterpretations{\Sigma}{V} \with \ivaluation{\I}{\theta_1}\sim\ivaluation{\I}{\theta_2}\} \text{ where }{\sim}\in\{<,\leq,=,\geq,>\}\)
\item \(\imodel{\I}{\lnot\phi} = \scomplement{(\imodel{\I}{\phi})}\)
\item \(\imodel{\I}{\phi\land\psi} = \imodel{\I}{\phi} \cap \imodel{\I}{\psi}\)
\item
{\def\Im{\imodif[state]{\I}{x}{r}}%
\(\imodel{\I}{\lexists{x}{\phi}} =  \{\iportray{\I} \in \linterpretations{\Sigma}{V} \with \iget[state]{\Im} \in \imodel{\I}{\phi} ~\text{for some}~r\in\reals\}\)
}
\item \(\imodel{\I}{\ddiamond{\alpha}{\phi}} = \strategyfor[\alpha]{\imodel{\I}{\phi}}\)
\item \(\imodel{\I}{\dbox{\alpha}{\phi}} = \dstrategyfor[\alpha]{\imodel{\I}{\phi}}\)
\end{enumerate}
A \dGL formula $\phi$ is \emph{valid}, written \m{\entails{\phi}}, iff it is true in all states, i.e.\ \m{\imodel{\I}{\phi}=\linterpretations{\Sigma}{V}}.
\end{definition}

\begin{definition}[Semantics of hybrid games] \label{def:HG-semantics}
The \emph{semantics of a hybrid game} $\alpha$ is a function \m{\strategyfor[\alpha]{\cdot}} that, for each set of Angel's winning states \m{X\subseteq\linterpretations{\Sigma}{V}}, gives the \emph{winning region}, i.e.\ the set of states \m{\strategyfor[\alpha]{X}} from which Angel has a winning strategy to achieve $X$ in $\alpha$ (whatever strategy Demon chooses). It is defined inductively as follows
\begin{enumerate}
\item \(\strategyfor[\pupdate{\pumod{x}{\theta}}]{X} = \{\iportray{\I} \in \linterpretations{\Sigma}{V} \with \modif{\iget[state]{\I}}{x}{\ivaluation{\I}{\theta}} \in X\}\)
\item \(\strategyfor[\pevolvein{\D{x}=\genDE{x}}{\ivr}]{X} = \{\varphi(0) \in \linterpretations{\Sigma}{V} \with
      \varphi(r)\in X\)
      for some $r\in\reals_{\geq0}$ and (differentiable)
      \m{\varphi:[0,r]\to\linterpretations{\Sigma}{V}}
      such that
      \(\varphi(\zeta)\in\imodel{\I}{\ivr}\)
      and
      \m{\D[t]{\,\varphi(t)(x)} (\zeta) =
      \ivaluation{\iconcat[state=\varphi(\zeta)]{\I}}{f(x)}}
      for all \(0\leq\zeta\leq r\}\)
\item \(\strategyfor[\ptest{\ivr}]{X} = \imodel{\I}{\ivr}\cap X\)
\item \(\strategyfor[\pchoice{\alpha}{\beta}]{X} = \strategyfor[\alpha]{X}\cup\strategyfor[\beta]{X}\)
\item \(\strategyfor[\alpha;\beta]{X} = \strategyfor[\alpha]{\strategyfor[\beta]{X}}\)
\item \(\strategyfor[\prepeat{\alpha}]{X} = \capfold\{Z\subseteq\linterpretations{\Sigma}{V} \with X\cup\strategyfor[\alpha]{Z}\subseteq Z\}\)

\item \(\strategyfor[\pdual{\alpha}]{X} = \scomplement{(\strategyfor[\alpha]{\scomplement{X}})}\)
\end{enumerate}
The \emph{winning region} of Demon, i.e.\ the set of states \m{\dstrategyfor[\alpha]{X}} from which Demon has a winning strategy to achieve $X$ in $\alpha$ (whatever strategy Angel chooses) is defined inductively as follows
\begin{enumerate}
\item \(\dstrategyfor[\pupdate{\pumod{x}{\theta}}]{X} = \{\iportray{\I} \in \linterpretations{\Sigma}{V} \with \modif{\iget[state]{\I}}{x}{\ivaluation{\I}{\theta}} \in X\}\)
\item \(\dstrategyfor[\pevolvein{\D{x}=\genDE{x}}{\ivr}]{X} = \{\varphi(0) \in \linterpretations{\Sigma}{V} \with
      \varphi(r)\in X\)
      for all $r\in\reals_{\geq0}$ and (differentiable)
      \m{\varphi:[0,r]\to\linterpretations{\Sigma}{V}}
      such that
      \(\varphi(\zeta)\in\imodel{\I}{\ivr}\)
      and
      \m{\D[t]{\,\varphi(t)(x)} (\zeta) =
      \ivaluation{\iconcat[state=\varphi(\zeta)]{\I}}{\theta}}
      for all $0\leq\zeta\leq r\}$
\item \(\dstrategyfor[\ptest{\ivr}]{X} = \scomplement{(\imodel{\I}{\ivr})}\cup X\)
\item \(\dstrategyfor[\pchoice{\alpha}{\beta}]{X} = \dstrategyfor[\alpha]{X}\cap\dstrategyfor[\beta]{X}\)
\item \(\dstrategyfor[\alpha;\beta]{X} = \dstrategyfor[\alpha]{\dstrategyfor[\beta]{X}}\)
\item \(\dstrategyfor[\prepeat{\alpha}]{X} = \cupfold\{Z\subseteq\linterpretations{\Sigma}{V} \with Z\subseteq X\cap\dstrategyfor[\alpha]{Z}\}\)
\item \(\dstrategyfor[\pdual{\alpha}]{X} = \scomplement{(\dstrategyfor[\alpha]{\scomplement{X}})}\)
\end{enumerate}
\end{definition}
}

\newcommand{\leftrule}{L}
\newcommand{\rightrule}{R}
\newsavebox{\Rval}%
\sbox{\Rval}{$\scriptstyle\mathbb{R}$}
\newcommand{\solf}{y}

\providecommand{\axkey}[1]{\textcolor{vblue}{#1}}%
\cinferenceRuleStore[diamond|$\didia{\cdot}$]{diamond axiom}
{\linferenceRule[equiv]
  {\lnot\dbox{\ausprg}{\lnot \ausfml}}
  {\axkey{\ddiamond{\ausprg}{\ausfml}}}
}
{}
\cinferenceRuleStore[diamondax|$\didia{\cdot}$]{diamond axiom}
{\linferenceRule[equiv]
  {\lnot\dbox{\ausprgax}{\lnot \ausfmlax}}
  {\axkey{\ddiamond{\ausprgax}{\ausfmlax}}}
}
{}
\cinferenceRuleStore[assignb|$\dibox{:=}$]{assignment / substitution axiom}
{\linferenceRule[equiv]
  {p(\genDJ{x})}
  {\axkey{\dbox{\pupdate{\umod{x}{\genDJ{x}}}}{p(x)}}}
}
{}%
\cinferenceRuleStore[assignbax|$\dibox{:=}$]{assignment / substitution axiom}
{\linferenceRule[equiv]
  {p(\aconst)}
  {\axkey{\dbox{\pupdate{\umod{x}{\aconst}}}{p(x)}}}
}
{}%
\cinferenceRuleStore[Dassignb|$\dibox{:=}$]{differential assignment}
{\linferenceRule[equiv]
{p(\astrm)}
{\axkey{\dbox{\Dupdate{\Dumod{\D{x}}{\astrm}}}{p(\D{x})}}}
}
{}%
\cinferenceRuleStore[testb|$\dibox{?}$]{test}
{\linferenceRule[equiv]
  {(\ivr \limply \ausfml)}
  {\axkey{\dbox{\ptest{\ivr}}{\ausfml}}}
}{}%
\cinferenceRuleStore[testbax|$\dibox{?}$]{test}
{\linferenceRule[equiv]
  {(q \limply p)}
  {\axkey{\dbox{\ptest{q}}{p}}}
}{}%
\cinferenceRuleStore[evolveb|$\dibox{'}$]{evolve}
{\linferenceRule[equiv]
  {\lforall{t{\geq}0}{\dbox{\pupdate{\pumod{x}{\solf(t)}}}{p(x)}}}
  {\axkey{\dbox{\pevolve{\D{x}=\genDE{x}}}{p(x)}}}
}{\m{\D{\solf}(t)=\genDE{\solf}}}%
\cinferenceRuleStore[choiceb|$\dibox{\cup}$]{axiom of nondeterministic choice}
{\linferenceRule[equiv]
  {\dbox{\ausprg}{\ausfml} \land \dbox{\busprg}{\ausfml}}
  {\axkey{\dbox{\pchoice{\ausprg}{\busprg}}{\ausfml}}}
}{}%
\cinferenceRuleStore[choicebax|$\dibox{\cup}$]{axiom of nondeterministic choice}
{\linferenceRule[equiv]
  {\dbox{\ausprgax}{\ausfmlax} \land \dbox{\busprgax}{\ausfmlax}}
  {\axkey{\dbox{\pchoice{\ausprgax}{\busprgax}}{\ausfmlax}}}
}{}%
\cinferenceRuleStore[evolveinb|$\dibox{'}$]{evolve}
{\linferenceRule[equiv]
  {
        \lforall{t{\geq}0}{\big(
          (\lforall{0{\leq}s{\leq}t}{q(\solf(s))})
          \limply
          \dbox{\pupdate{\pumod{x}{\solf(t)}}}{p(x)}
        \big)}
      }
  {
        \dbox{\pevolvein{\D{x}=\genDE{x}}{q(x)}}{p(x)}
  }
}{}%
\cinferenceRuleStore[composeb|$\dibox{{;}}$]{composition} %
{\linferenceRule[equiv]
  {\dbox{\ausprg}{\dbox{\busprg}{\ausfml}}}
  {\axkey{\dbox{\ausprg;\busprg}{\ausfml}}}
}{}%
\cinferenceRuleStore[composebax|$\dibox{{;}}$]{composition} %
{\linferenceRule[equiv]
  {\dbox{\ausprgax}{\dbox{\busprgax}{\ausfmlax}}}
  {\axkey{\dbox{\ausprgax;\busprgax}{\ausfmlax}}}
}{}%
\cinferenceRuleStore[iterateb|$\dibox{{}^*}$]{iteration/repeat unwind} %
{\linferenceRule[equiv]
  {\ausfml \land \dbox{\ausprg}{\dbox{\prepeat{\ausprg}}{\ausfml}}}
  {\axkey{\dbox{\prepeat{\ausprg}}{\ausfml}}}
}{}%
\cinferenceRuleStore[iteratebax|$\dibox{{}^*}$]{iteration/repeat unwind} %
{\linferenceRule[equiv]
  {\ausfmlax \land \dbox{\ausprgax}{\dbox{\prepeat{\ausprgax}}{\ausfmlax}}}
  {\axkey{\dbox{\prepeat{\ausprgax}}{\ausfmlax}}}
}{}%
\cinferenceRuleStore[K|K]{K axiom / modal modus ponens}
{\linferenceRule[impl]
  {\dbox{\ausprg}{(\ausfml\limply\busfml)}}
  {(\dbox{\ausprg}{\ausfml}\limply\axkey{\dbox{\ausprg}{\busfml}})}
}{}%
\cinferenceRuleStore[Kax|K]{K axiom / modal modus ponens}
{\linferenceRule[impl]
  {\dbox{\ausprgax}{(\ausfmlax\limply\busfmlax)}}
  {(\dbox{\ausprgax}{\ausfmlax}\limply\axkey{\dbox{\ausprgax}{\busfmlax}})}
}{}%
\cinferenceRuleStore[I|II]{loop induction}
{\linferenceRule[impl]
  {\dbox{\prepeat{\ausprg}}{(\ausfml\limply\dbox{\ausprg}{\ausfml})}}
  {(\ausfml\limply\axkey{\dbox{\prepeat{\ausprg}}{\ausfml}})}
}{}%
\cinferenceRuleStore[Ieq|I]{loop induction}
{\linferenceRule[equiv]
  {\ausfml \land \dbox{\prepeat{\ausprg}}{(\ausfml\limply\dbox{\ausprg}{\ausfml})}}
  {\axkey{\dbox{\prepeat{\ausprg}}{\ausfml}}}
}{}%
\cinferenceRuleStore[Ieqax|I]{loop induction}
{\linferenceRule[equiv]
  {\ausfmlax \land \dbox{\prepeat{\ausprgax}}{(\ausfmlax\limply\dbox{\ausprgax}{\ausfmlax})}}
  {\axkey{\dbox{\prepeat{\ausprgax}}{\ausfmlax}}}
}{}%
\dinferenceRuleStore[iterateiterateb|$\dibox{{}^*{}^*}$]{double iteration}
{\linferenceRule[equiv]
  {\dbox{\prepeat{\ausprg}}{\ausfml}}
  {\axkey{\dbox{\prepeat{\ausprg};\prepeat{\ausprg}}{\ausfml}}}
}{}%
\dinferenceRuleStore[iterateiterated|$\didia{{}^*{}^*}$]{double iteration}
{\linferenceRule[equiv]
  {\ddiamond{\prepeat{\ausprg}}{\ausfml}}
  {\axkey{\ddiamond{\prepeat{\ausprg};\prepeat{\ausprg}}{\ausfml}}}
}{}%
\cinferenceRuleStore[B|B]{Barcan and converse}
{\linferenceRule[equiv]
        {\ddiamond{\ausprg}{\lexists{x}{\ausfml}}}
        {\lexists{x}{\ddiamond{\ausprg}{\ausfml}}}
}{\m{x{\not\in}\ausprg}}
\cinferenceRuleStore[V|V]{vacuous $\dbox{}{}$}
{\linferenceRule[impl]
  {p}
  {\axkey{\dbox{\ausprg}{p}}}
}{\m{FV(p)\cap BV(\ausprg)=\emptyset}}%
\cinferenceRuleStore[Vax|V]{vacuous $\dbox{}{}$}
{\linferenceRule[impl]
  {p}
  {\axkey{\dbox{a}{p}}}
}{}%
\cinferenceRuleStore[G|G]{$\dbox{}{}$ generalization} %
{\linferenceRule[formula]
  {\ausfml}
  {\dbox{\ausprg}{\ausfml}}
}{}%
\cinferenceRuleStore[Gax|G]{$\dbox{}{}$ generalization} %
{\linferenceRule[formula]
  {\ausfmlax}
  {\dbox{\ausprgax}{\ausfmlax}}
}{}%
\cinferenceRuleStore[genaax|$\forall{}$]{$\forall{}$ generalisation}
{\linferenceRule[formula]
  {p(x)}
  {\lforall{x}{p(x)}}
}{}%
\cinferenceRuleStore[MPax|MP]{modus ponens}
{\linferenceRule[formula]
  {p\limply q \quad p}
  {q}
}{}%
\cinferenceRuleStore[Mb|M${\dibox{\cdot}}$]{$\dbox{}{}$ monotone}
{\linferenceRule[formula]
  {\ausfml\limply \busfml}
  {\dbox{\ausprg}{\ausfml}\limply\dbox{\ausprg}{\busfml}}
}{}%
\cinferenceRuleStore[M|M]{$\ddiamond{}{}$ monotone / $\ddiamond{}{}$-generalization}
{\linferenceRule[formula]
  {\ausfml\limply\busfml}
  {\ddiamond{\ausprg}{\ausfml}\limply\ddiamond{\ausprg}{\busfml}}
}{}%

\dinferenceRuleStore[Mbr|M\rightrule]%
{$\ddiamond{}{}/\dbox{}{}$ generalization=M=G+K} 
{\linferenceRule[sequent]
  {\lsequent[L]{} {\dbox{\ausprg}{\busfml}} 
  &\lsequent[g]{\busfml} {\ausfml}}
  {\lsequent[L]{} {\dbox{\ausprg}{\ausfml}}}
}{}%

\dinferenceRuleStore[loop|loop]{inductive invariant}
{\linferenceRule[sequent]
  {\lsequent[L]{} {\inv}
  &\lsequent[g]{\inv} {\dbox{\ausprg}{\inv}}
  &\lsequent[g]{\inv} {\ausfml}}
  {\lsequent[L]{} {\dbox{\prepeat{\ausprg}}{\ausfml}}}
}{}%
\dinferenceRuleStore[invind|ind]{inductive invariant}
{\linferenceRule[sequent]
  {\lsequent[\globalrule]{\ausfml}{\dbox{\ausprg}{\ausfml}}}
  {\lsequent{\ausfml}{\dbox{\prepeat{\ausprg}}{\ausfml}}}
}{}%
\cinferenceRuleStore[con|con]{loop convergence right} 
{\linferenceRule[formula]
  {\lsequent[G]{\mapply{\var}{v}\land v>0}{\ddiamond{\ausprg}{\mapply{\var}{v-1}}}}
  {\lsequent[L]{\lexists{v}{\mapply{\var}{v}}}
      {\axkey{\ddiamond{\prepeat{\ausprg}}{\lexists{v{\leq}0}{\mapply{\var}{v}}}}}}
}{v\not\in\ausprg}%
\dinferenceRuleStore[congen|con]{loop convergence}
{\linferenceRule[sequent]
  {\lsequent[L]{}{\lexists{v}{\mapply{\var}{v}}}
  &\lsequent[G]{}{\lforall{v{>}0}{({\mapply{\var}{v}}\limply{\ddiamond{\ausprg}{\mapply{\var}{v-1}})}}}
  &\lsequent[G]{\lexists{v{\leq}0}{\mapply{\var}{v}}}{\busfml}
  }
  {\lsequent[L]{}{\ddiamond{\prepeat{\ausprg}}{\busfml}}}
}{v\not\in\ausprg}

\dinferenceRuleStore[band|${[]\land}$]{$\dbox{\cdot}{\land}$}
{\linferenceRule[equiv]
  {\dbox{\ausprg}{\ausfml} \land \dbox{\ausprg}{\busfml}}
  {\axkey{\dbox{\ausprg}{(\ausfml\land\busfml)}}}
}{}%

\dinferenceRuleStore[Hoarecompose|H${;}$]{Hoare $;$}
{\linferenceRule
  {A\limply\dbox{\ausprg}{E} & E\limply\dbox{\busprg}{B}}
  {A \limply \dbox{\ausprg;\busprg}{B}}
}{}%
\dinferenceRuleStore[composebrexplicit|$\dibox{{;}}$\rightrule]{$;$}
{\linferenceRule
  {A\limply\dbox{\ausprg}{\dbox{\busprg}{B}}}
  {A \limply \dbox{\ausprg;\busprg}{B}}
}{}
\cinferenceRuleStore[notr|$\lnot$\rightrule]{$\lnot$ right}
{\linferenceRule[sequent]
  {\lsequent[L]{\asfml}{}}
  {\lsequent[L]{}{\lnot \asfml}}
}{}%
\cinferenceRuleStore[notl|$\lnot$\leftrule]{$\lnot$ left}
{\linferenceRule[sequent]
  {\lsequent[L]{}{\asfml}}
  {\lsequent[L]{\lnot \asfml}{}}
}{}%
\cinferenceRuleStore[andr|$\land$\rightrule]{$\land$ right}
{\linferenceRule[sequent]
  {\lsequent[L]{}{\asfml}
    & \lsequent[L]{}{\bsfml}}
  {\lsequent[L]{}{\asfml \land \bsfml}}
}{}%
\cinferenceRuleStore[andl|$\land$\leftrule]{$\land$ left}
{\linferenceRule[sequent]
  {\lsequent[L]{\asfml , \bsfml}{}}
  {\lsequent[L]{\asfml \land \bsfml}{}}
}{}%
\cinferenceRuleStore[orr|$\lor$\rightrule]{$\lor$ right}
{\linferenceRule[sequent]
  {\lsequent[L]{}{\asfml, \bsfml}}
  {\lsequent[L]{}{\asfml \lor \bsfml}}
}{}%
\cinferenceRuleStore[orl|$\lor$\leftrule]{$\lor$ left}
{\linferenceRule[sequent]
  {\lsequent[L]{\asfml}{}
    & \lsequent[L]{\bsfml}{}}
  {\lsequent[L]{\asfml \lor \bsfml}{}}
}{}%
\cinferenceRuleStore[implyr|$\limply$\rightrule]{$\limply$ right}
{\linferenceRule[sequent]
  {\lsequent[L]{\asfml}{\bsfml}}
  {\lsequent[L]{}{\asfml \limply \bsfml}}
}{}%
\cinferenceRuleStore[implyl|$\limply$\leftrule]{$\limply$ left}
{\linferenceRule[sequent]
  {\lsequent[L]{}{\asfml}
    & \lsequent[L]{\bsfml}{}}
  {\lsequent[L]{\asfml \limply \bsfml}{}}
}{}%
\cinferenceRuleStore[equivr|$\lbisubjunct$\rightrule]{$\lbisubjunct$ right}
{\linferenceRule[sequent]
  {\lsequent[L]{\asfml}{\bsfml}
   & \lsequent[L]{\bsfml}{\asfml}}
  {\lsequent[L]{}{\asfml \lbisubjunct \bsfml}}
}{}%
\cinferenceRuleStore[equivl|$\lbisubjunct$\leftrule]{$\lbisubjunct$ left}
{\linferenceRule[sequent]
  {\lsequent[L]{\asfml\limply\bsfml, \bsfml\limply\asfml}{}}
  {\lsequent[L]{\asfml \lbisubjunct \bsfml}{}}
}{}%
\cinferenceRuleStore[id|id]{identity}
{\linferenceRule[sequent]
  {}
  {\lsequent[L]{\asfml}{\asfml}}
}{}%
\cinferenceRuleStore[cut|cut]{cut}
{\linferenceRule[sequent]
  {\lsequent[L]{}{\cusfml}
  &\lsequent[L]{\cusfml}{}}
  {\lsequent[L]{}{}}
}{}%
\cinferenceRuleStore[weakenr|W\rightrule]{weakening right}
{\linferenceRule[sequent]
  {\lsequent[L]{}{}}
  {\lsequent[L]{}{\asfml}}
}{}%
\cinferenceRuleStore[weakenl|W\leftrule]{weakening left}
{\linferenceRule[sequent]
  {\lsequent[L]{}{}}
  {\lsequent[L]{\asfml}{}}
}{}%
\cinferenceRuleStore[exchanger|P\rightrule]{exchange right}
{\linferenceRule[sequent]
  {\lsequent[L]{}{\bsfml,\asfml}}
  {\lsequent[L]{}{\asfml,\bsfml}}
}{}%
\cinferenceRuleStore[exchangel|P\leftrule]{exchange left}
{\linferenceRule[sequent]
  {\lsequent[L]{\bsfml,\asfml}{}}
  {\lsequent[L]{\asfml,\bsfml}{}}
}{}%
\cinferenceRuleStore[contractr|c\rightrule]{contract right}
{\linferenceRule[sequent]
  {\lsequent[L]{}{\asfml,\asfml}}
  {\lsequent[L]{}{\asfml}}
}{}%
\cinferenceRuleStore[contractl|c\leftrule]{contract left}
{\linferenceRule[sequent]
  {\lsequent[L]{\asfml,\asfml}{}}
  {\lsequent[L]{\asfml}{}}
}{}
\cinferenceRuleStore[closeTrue|$\top$\rightrule]{close by true}
{\linferenceRule[sequent]
  {}
  {\lsequent[L]{}{\ltrue}}
}{}%
\cinferenceRuleStore[closeFalse|$\bot$\leftrule]{close by false}
{\linferenceRule[sequent]
  {}
  {\lsequent[L]{\lfalse}{}}
}{}%

\cinferenceRuleStore[CE|CE]{congequiv congruence of equivalences on formulas}
{\linferenceRule[formula]
  {\ausfml \lbisubjunct \busfml}
  {\contextapp{C}{\ausfml} \lbisubjunct \contextapp{C}{\busfml}}
}{}%
\dinferenceRuleStore[CEr|CE\rightrule]{congequiv congruence of equivalences on formulas}
{\linferenceRule[formula]
  {\lsequent[L]{} {\contextapp{C}{\busfml}}
  &\lsequent[g]{} {\ausfml \lbisubjunct \busfml}}
  {\lsequent[L]{} {\contextapp{C}{\ausfml}}}
}{}%
\dinferenceRuleStore[CEl|CE\leftrule]{congequiv congruence of equivalences on formulas}
{\linferenceRule[formula]
  {\lsequent[L]{\contextapp{C}{\busfml}} {}
  &\lsequent[g]{} {\ausfml \lbisubjunct \busfml}}
  {\lsequent[L]{\contextapp{C}{\ausfml}} {}}
}{}%

\cinferenceRuleStore[allr|$\forall$\rightrule]{$\lforall{}{}$ right}
{\linferenceRule[sequent]
  {\lsequent[L]{}{p(y)}}
  {\lsequent[L]{}{\lforall{x}{p(x)}}}
}{\m{y\not\in\Gamma{,}\Delta{,}\lforall{x}{p(x)}}}%
\cinferenceRuleStore[alll|$\forall$\leftrule]{$\lforall{}{}$ left instantiation}
{\linferenceRule[sequent]
  {\lsequent[L]{p(\astrm)}{}}
  {\lsequent[L]{\lforall{x}{p(x)}}{}}
}{arbitrary term $\astrm$}%
\cinferenceRuleStore[existsr|$\exists$\rightrule]{$\lexists{}{}$ right}
{\linferenceRule[sequent]
  {\lsequent[L]{}{p(\astrm)}}
  {\lsequent[L]{}{\lexists{x}{p(x)}}}
}{arbitrary term $\astrm$}%
\cinferenceRuleStore[existsl|$\exists$\leftrule]{$\lexists{}{}$ left}
{\linferenceRule[sequent]
  {\lsequent[L]{p(y)} {}}
  {\lsequent[L]{\lexists{x}{p(x)}} {}}
}{\m{y\not\in\Gamma{,}\Delta{,}\lexists{x}{p(x)}}}%

\cinferenceRuleStore[qear|\usebox{\Rval}]{quantifier elimination real arithmetic}
{\linferenceRule[sequent]
  {}
  {\lsequent[g]{\Gamma}{\Delta}}
}{\text{if}~\landfold_{\ausfml\in\Gamma} \ausfml \limply \lorfold_{\busfml\in\Delta} \busfml ~\text{is valid in \LOS[\reals]}}%

\dinferenceRuleStore[allGi|i$\forall$]{inverse universal generalization / universal instantiation}
{\linferenceRule[sequent]
  {\lsequent[L]{} {\lforall{x}{\ausfml}}}
  {\lsequent[L]{} {\ausfml}}
}{}%

\cinferenceRuleStore[applyeqr|=\rightrule]{apply equation}
{\linferenceRule[sequent]
  {\lsequent[L]{x=\astrm}{p(\astrm)}}
  {\lsequent[L]{x=\astrm}{p(x)}}
}{}%
\cinferenceRuleStore[applyeql|=\leftrule]{apply equation}
{\linferenceRule[sequent]
  {\lsequent[L]{x=\astrm,p(\astrm)}{}}
  {\lsequent[L]{x=\astrm,p(x)}{}}
}{}%

\dinferenceRuleStore[alldupl|$\forall\forall$\leftrule]{$\lforall{}{}$ left instantiation retaining duplicates}
{\linferenceRule[sequent]
  {\lsequent[L]{\lforall{x}{p(x)},p(\astrm)}{}}
  {\lsequent[L]{\lforall{x}{p(x)}}{}}
}{}%

\dinferenceRuleStore[choicebrinsist|$\dibox{\cup}\rightrule$]{}
{\linferenceRule
  {\lsequent[L]{}{\dbox{\asprg}{\ausfml}\land\dbox{\bsprg}{\ausfml}}}
  {\lsequent[L]{}{\dbox{\pchoice{\asprg}{\bsprg}}{\ausfml}}}
}{}
\dinferenceRuleStore[choiceblinsist|$\dibox{\cup}\leftrule$]{}
{\linferenceRule
  {\lsequent[L]{\dbox{\asprg}{\ausfml}\land\dbox{\bsprg}{\ausfml}}{}}
  {\lsequent[L]{\dbox{\pchoice{\asprg}{\bsprg}}{\ausfml}}{}}
}{}
\dinferenceRuleStore[choicebrinsist2|$\dibox{\cup}\rightrule2$]{}
{\linferenceRule
  {\lsequent[L]{}{\dbox{\asprg}{\ausfml}}
  &\lsequent[L]{}{\dbox{\bsprg}{\ausfml}}}
  {\lsequent[L]{}{\dbox{\pchoice{\asprg}{\bsprg}}{\ausfml}}}
}{}
\dinferenceRuleStore[choiceblinsist2|$\dibox{\cup}\leftrule2$]{}
{\linferenceRule
  {\lsequent[L]{\dbox{\asprg}{\ausfml},\dbox{\bsprg}{\ausfml}}{}}
  {\lsequent[L]{\dbox{\pchoice{\asprg}{\bsprg}}{\ausfml}}{}}
}{}
\dinferenceRuleStore[cutr|cut\rightrule]{cut right}
{\linferenceRule[sequent]
  {\lsequent[L]{}{\bsfml}
  &\lsequent[L]{}{\bsfml\limply\asfml}}
  {\lsequent[L]{}{\asfml}}
}{}
\dinferenceRuleStore[cutl|cut\leftrule]{cut left}
{\linferenceRule[sequent]
  {\lsequent[L]{\bsfml} {}
  &\lsequent[L]{}{\asfml\limply\bsfml}}
  {\lsequent[L]{\asfml} {}}
}{}

\cinferenceRuleStore[Dplus|$+'$]{derive sum}
{\linferenceRule[eq]
  {\der{\asdtrm}+\der{\bsdtrm}}
  {\axkey{\der{\asdtrm+\bsdtrm}}}
}
{}
\cinferenceRuleStore[Dplusax|$+'$]{derive sum}
{\linferenceRule[eq]
  {\der{\asdtrmax}+\der{\bsdtrmax}}
  {\axkey{\der{\asdtrmax+\bsdtrmax}}}
}
{}
\cinferenceRuleStore[Dminus|$-'$]{derive minus}
{\linferenceRule[eq]
  {\der{\asdtrm}-\der{\bsdtrm}}
  {\axkey{\der{\asdtrm-\bsdtrm}}}
}
{}
\cinferenceRuleStore[Dminusax|$-'$]{derive minus}
{\linferenceRule[eq]
  {\der{\asdtrmax}-\der{\bsdtrmax}}
  {\axkey{\der{\asdtrmax-\bsdtrmax}}}
}
{}
\cinferenceRuleStore[Dtimes|$\cdot'$]{derive product}
{\linferenceRule[eq]
  {\der{\asdtrm}\cdot \bsdtrm+\asdtrm\cdot\der{\bsdtrm}}
  {\axkey{\der{\asdtrm\cdot \bsdtrm}}}
}
{}
\cinferenceRuleStore[Dtimesax|$\cdot'$]{derive product}
{\linferenceRule[eq]
  {\der{\asdtrmax}\cdot \bsdtrmax+\asdtrmax\cdot\der{\bsdtrmax}}
  {\axkey{\der{\asdtrmax\cdot \bsdtrmax}}}
}
{}
\cinferenceRuleStore[Dquotient|$/'$]{derive quotient}
{\linferenceRule[eq]
  {\big(\der{\asdtrm}\cdot \bsdtrm-\asdtrm\cdot\der{\bsdtrm}\big) / \bsdtrm^2}
  {\axkey{\der{\asdtrm/\bsdtrm}}}
}
{}
\cinferenceRuleStore[Dquotientax|$/'$]{derive quotient}
{\linferenceRule[eq]
  {\big(\der{\asdtrmax}\cdot \bsdtrmax-\asdtrmax\cdot\der{\bsdtrmax}\big) / \bsdtrmax^2}
  {\axkey{\der{\asdtrmax/\bsdtrmax}}}
}
{}
\cinferenceRuleStore[Dconst|$c'$]{derive constant}
{\linferenceRule[eq]
  {0}
  {\axkey{\der{\aconst}}}
  \hspace{3cm}
}
{\text{for numbers or constants~$\aconst$}}%
\cinferenceRuleStore[Dvar|$x'$]{derive variable}
{\linferenceRule[eq]
  {\D{x}}
  {\axkey{\der{x}}}
}
{\text{for variable~$x\in\allvars$}}%

\cinferenceRuleStore[DE|DE]{differential effect} %
{\linferenceRule[viuqe]
  {\axkey{\dbox{\pevolvein{\D{x}=\genDE{x}}{\ivr}}{\ausfml}}}
  {\dbox{\pevolvein{\D{x}=\genDE{x}}{\ivr}}{\dbox{\axeffect{\Dupdate{\Dumod{\D{x}}{\genDE{x}}}}}{\ausfml}}}
}
{}%
\cinferenceRuleStore[DEax|DE]{differential effect} %
{\linferenceRule[viuqe]
  {\axkey{\dbox{\pevolvein{\D{x}=\genDE{x}}{q(x)}}{\ausfmlax}}}
  {\dbox{\pevolvein{\D{x}=\genDE{x}}{q(x)}}{\dbox{\axeffect{\Dupdate{\Dumod{\D{x}}{\genDE{x}}}}}{\ausfmlax}}}
}
{}%

\cinferenceRuleStore[Dand|${\land}'$]{derive and}
{\linferenceRule[equiv]
  {\der{\asfml}\land\der{\bsfml}}
  {\axkey{\der{\asfml\land\bsfml}}}
}
{}
\cinferenceRuleStore[Dor|${\lor}'$]{derive or}
{\linferenceRule[equiv]
  {\der{\asfml}\land\der{\bsfml}}
  {\axkey{\der{\asfml\lor\bsfml}}}
}
{}
\cinferenceRuleStore[diffweaken|DW]{differential evolution domain} %
{\linferenceRule[viuqe]
  {\axkey{\dbox{\pevolvein{\D{x}=\genDE{x}}{\ivr}}{\ousfml[x]}}}
  {\dbox{\pevolvein{\D{x}=\genDE{x}}{\ivr}}{(\axeffect{\ivr}\limply \ousfml[x])}}
}
{}%
\cinferenceRuleStore[diffweakenax|DW]{differential evolution domain} 
{\linferenceRule[viuqe]
  {\axkey{\dbox{\pevolvein{\D{x}=\genDE{x}}{q(x)}}{p(x)}}}
  {\dbox{\pevolvein{\D{x}=\genDE{x}}{q(x)}}{(\axeffect{q(x)}\limply p(x))}}
}
{}%
\cinferenceRuleStore[dW|dW]{differential weakening}
{\linferenceRule[sequent]
  {\lsequent[g]{\ivr} {\ousfml[x]}}
  {\lsequent[g]{\Gamma} {\dbox{\pevolvein{\D{x}=f(x)}{\ivr}}{\ousfml[x]},\Delta}}
}
{}%
\cinferenceRuleStore[DI|DI]{differential induction}
{\linferenceRule[lpmi]
  {\big(\axkey{\dbox{\pevolvein{\D{x}=\genDE{x}}{\ivr}}{\ousfml[x]}}
  \lbisubjunct \dbox{\ptest{\ivr}}{\ousfml[x]}\big)}
  {(\ivr\limply\dbox{\pevolvein{\D{x}=\genDE{x}}{\ivr}}{\axeffect{\der{\ousfml[x]}}})}
}
{}%
\cinferenceRuleStore[DIax|DI]{differential induction}
{\linferenceRule[lpmi]
  {\big(\axkey{\dbox{\pevolvein{\D{x}=\genDE{x}}{q(x)}}{p(x)}}
  \lbisubjunct \dbox{\ptest{q(x)}}{p(x)}\big)}
  {(q(x)\limply\dbox{\pevolvein{\D{x}=\genDE{x}}{q(x)}}{\axeffect{\der{p(x)}}})}
}
{}%
\cinferenceRuleStore[DIlight|DI]{differential induction}
{\linferenceRule[lpmi]
  {\big(\axkey{\dbox{\pevolvein{\D{x}=\genDE{x}}{\ivr}}{\ousfml[x]}}
  \lbisubjunct \dbox{\ptest{\ivr}}{\ousfml[x]}\big)}
  {\dbox{\pevolvein{\D{x}=\genDE{x}}{\ivr})}{\axeffect{\der{\ousfml[x]}}}}
}
{}%

\cinferenceRuleStore[dI|dI]{differential invariant}
{\linferenceRule[sequent]
  {\lsequent[g]{\ivr}{\Dusubst{\D{x}}{\genDE{x}}{\der{F}}}}
  {\lsequent{F}{\dbox{\pevolvein{\D{x}=\genDE{x}}{\ivr}}{F}}}
}{}
\cinferenceRuleStore[DC|DC]{differential cut}
{\linferenceRule[lpmi]
  {\big(\axkey{\dbox{\pevolvein{\D{x}=\genDE{x}}{\ivr}}{\ousfml[x]}} \lbisubjunct \dbox{\pevolvein{\D{x}=\genDE{x}}{\ivr\land \axeffect{\ousfmlc[x]}}}{\ousfml[x]}\big)}
  {\dbox{\pevolvein{\D{x}=\genDE{x}}{\ivr}}{\axeffect{\ousfmlc[x]}}}
}
{}%
\cinferenceRuleStore[DCax|DC]{differential cut}
{\linferenceRule[lpmi]
  {\big(\axkey{\dbox{\pevolvein{\D{x}=\genDE{x}}{q(x)}}{p(x)}} \lbisubjunct \dbox{\pevolvein{\D{x}=\genDE{x}}{q(x)\land \axeffect{r(x)}}}{p(x)}\big)}
  {\dbox{\pevolvein{\D{x}=\genDE{x}}{q(x)}}{\axeffect{r(x)}}}
}
{}%
\cinferenceRuleStore[dC|dC]{differential cut}%
{\linferenceRule[sequent]
  {\lsequent[L]{}{\dbox{\pevolvein{\D{x}=\genDE{x}}{\ivr}}{\axeffect{\cusfml}}}
  &\lsequent[L]{}{\dbox{\pevolvein{\D{x}=\genDE{x}}{(\ivr\land \axeffect{\cusfml})}}{\ousfml[x]}}}
  {\lsequent[L]{}{\dbox{\pevolvein{\D{x}=\genDE{x}}{\ivr}}{\ousfml[x]}}}
}{}
\cinferenceRuleStore[DGanyode|DG]{differential ghost variables (unsound!)}
{\linferenceRule[viuqe]
  {\axkey{\dbox{\pevolvein{\D{x}=\genDE{x}}{\ivr}}{\ousfml[x]}}}
  {\lexists{y}{\dbox{\pevolvein{\D{x}=\genDE{x}\syssep\axeffect{\D{y}=g(x,y)}}{\ivr}}{\ousfml[x]}}}
}
{}
\cinferenceRuleStore[DG|DG]{differential ghost variables}
{\linferenceRule[viuqe]
  {\axkey{\dbox{\pevolvein{\D{x}=\genDE{x}}{\ivr}}{\ousfml[x]}}}
  {\lexists{y}{\dbox{\pevolvein{\D{x}=\genDE{x}\syssep\axeffect{\D{y}=a(x)\cdot y+b(x)}}{\ivr}}{\ousfml[x]}}}
}
{}
\cinferenceRuleStore[DGax|DG]{differential ghost variables}
{\linferenceRule[viuqe]
  {\axkey{\dbox{\pevolvein{\D{x}=\genDE{x}}{q(x)}}{p(x)}}}
  {\lexists{y}{\dbox{\pevolvein{\D{x}=\genDE{x}\syssep\axeffect{\D{y}=a(x)\cdot y+b(x)}}{q(x)}}{p(x)}}}
}
{}
\cinferenceRuleStore[dG|dG]{dG}%
{\linferenceRule[sequent]
  {\lsequent[L]{} {\lexists{y}{\dbox{\pevolvein{\D{x}=f(x)\syssep\axeffect{\D{y}=a(x)\cdot y+b(x)}}{\oivr[x]}}{\ousfml[x]}}}
  }
  {\lsequent[L]{} {\dbox{\pevolvein{\D{x}=f(x)}{\oivr[x]}}{\ousfml[x]}}}
}
{}%

\dinferenceRuleStore[assignbeqr|$\dibox{:=}_=$]{assignb}%
  {\linferenceRule[sequent]
    {\lsequent[L]{y=\austrm} {p(y)}}
    {\lsequent[L]{} {\dbox{\pupdate{\umod{x}{\austrm}}}{p(x)}}}
  }
  {\text{$y$ new}}

\cinferenceRuleStore[DSax|DS]{(constant) differential equation solution} 
{\linferenceRule[viuqe]
  {\axkey{\dbox{\pevolvein{\D{x}=\aconst}{q(x)}}{p(x)}}}
  {\lforall{t{\geq}0}{\big((\lforall{0{\leq}s{\leq}t}{q(x+\aconst\itimes s)}) \limply \dbox{\pupdate{\pumod{x}{x+\aconst\itimes t}}}{p(x)}\big)}}
}
{}

\dinferenceRuleStore[DIeq0|DI]{differential invariant axiom}
{\linferenceRule[lpmi]
  {\big(\axkey{\dbox{\pevolve{\D{x}=\genDE{x}}}{\,\astrm=0}} \lbisubjunct \astrm=0\big)}
  {\dbox{\pevolve{\D{x}=\genDE{x}}}{\,\axeffect{\der{\astrm}=0}}}
}
{}
\dinferenceRuleStore[diffindeq0|dI]{differential invariant $=0$ case}
{\linferenceRule[sequent]
  {\lsequent{~}{\Dusubst{\D{x}}{\genDE{x}}{\der{\astrm}}=0}}
  {\lsequent{\astrm=0}{\dbox{\pevolve{\D{x}=\genDE{x}}}{\astrm=0}}}
}{}
\cinferenceRuleStore[Liec|dI$_c$]{}
{\linferenceRule
  {\lsequent{\ivr}{\Dusubst{\D{x}}{\genDE{x}}{\der{\astrm}}=0}}
  {\lsequent{}{\lforall{c}{\big(\astrm=c \limply \dbox{\pevolvein{\D{x}=\genDE{x}}{\ivr}}{\astrm=c}\big)}}}
}{}

\cinferenceRuleStore[DIeq|DI$_=$]{differential induction $=$ case}
{\linferenceRule[lpmi]
  {\big(\axkey{\dbox{\pevolvein{\D{x}=\genDE{x}}{\ivr}}{\asdtrm=\bsdtrm}}
  \lbisubjunct \dbox{\ptest{\ivr}}{\asdtrm=\bsdtrm}\big)}
  {\dbox{\pevolvein{\D{x}=\genDE{x}}{\ivr})}{\axeffect{\der{\asdtrm}=\der{\bsdtrm}}}}
}
{}%

\dinferenceRuleStore[diffindgen|dI']{differential invariant}
{\linferenceRule[sequent]
  {\lsequent[L]{}{\inv}
  &\lsequent[g]{\ivr}{\Dusubst{\D{x}}{\genDE{x}}{\der{\inv}}}
  &\lsequent[g]{\inv}{\psi}
  }
  {\lsequent[L]{}{\dbox{\pevolvein{\D{x}=\genDE{x}}{\ivr}}{\psi}}}
}{}
\cinferenceRuleStore[diffindunsound|dI$_{??}$]{unsound}
{\linferenceRule[sequent]
  {\lsequent{\ivr\land\inv}{\Dusubst{\D{x}}{\genDE{x}}{\der{\inv}}}}
  {\lsequent{\inv}{\dbox{\pevolvein{\D{x}=\genDE{x}}{\ivr}}{\inv}}}
}{}

\dinferenceRuleStore[introaux|iG]{introduce discrete ghost variable}
{\linferenceRule[sequent]
  {\lsequent[L]{}{\dbox{\axeffect{\pupdate{\pumod{y}{\astrm}}}}{p}}}
  {\lsequent[L]{} {p}}
}{\text{$y$ new}}%
\dinferenceRuleStore[diffaux|dA]{differential auxiliary variables}
{\linferenceRule[sequent]
  {\lsequent[\globalrule]{}{\inv\lbisubjunct\lexists{y}{G}}
  &\lsequent{G} {\dbox{\pevolvein{\D{x}=\genDE{x}\syssep\axeffect{\D{y}=a(x)\cdot y+b(x)}}{\ivr}}{G}}}
  {\lsequent{\inv} {\dbox{\pevolvein{\D{x}=\genDE{x}}{\ivr}}{\inv}}}
}{}%
\cinferenceRuleStore[randomd|$\didia{{:}*}$]{nondeterministic assignment}
{\linferenceRule[equiv]
  {\lexists{x}{\ousfml[x]}}
  {\axkey{\ddiamond{\prandom{x}}{\ousfml[x]}}}
}{}
\cinferenceRuleStore[randomb|$\dibox{{:}*}$]{nondeterministic assignment}
{\linferenceRule[equiv]
  {\lforall{x}{\ousfml[x]}}
  {\axkey{\dbox{\prandom{x}}{\ousfml[x]}}}
}{}

\cinferenceRuleStore[box|$\dibox{\cdot}$]{box axiom}
{\linferenceRule[equiv]
  {\lnot\ddiamond{\ausprg}{\lnot\ausfml}}
  {\axkey{\dbox{\ausprg}{\ausfml}}}
}
{}
\cinferenceRuleStore[assignd|$\didia{:=}$]{assignment / substitution axiom}
{\linferenceRule[equiv]
  {p(\genDJ{x})}
  {\axkey{\ddiamond{\pupdate{\umod{x}{\genDJ{x}}}}{p(x)}}}
}
{}%
\cinferenceRuleStore[evolved|$\didia{'}$]{evolve}
{\linferenceRule[equiv]
  {\lexists{t{\geq}0}{\ddiamond{\pupdate{\pumod{x}{\solf(t)}}}{p(x)}}\hspace{1cm}}
  {\axkey{\ddiamond{\pevolve{\D{x}=\genDE{x}}}{p(x)}}}
}{\m{\D{\solf}(t)=\genDE{\solf}}}%
\cinferenceRuleStore[evolveind|$\didia{'}$]{evolve}
{\linferenceRule[equiv]
  {\lexists{t{\geq}0}{\big((\lforall{0{\leq}s{\leq}t}{q(\solf(s))}) \land 
  \ddiamond{\pupdate{\pumod{x}{\solf(t)}}}{p(x)}\big)}}
  {\axkey{\ddiamond{\pevolvein{\D{x}=\genDE{x}}{q(x)}}{p(x)}}}
}{\m{\D{\solf}(t)=\genDE{\solf}}}%
\cinferenceRuleStore[testd|$\didia{?}$]{test}
{\linferenceRule[equiv]
  {\ivr \land \ausfml}
  {\axkey{\ddiamond{\ptest{\ivr}}{\ausfml}}}
}{}
\cinferenceRuleStore[choiced|$\didia{\cup}$]{axiom of nondeterministic choice}
{\linferenceRule[equiv]
  {\ddiamond{\ausprg}{\ausfml} \lor \ddiamond{\busprg}{\ausfml}}
  {\axkey{\ddiamond{\pchoice{\ausprg}{\busprg}}{\ausfml}}}
}{}
\cinferenceRuleStore[composed|$\didia{{;}}$]{composition}
{\linferenceRule[equiv]
  {\ddiamond{\ausprg}{\ddiamond{\busprg}{\ausfml}}}
  {\axkey{\ddiamond{\ausprg;\busprg}{\ausfml}}}
}{}
\cinferenceRuleStore[iterated|$\didia{{}^*}$]{iteration/repeat unwind pre-fixpoint, even fixpoint}
{\linferenceRule[equiv]
  {\ausfml \lor \ddiamond{\ausprg}{\ddiamond{\prepeat{\ausprg}}{\ausfml}}}
  {\axkey{\ddiamond{\prepeat{\ausprg}}{\ausfml}}}
}{}
\cinferenceRuleStore[duald|$\didia{{^d}}$]{dual}
{\linferenceRule[equiv]
  {\lnot\ddiamond{\ausprg}{\lnot\ausfml}}
  {\axkey{\ddiamond{\pdual{\ausprg}}{\ausfml}}}
}{}
\cinferenceRuleStore[dualb|$\dibox{{^d}}$]{dual}
{\linferenceRule[equiv]
  {\lnot\dbox{\ausprg}{\lnot\ausfml}}
  {\axkey{\dbox{\pdual{\ausprg}}{\ausfml}}}
}{}
\cinferenceRuleStore[FP|FP]{iteration is least fixpoint / reflexive transitive closure RTC, equivalent to invind in the presence of M}
{\linferenceRule[formula]
  {\ausfml \lor \ddiamond{\ausprg}{\busfml} \limply \busfml}
  {\ddiamond{\prepeat{\ausprg}}{\ausfml} \limply \busfml}
}{}
\cinferenceRuleStore[invindg|ind]{inductive invariant for games}
{\linferenceRule[formula]
  {\ausfml\limply\dbox{\ausprg}{\ausfml}}
  {\ausfml\limply\dbox{\prepeat{\ausprg}}{\ausfml}}
}{}%

\dinferenceRuleStore[dchoiced|$\didia{{\cap}}$]{Demon's choice}
{
\axkey{\ddiamond{\dchoice{\ausprg}{\busprg}}{\ausfml}} \lbisubjunct \ddiamond{\ausprg}{\ausfml} \land \ddiamond{\busprg}{\ausfml}
}{}
\dinferenceRuleStore[dchoiceb|$\dibox{{\cap}}$]{Demon's choice}
{
\axkey{\dbox{\dchoice{\ausprg}{\busprg}}{\ausfml}} \lbisubjunct \dbox{\ausprg}{\ausfml} \lor \dbox{\busprg}{\ausfml}
}{}
\dinferenceRuleStore[diterateb|$\dibox{\drepeat{}}$]{Demon's repetition}
{\linferenceRule[equiv]
  {\ausfml \lor \dbox{\ausprg}{\dbox{\drepeat{\ausprg}}{\ausfml}}}
  {\axkey{\dbox{\drepeat{\ausprg}}{\ausfml}}}
}{}
\dinferenceRuleStore[diterated|$\didia{\drepeat{}}$]{Demon's repetition}
{\linferenceRule[equiv]
  {\ausfml \land \ddiamond{\ausprg}{\ddiamond{\drepeat{\ausprg}}}{\ausfml}}
  {\axkey{\ddiamond{\drepeat{\ausprg}}{\ausfml}}}
}{}
\dinferenceRuleStore[dinvindg|ind$\drepeat{}$]{inductive invariant for games}
{\linferenceRule[formula]
  {\ausfml\limply\ddiamond{\ausprg}{\ausfml}}
  {\ausfml\limply\ddiamond{\drepeat{\ausprg}}{\ausfml}}
}{}
\dinferenceRuleStore[dFP|FP$\drepeat{}$]{dual iteration is least fixpoint in Demon's winning strategy}
{\linferenceRule[formula]
  {\ausfml \lor \dbox{\ausprg}{\busfml} \limply \busfml}
  {\dbox{\drepeat{\ausprg}}{\ausfml} \limply \busfml}
}{}

\cinferenceRuleStore[US|US]{uniform substitution}
{\linferenceRule[formula]
  {\phi}
  {\applyusubst{\sigma}{\phi}}
}{}%

\cinferenceRuleStore[linequs|$\exists$lin]{linear equation uniform substitution}
{\linferenceRule[impl]
  {b\neq0}
  {\big(\lexists{x}{(b\cdot x+c=0 \land q(x))}
  \lbisubjunct {q(-c/b)}\big)}
}{}

\dinferenceRuleStore[FA|FA]{First arrival}
{\ddiamond{\prepeat{\ausprg}}{\ausfml} \limply \ausfml \lor \ddiamond{\prepeat{\ausprg}}{(\lnot\ausfml\land\ddiamond{\ausprg}{\ausfml})}
}{}
\dinferenceRuleStore[Mor|M]{monotonicity axiom}
{\ddiamond{\ausprg}{(\ausfml\lor\busfml)}
\lbisubjunct
\ddiamond{\ausprg}{\ausfml} \lor \ddiamond{\ausprg}{\busfml}
}{}
\dinferenceRuleStore[VK|VK]{vacuous possible $\dbox{}{}$}
{\linferenceRule[impl]
  {p}
  {(\dbox{\ausprg}{\ltrue}{\limply}\dbox{\ausprg}{p})}
  \qquad
}{\m{\freevars{p}\cap \boundvars{\ausprg}=\emptyset}}
\dinferenceRuleStore[R|R]{Regular}
{\linferenceRule[formula]
  {\ausfml_1\land\ausfml_2\limply\busfml}
  {\dbox{\ausprg}{\ausfml_1} \land \dbox{\ausprg}{\ausfml_2} \limply \dbox{\ausprg}{\busfml}}
}{}

\begin{figure}[t]
\centering
\cinferenceRuleQuoteDef{loop}
\hfill
\cinferenceRuleQuoteDef{existsr}
\\[1em]
\cinferenceRuleQuoteDef{dC}\\[1em]
\cinferenceRuleQuoteDef{dI}
\caption{\dGL Proof Rules and Axioms relevant to the proof of \rref{model:lotka-volterra}. The rules are as follows: differential invariant (dI), differential cut (dC), loop (loop), and exists right (existsr).}
\label{fig:lotka-volterra-rules}
\end{figure}

\begin{figure}[tbhp]
  \centering
  \begin{calculuscollections}{\columnwidth}
    \begin{calculus}
      \cinferenceRuleQuote{box}
      \cinferenceRuleQuote{assignd}
      \cinferenceRuleQuote{evolved}
      \cinferenceRuleQuote{testd}
      \cinferenceRuleQuote{choiced}
      \cinferenceRuleQuote{composed}
      \cinferenceRuleQuote{iterated}
      \cinferenceRuleQuote{duald}
      \cinferenceRuleQuote{dchoiced}
      \cinferenceRuleQuote{diterated}
      \cinferenceRuleQuote{assignb}
      \cinferenceRuleQuote{evolveb}
      \cinferenceRuleQuote{testb}
      \cinferenceRuleQuote{choiceb}
      \cinferenceRuleQuote{composeb}
      \cinferenceRuleQuote{iterateb}
      \cinferenceRuleQuote{dualb}
      \cinferenceRuleQuote{dchoiceb}
      \cinferenceRuleQuote{diterateb}
    \end{calculus}

    \begin{calculus}
      \cinferenceRuleQuote{loop}
      \cinferenceRuleQuote{M}
      \cinferenceRuleQuote{Mb}
    \end{calculus}
    \begin{calculus}
      \dinferenceRuleQuote{invindg}
      \cinferenceRuleQuote{FP}
      \cinferenceRuleQuote{dFP}
    \end{calculus}
  \end{calculuscollections}
  \caption{\dGL axiomatization and derived axioms and rules}
  \label{fig:dGL-calculus}
\end{figure}

\section{Verification Prompts}
\label{app:verification-prompts}

This section lists all the prompts used for the verification case studies.
The prompts use the OpenAI api format with system and user queries, and sometimes include few-shot prompts. Here we have shown the queries for the chemical reaction case study as an example. Some Unicode characters have been replaced with LaTeX commands and line breaks have been added for compilation. Prompts are listed in the order in which they are called in the verification progress, and have a one-to-one correspondence with the LLM queries depicted in \rref{fig:atp-pipeline}.

\noindent 1. AnalyzeGame is the first query in the pipeline that produces an analysis of the game used as context in later prompts.
One example is shown as a few-shot example.

\begin{verbatim}
role: system
content: "Your task is to analyze a game and identify which
actions are taken by the Angel (controllable player) versus
the Demon (adversarial player).

## Background Information

In differential game logic (dGL), there are two players:
- **Angel**: The controllable player who makes choices we can
  control. Angel constructs include:
  - Angel choices: `a ++ b` (we choose between a and b)
  - Angel loops: `{a}*` (we choose how many iterations)
  - Angel tests: `?P` (we can proceed if P holds, otherwise we
    immediately lose)
  - Angel ODEs: `{x'=f(x) & Q}` (we control the duration)
  - Angel nondeterministic assignments: `x := *` (we choose the
    value)

There is the dual operator `^@` that switches Angel and Demon
roles in a game construct.
Duals can be nested arbitrarily, so carefully check nesting
levels to determine which player controls each action.
Being within an odd number of `^@` operators indicates Demon
control, while an even number (including zero) indicates Angel
control.
For example, in the game `{(a ++ b)^@; (c++d)}^@`, the choice
between `a` and `b` is controlled by Angel, being within two
`^@` operators, while the choice between `c` and `d` is
controlled by Demon.
Using this notation, we go over the Demon constructs below,
indicating that they are Demon-controlled by wrapping them in a
single `^@` operator.

- **Demon**: The adversarial player who makes choices we cannot
  control. Demon has all the same constructs as Angel.
  - Demon choices: `{a ++ b}^@` (adversary chooses between a and
    b)
  - Demon loops: `{a*}^@` (adversary chooses how many iterations)
  - Demon tests: `{?P}^@` (adversary can proceed if P holds, and
    immediately loses otherwise. So Angel can *assume* in
    subsequent game play that P held at this point; otherwise
    she would already have won.)
  - Demon ODEs: `{x'=f(x) & Q}^@` (adversary controls the
    duration)
  - Demon nondeterministic assignments: `{x := *}^@` (adversary
    chooses the value)

Additionally, the non-controlled operations are:
- Assignment: `x := e` (deterministic assignment like in regular
  programs)
- Sequential composition: `a; b` (first play a, then b)

## Notes
- ODEs can sometimes introduce unmatched single quotes, so
  account for that when parsing.
- In syntax such as { a++b; x:=* }*^@, the `^@` applies to the
  entire loop including its body. In this example, Demon
  controls the choice within the loop between a and b, as well as
  the assignment to x, because both are inside one `^@`.

Some subgames are labeled with unique identifiers with the
following syntax:
    <subgame_id>: <subgame>

These subgame ids will help make your answers more specific.

## Input

You will be provided a dGL game.

## Output

Your analysis should identify:
1. What actions or decisions are under Angel's control in the
   game.
2. What actions or decisions are under Demon's control in the
   game.
3. The control modes in the game, that is, regions of different
   qualitative behavior (e.g. growth vs. oscillation vs. decay
   vs. equilibrium).
The modes are typically defined by two things: (1) initial
conditions inducing different qualitative behavior in the
dynamics, and (2) control actions leading to different possible
values of variables. Comment about the asymptotes or polynomial
invariants of each mode, when relevant.
4. The overall control pattern in the game.

## Further Instructions

Provide your analysis as clear, readable text describing the
Angel and Demon actions, and control modes. Be specific about
what choices each player makes and how they might interact.
Structure it into the sections corresponding to the points
above.
Be as concise as possible while still being clear, with no
superfluous or filler text or hedging.

- Carefully consider the differential equation. There can be
  coupling between variables that affects your reasoning.
- Based on the dynamics, you may some times need to come up with
  upper/lower bounds of quantities and use that to derive sound
  preconditions. It's best to use simplistic, polynomial bounds
  so that they verify more easily.

You will now be provided with 0 or more examples of games and
their analysis, followed by a new game that you must analyze."
- role: user
  content: |-
    Game to analyze:
    {? x >= 0 & g > 0 & air > 0 & parachute > air & T > 0 & m > 0;^@};
    {{subgame_c: {{{{{r := air;} ++ {r := parachute;}}^@};
    {t := 0;}}; {subgame_j: {x' = v, v' = r * v ^ 2 - g, t' = 1 &
    x >= 0 & v < 0 & t <= T}
    }}*}^@}

    Please analyze this game and identify the requested items
    (Angel vs. Demon control actions, control modes, overall
    control pattern).
- role: assistant
  answer:
    mode: null
    content: |
      1) Angel-controlled actions
      - Choice {{{r := air;} ++ {r := parachute;}}^@ — Angel chooses
        drag coefficient r each iteration, either r := air (low
        drag) or r := parachute (high drag).

      2) Demon-controlled actions
      - Initial test {? x >= 0 & g > 0 & air > 0 & parachute > air &
        T > 0 & m > 0}^@ — Demon test. Angel can assume these
        preconditions hold initially; otherwise, play won't
        proceed.
      - Looping subgame_c — Demon chooses how many iterations of
        subgame_c to perform (including possibly zero; finite but
        unbounded). He chooses whether to repeat for one more cycle
        or stop at the end of each iteration.
      - ODE duration: subgame_j - Demon controls how long to flow
        along the ODE x' = v, v' = r v^2 - g, t' = 1 within the
        evolution domain x >= 0, v < 0, t <= T. He may stop at any
        time before the domain boundary is reached, including zero
        time; he cannot continue past any boundary (for example x
        cannot go below 0). Since t is reset to 0 at the start of
        each iteration, the duration is always capped by T (though
        the other boundary conditions may force an earlier stop).

      3) Control modes (qualitative behaviors)
      - Mode AIR HIGH VEL (Angel chose r = air, low drag, v<-(g/air)^(1/2)):
        - air * v^2 > g, so v' > 0. Velocity increases (becomes less
          negative) over time, approaching terminal velocity
          -((g / air)^(1/2)) from below. Will not cross it.
      - Mode PARACHUTE HIGH VEL (Angel chose r = parachute, high drag,
        v<-(g/parachute)^(1/2)):
        - parachute * v^2 > g, so v' > 0. Velocity increases (becomes
          less negative) over time, approaching terminal velocity
          -((g / parachute)^(1/2)) from below, which is a higher (less
          negative) terminal velocity than in AIR mode.
      - Mode AIR EQU VEL (Angel chose r = air, low drag, and v is
        terminal velocity -((g / air)^(1/2))):
        - v' = air * v^2 - g = 0, so velocity remains constant at
          terminal velocity.
      - Mode PARACHUTE EQU VEL (Angel chose r = parachute, high drag,
        and v is terminal velocity -((g / parachute)^(1/2))):
        - v' = parachute * v^2 - g = 0, so velocity remains constant
          at terminal velocity, slower than in AIR mode.
      - Mode AIR LOW VEL (Angel chose r = air, low drag, and 0 > v >
        -((g / air)^(1/2))):
        - air * v^2 < g, so v' < 0. Velocity decreases (becomes more
          negative) over time. It approaches terminal velocity
          -((g / air)^(1/2)) from above but will not cross it.
      - Mode PARACHUTE LOW VEL (Angel chose r = parachute, high drag,
        and 0 > v > -((g / parachute)^(1/2))):
        - parachute * v^2 < g, so v' < 0. Velocity decreases (becomes
          more negative) over time, approaching terminal velocity
          -((g / parachute)^(1/2)) from above, but will not cross it.

      4) Overall control pattern
      - Some global preconditions hold initially (x >= 0, g > 0,
        air > 0, parachute > air, T > 0, m > 0).
      - Per iteration:
        1) Demon decides whether to continue for another iteration or
           stop (Demon loop).
        2) Angel picks drag coefficient r \in {air, parachute}
           (Angel choice).
        3) Reset timer t := 0 (deterministic assignment).
        4) Demon chooses ODE flow duration within domain (Demon ODE).
      - Across iterations: Demon decides how many iterations to
        perform and length of evolution per iteration. Angel's
        control lever is her binary choice of drag coefficient.
- role: user
  content: |-
    Game to analyze:
    (kA > 0 & kB > 0 & kC > 0 & kT >= 0 & T > 0 & A >= 0 & B >= 0 &
    C >= 0 & Temp >= 0 & (kB*A <= kA*B & kA*Temp + kT*A <= kA*Tmax |
    kA*B <= kB*A & kB*Temp + kT*B <= kB*Tmax)) -> < { { { isOn:=0;
    ++ isOn:=1; }; t:=0; { A' = - A*B*Temp*isOn*kA, B' = - A*B*Temp
    *isOn*kB, C' = A*B*Temp*isOn*kC, Temp' = A*B*Temp*isOn*kT, t' = 1
    & t<=T & A>=0 & B>=0 & C>=0 & Temp>=0 }; }* }^@ > Temp<=Tmax

    Please analyze this game and identify the requested items (Angel
    vs. Demon control actions, control modes, overall control
    pattern).
\end{verbatim}

\noindent 2. GetTactic is the second prompt in the pipeline (first in the proof loop). It gets a Bellerophon proof tactic from the LLM.

\begin{verbatim}
- role: system
  content: |
    You are trying to prove a differential game logic (dGL)
    theorem in KeYmaera X.

    You must provide a Bellerophon tactic to complete the
    proof.

    ## Background Information

    Here is some information about dGL.
    In differential game logic (dGL), there are two players:
    - **Angel**: The controllable player who makes choices we
      can control. Angel constructs include:
      - Angel choices: `a ++ b` (we choose between a and b)
      - Angel loops: `{a}*` (we choose how many iterations)
      - Angel tests: `?P` (we can proceed if P holds,
        otherwise we immediately lose)
      - Angel ODEs: `{x'=f(x) & Q}` (we control the duration)
      - Angel nondeterministic assignments: `x := *` (we
        choose the value)

    There is the dual operator `^@` that switches Angel and
    Demon roles in a game construct.
    Duals can be nested arbitrarily, so carefully check
    nesting levels to determine which player controls each
    action.
    Being within an odd number of `^@` operators indicates
    Demon control, while an even number (including zero)
    indicates Angel control.
    For example, in the game `{(a ++ b)^@; (c++d)}^@`, the
    choice between `a` and `b` is controlled by Angel, being
    within two `^@` operators, while the choice between `c`
    and `d` is controlled by Demon.
    Using this notation, we go over the Demon constructs
    below, indicating that they are Demon-controlled by
    wrapping them in a single `^@` operator.

    - **Demon**: The adversarial player who makes choices we
      cannot control. Demon has all the same constructs as
      Angel.
      - Demon choices: `{a ++ b}^@` (adversary chooses
        between a and b)
      - Demon loops: `{a*}^@` (adversary chooses how many
        iterations)
      - Demon tests: `{?P}^@` (adversary can proceed if P
        holds, and immediately loses otherwise. So Angel can
        *assume* in subsequent game play that P held at this
        point; otherwise she would already have won.)
      - Demon ODEs: `{x'=f(x) & Q}^@` (adversary controls the
        duration)
      - Demon nondeterministic assignments: `{x := *}^@`
        (adversary chooses the value)

    Additionally, the non-controlled operations are:
    - Assignment: `x := e` (deterministic assignment like in
      regular programs)
    - Sequential composition: `a; b` (first play a, then b)

    ## Notes
    - ODEs can sometimes introduce unmatched single quotes,
      so account for that when parsing.
    - In syntax such as { a++b; x:=* }*^@, the `^@` applies
      to the entire loop including its body. In this
      example, Demon controls the choice within the loop
      between a and b, as well as the assignment to x,
      because both are inside one `^@`.

    Here is some information about Bellerophon, the tactic
    language in which you must output.
    # Bellerophon Tactic Guide (KeYmaera X 5.1)

    ## 1. Overview

    **KeYmaera X** proves properties of hybrid systems using
    differential dynamic logic (dL) or hybrid games using
    differential game logic (dGL).  
    A **tactic** is a small program that automatically applies
    proof rules.  
    **Bellerophon** is the language for writing these tactics.  
    The following are instructions to write one Bellerophon
    script, ideally that closes the proof in one run, or at
    least provides new information to write such a tactic
    later.

    ---

    ## 2. Proof State Basics

    To write good tactics, you should internally keep track
    of proof state. A tactic maps one proof state to another.
    A proof state consists of zero or more goals (subgoals).
    Each goal is a sequent consisting of an antecedent
    (assumptions) and a succedent (formulas to prove).

    ### 2.1 Sequents and Antecedents in the KeYmaera X CLI

    A **sequent** in the CLI is written as:

    ```
    -1:  x>=0	GreaterEqual
    -2:  y=0	Equal
    ==> 1:  x>=0 GreaterEqual	
    ```

    which corresponds to the following sequent in
    mathematical notation:
    ```
    x>=0, y=0  |-  x>=0
    ```

    Formulas are indexed.
    Negative indices indicate antecedent formulas, and
    positive indices indicate succedent formulas.

    The antecedent lists assumptions.
    The succedent lists formulas to prove.

    We can assume all antecedent formulas are true and must
    use them to prove at least one succedent formula.
    You just need to prove one succedent formula to close the
    goal.

    ### 2.2 Goals and Subgoals
        
    Tactics transform the current goal into zero or more
    subgoals.
    You apply tactics until no subgoals remain, which would
    indicate the theorem is proved.

    ### 2.3 Formula Indexing

    Many tactics require a formula index.
    A common failure mode is using the wrong index.
    Use printing (Section 3.2) to inspect the state and
    indices while debugging.

    ---

    ## 3. Proving Interactions

    ### 3.1 Branches (Multiple Goals)

    When a tactic creates multiple goals (e.g., from orL), the
    proof state has separate branches inside one printed
    block.

    Example pattern:

    ===== State of Hard Cases ==== ElidingProvable(Provable{
            -1:  <formula for case 1>  Or
            -2:  <assumption>          Greater
            ...
          ==> 1:  <modal or post-condition>  Box
            from
            -1:  <formula for first disjunct>  And
            ...
          ==> 1:  <some goal 1>  Box
            with
            -1:  <formula for remaining disjuncts>  Or
            ...
          ==> 1:  <some goal 2>  Box}) =====
    print done (no progress, 1s)
    Done .

    Here:
      •	The from part shows the first branch (e.g., first
        disjunct, such as a>=0 & ...).
      •	The with part shows the other branch.
      •	Each branch has its own sequent (==> line) with
        indexed formulas.

    Each branch is a separate goal that must be closed.

    You should check if a subgoal is closed by printing it. If
    the subgoal is proved, you should see something like:

    ===== subgoal state ==== ElidingProvable(Provable{
        -1:  x>0	Greater
        ==> 1:  x>=0	GreaterEqual proved}) =====
        print done (proved, 1s)

    ### 3.2 Running and Rerunning Tactics; Printing State

    If you know of a tactic that you want to slightly modify,
    simply reuse the same text, and edit it as needed.

    To inspect the current proof state at any point, use the
    tactic:

    print("State of Hard Cases")

    This prints the current provable state (all goals and
    their indexed formulas) at the point where the tactic is
    invoked.
    Multiple print statements can be used in one tactic, to
    print the state at different points in the proof.
    The print message helps identify which part of the script
    produced that state.

    ### 3.3 Printing for Debugging

    For debugging, when the proof state has been reduced to
    one or more purely arithmetic subgoals, you can:
      1.	Use print to display the current arithmetic
        constraints.
      2.	Copy these constraints into an external tool such as
        Z3 to search for counterexamples by checking if the
        formula (AND of all antecedent constraints) -> (OR of
        all succedent constraints) is valid (i.e., its
        negation is unsatisfiable).

    ---

    ## 4. Bellerophon Essentials
      •	A script is a sequence of tactic expressions.
      •	t1 ; t2 runs t1 then t2.
      •	<(t1, t2) applies t1 to the first branch and t2 to
        the second branch. Note the absence of a closing angle
        bracket.
      •	A tactic either:
          •	succeeds and may produce subgoals,
          •	fails, or
          •	closes the current goal.

    ---

    ## 5. Primitive Tactics

    ### 5.1 Logical Tactics

    Common examples (indices refer to formulas in the sequent):
      - Implication: implyR(i)
            This transforms a goal of form |- A -> B into
            A |- B.
      - Conjunction: andR(i),
            This transforms a goal of form |- A & B into two
            subgoals: |- A and |- B.
        - andL(i)
            This transforms a goal of form 
              A & B, \Gamma |- \Delta into
            A, B, \Gamma |- \Delta, where the i-th antecedent formula in
            the original subgoal was A & B.
      - Disjunction: orR
            This transforms a goal of form |- A | B into two
            subgoals: |- A and |- B.
        - orL(i)
            This transforms a goal of form 
              A | B, \Gamma |- \Delta into
            two subgoals: 
              A, \Gamma |- \Delta and B, \Gamma |- \Delta,
            where the i-th antecedent formula in the original
            subgoal was A | B.
      - Universal: allR
            This transforms a goal of form |- \forallx. P(x) into
            |- P(y) for a fresh variable y.
        - allL(i):
            This transforms a goal of form 
              \forallx. P(x), \Gamma |- \Delta
            into P(t), 
              \Gamma |- \Delta for a term t, where the i-th
            antecedent formula in the original subgoal was
            \forallx. P(x).
        - allL2R(i)
            This transforms a goal of form 
              x=e, \Gamma |- \Delta, \forallx.
            P(x) into \Gamma |- \Delta, P(e),
            where the i-th antecedent formula in the
            original subgoal was x=e.
        - cut(formula)
            This introduces formula as an intermediate lemma
            to prove.
            It transforms a goal of form \Gamma |- \Delta into two
            subgoals:
            (1) Use subgoal: formula, \Gamma |- \Delta
            (2) Show subgoal: \Gamma |- formula.

    ### 5.2 Program / Modal Tactics

    For modal formulas like [\alpha]\varphi and ⟨\alpha⟩\varphi, 
    tactics handle:
    - assignments,
    - tests,
    - nondeterministic choice,
    - sequential composition,
    - duals.
    These reduce program structure to simpler logical or
    arithmetic goals. You will almost never need to use these
    low-level tactics directly, as they are all handled by the
    high-level `unfold` tactic.

    A structure not handled by unfold, that you will need to
    handle directly is loops. Usually this needs the loop
    rule, `loop("invariant", i)`, which transforms a goal of
    form  \Gamma |- [\alpha*]\varphi, \Delta where 
    [\alpha*]\varphi is at index i into three
    subgoals:
        1. init subgoal: \Gamma |- invariant, \Delta
        2. step subgoal: invariant |- [\alpha]invariant
        3. post subgoal: invariant |- \varphi
    Here, invariant is the (literal) formula you provide as a
    loop invariant.

    There is a second class of modality tactics used to
    simplify not the program inside the modality, but the
    postcondition after the modality or otherwise modify the
    modality. The most common are:
    - boxAnd(i)
        Transforms a goal of form [a](P&Q) to [a]P & [a]Q.
        Usually followed by andR(i) to split into two
        subgoals.
    - GV(i)
        Transforms a goal of form 
        \Gamma |- [a]P, \Delta to \Gammaconst |- P, \Deltaconst,
        where \Gammaconst and \Deltaconst are the subsets of \Gamma
        and \Delta that contain only formulas with constants (no
        variables).
    - discreteGhost("x", "x0", i)
        Adds a new variable x0 and the antecedent formula
        x = x0. The position, i, should be that of the
        modality formula (usually 1).

    ### 5.3 Differential Equation Tactics

    For goals involving differential equations of the form:

    ```
    [{x' = f(x) & Q}] P
    ```

    you can use the following high-level ODE tactics:
    - dIRule(i) (Differential Invariant)
        Applies a differential invariant rule to the ODE
        formula at index i.
        This proves formulas like [ODE] P where P may
        contain any comparison operator (<, <=, =, >=, >).
        The rule generates two proof obligations:
      1.	The postcondition P holds initially, and
      2.	The derivative condition P' satisfies the correct
          sign constraint (depending on the operator)
          throughout the ODE.
          For example, if in `[{x' = f(x) & Q}] P`,  P is
          p <= 0, the check is
          `Q |= [x':=f(x)] p' <= 0`;
            if P is p > 0, the check is
            `Q |= [x':=f(x)] p' > 0`;
            and similarly for the other operators.
            Along with Q, any assumptions from the
            antecedents that had only constants (no
            variables) are also kept.
            Note that you cannot use the invariant that
            you are trying to prove as an assumption in
            this subgoal, that is, you will not be able to
            assume p > 0 to prove p' > 0.
        If these obligations are proved, the ODE is
        handled without solving it explicitly.
        Note how to be able to prove subgoal 2, we may
        want to strengthen the domain constraint Q using
        dC first before applying dIRule.

    - dC(“\varphi”, i) (Differential Cut)
    Adds an additional domain constraint \varphi to the ODE at
    index i.
    This transforms a goal with `[{x'=f(x) & Q}] P` at
    position i into two subgoals:
        1. Use:
        The formula at position i becomes
        `[{x'=f(x) & Q & \varphi}] P`
        2. Show:
        The formula at position i becomes
        `[{x'=f(x) & Q}] \varphi`

    - dW(i) (Differential Weakening)
    Applies weakening to an ODE goal at index i.
    If the formula at index i was `[{x'=f(x) & Q}] P`, the
    new goal becomes `Q |- P`, which removes the ODE
    entirely. Along with Q, any assumptions from the
    antecedents that had only constants (no variables) are
    also kept.
    It makes sense to apply dW only after sufficiently
    strengthening the domain constraint Q using dC first.

    These tactics reduce ODE proof obligations to simpler
    goals.
    A common pattern is to cut in *solutions* or *bounds*
    of the variables involved in the ODE and postcondition
    using dC.
    In the branch where we need to show that the bound
    holds, it is common to use dIRule.
    In the other, "use" branch, we typically apply further
    dC cuts to accumulate more bounds.
    Then after accumulating sufficiently many such cuts, we
    apply dW followed by QE to finish the proof.
    Sometimes the order of cuts really matters, as earlier
    cuts may become important to closing the dIRule
    obligations of later cuts.

    Often, bounds are best expressed in terms of the
    initial values of variables before the ODE starts, e.g.,
    x <= x0 + v*t for a variable x with initial value x0
    and constant velocity v over time t.
    In this case, if x0 does not already exist, a useful
    tactic is `discreteGhost("x", "x0", i)`, which adds a
    new variable x0 and the antecedent formula x = x0. The
    position, i, should be that of the ODE formula (usually
    1).

    Besides dIRule, another useful tactic that is
    especially useful for ODEs with exponential solutions,
    is the Darboux tactic:
    `dbx("g", 1)`. Given a goal of form
    `A |- [x'=f(x) & Q] p>=0`, this tactic generates two
    subgoals: `A |- p>=0` and `Q |- p' >= g*p`, where p' is
    the Lie derivative of p along the ODE x'=f(x). Here,
    >= can be replaced by any comparison operator. dbx is
    usually the way to go when we want to prove something
    about a variable whose derivative is dependent on
    itself.
    In cases where you are trying to prove, e.g., p'>=0 and
    it is not immediate that p'>0, but it is true that p'
    >= g*p, you should use Darboux and not dIRule (here >=
    and > can be replaced by the other comparison
    operators).
    Unlike dIRule, dbx does not create further subgoals,
    and is itself a terminal tactic.

    ---

    ### 5.4 Arithmetic Tactics

    When there are no modalities left, the goal is purely
    arithmetic.
    Then, it is possible to use the `QE` tactic to close
    the goal.
    It is recommended to use `QE("Z3")` which runs QE with
    the Z3 SMT solver backend, as it is generally most
    effective.

    If QE times out, try hiding irrelevant formulas using
    `hide(i)` to remove formula with index i from the
    subgoal permanently.
    It can also sometimes help to use the `fullSimplify`
    tactic to make expressions simpler.

    If QE fails, there is likely an error in the proof or
    the theorem is not valid. It can help to look for a
    counterexample as described in Section 3.3.

    ---

    ## 6. Working With Indices

    ### 6.1 Index Semantics

    Indices refer to the current goal only:
      •	1 is the first succedent formula.
      •	2 is the second succedent formula, etc.
      •	-1 is the first antecedent formula.
        •	-2 is the second antecedent formula, etc.

    After some tactics, the sequent may change, so indices
    may point to different formulas.

    ### 6.2 Common Failure: Incorrect Positions

    A common reason for tactic failure is incorrect indices,
    e.g.:
      •	the index no longer refers to the formula you expect,
        or
      •	the selected formula does not match the rule (e.g.,
        andL on a non-& formula).

    Debugging pattern:
      1.	Insert print before the failing step.
      2.	Check the exact index and formula.
      3.	Update the index or tactic accordingly.

    ---

    ## 7. Combinators

    Useful control constructs:
      •	Sequential composition: `t1 ; t2`
        Run t1 then t2 on the resulting goals.
      •	Branching `<(t1, t2)`
        Apply t1 to the first branch and t2 to the second,
        after a rule that splits the goal.

    Combining these is generally enough to express proofs.

    ---

    ## 8. High-Level Tactics

    High-level tactics:
      •	perform several logical steps at once,
      •	handle common patterns in hybrid programs,
      •	push the proof toward arithmetic goals.

    The two most common ones are:
    - unfold
        Unfolds simpler constructs (assignments, tests,
        choices, sequences, antecedent conjunctions, dual,
        implication). Often changes sequent structure and
        might create multiple branches.
    - auto
        Applies a library of rules to try to automatically
        prove the goal. Generally succeeds on simpler goals.

    You can use these to make the tactic shorter, but the
    tradeoff is less control and harder debugging.

    ---

    ## 9. Debugging

    Common debugging workflows:
      1.	Insert print between suspicious steps to see the
        goal after each one. This can help identify where
        the proof diverges from expectations.
        2.	Remove steps until the error disappears. Once
        behavior is as expected, add more steps.

    _____

    ## 10. Additional Tips:

    ### 10.1 Syntactic:

    - Comments in Bellerophon are written using the syntax
      `/* This is a comment */`.
    - Do not use a concluding semicolon after the last
      tactic in your script. The semicolon is only a
      separator, not a terminator. Write `t1 ; t2`, not
      `t1 ; t2 ;`.
    - The branching combinator `<( ... )` does not have a
      closing angle bracket. Write `<(t1, t2)`, not
      `<(t1, t2)>`.
    - When using branching tactics like dC or cut, further
      development should be nested inside the branches using
      `<( ..., ... )` syntax. Write `dC("formula", pos);
      <(t1; dC("formula", pos); <(t2, t3), t4)`, not
      `dC("formula", pos); <(t1, t4) ; dC("formula", pos);
      <(t2 ; t3)`. Here, the first dC splits the goal into
      two branches. On the first branch we apply t1 followed
      by dC that further splits into two branches where we
      apply t2 and t3 respectively. On the second branch of
      the first dC, we apply t4.
    - Output along the lines of "Exception in thread "main"
      6:2 Lexer 6:2 Lexer does not recognize input at ..."
      indicates that there is a syntax error in the
      Bellerophon tactic. There can still be syntax errors
      that do not get caught by the lexer, though.
    - The Bellerophon parser supports only ASCII characters.
    - You can apply tactics like andL and orL only to
      formulas that have the corresponding logical
      connective at the top level. For example, andL would
      not apply to a | b & c because the top-level connective
      is |, since & binds more tightly. A good pattern is to
      use unfold to split up all top-level conjunctions at
      once, and then apply orL as needed.

    ### 10.2 Common Proof Patterns:
    - Sometimes, it is helpful to split out disjunctive cases
      in the antecedent using orL, because the disjuncts
      correspond to different control modes requiring
      different reasoning.
    - When there are top-level conjunctions in the
      antecedent, you can either split them one-by-one using
      andL, or use unfold to split all at once. The latter
      is often more concise, while the former gives more
      control. It is often necessary to split conjunctions
      to expose individual underlying assumptions to which
      further rules apply, e.g., disjunctions for orL case
      splitting, or equations on which to apply allL2R.
    - When you see errors about inapplicability of a tactic
      due to formula shape mismatch, try to work around by
      using a different tactic or by introducing cuts to
      reshape the goal into a more convenient form.
    - For time-triggered controllers, `t>=0` is often a
      useful invariant to cut in using dC, as it ensures we
      only have to reason about ODE properties after
      nonnegative time durations.

    ### 10.3 Other Advice:
    - Do not assume a branch is closed until you print and
      verify that it shows "proved". Therefore, after
      applying a terminal tactic like auto or QE, it is good
      practice to print the state right after to check that
      the tactic indeed closed the goal.
    - `unfold` steps are important and should not be
      forgotten. They often change the shape of the subgoal,
      and their can lead subsequent tactics to fail because
      the positions of sequents they expect are no longer
      present. It is a good idea to start tactic scripts
      with `unfold` to normalize the formula first.
    - For the full list of tactics, see https://github.com
    /LS-Lab/KeYmaeraX-release/tree/master/keymaerax-webui
    /src/main/resources/help/axiomsrules.
    - Duals are usually handled by `unfold`, which generally
      eliminates the outermost dual by flipping the modality
      (box to diamond or vice versa). Otherwise the
      individual tactics are `duald(i)` that handles diamond
      modality containing a dual game at position i and
      similarly and `dualb(i)` for box modality.
    - unfold generally handles implications, so implyR(1) is
      usually not necessary after unfold.
    - dbx is strict about the shape it allows: the
      postcondition formula after the box must be a single
      inequality or equation with 0 on the RHS. Rewrite the
      goal using a differential cut if needed to get it into
      this shape.
    - When you have two subsequent formulas, e.g., \Gamma |-
      [\alpha1*]\varphi, [\alpha2*]\varphi, 
      (i.e., [\alpha1*]\varphi at position 1 and
      [\alpha2*]\varphi at position 2), remember that you 
      only need to prove one of the subsequents.
      You can think carefully about which one is more 
      promising to prove, and get
      rid of the other using hide(2) or hide(1) as
      appropriate. But don't eliminate options too early: if
      you have a disjunction `f1 | f2` in your assumptions,
      it may be that [\alpha1*]\varphi is provable when f1 
      holds, and [\alpha2*]\varphi is provable when f2 
      holds, so you need to keep
      both options until you have split on the disjunction
      using orL.
    - If an arithmetic proof obligation times out on auto or
      QE, it is a good idea ot check if the goal is even
      valid by internally checking if a solver like Z3 can
      find a counterexample within some small timeout.

    ### 10.4 On Loop Invariants:

    identifying the right loop invariant for proving
    properties of loops (using tactic `loop("invariant", i)`)
    can be tricky, yet critical to allowing the subsequent
    proof branches succeed. A common pattern is to start with
    an invariant, and then strengthen or weaken it as
    required as one encounters counterexamples or
    difficulties in closing the subgoals, init, step and
    post, out of which step is usually the hardest to close.

    The invariant can be no stronger than the initial
    assumptions, that is, `\Gamma1 & \Gamma2` in 
    `\Gamma1, \Gamma2 |- [\alpha*]\varphi`.
    Anything stronger than `\Gamma1 & \Gamma2` would not
    allow the init subgoal to close.
    The invariant cannot be weaker than \varphi, the 
    postcondition. Anything weaker than \varphi would
    not allow the post subgoal to close.
    The challenge is then to weaken the assumptions enough,
    or strengthen the postcondition enough, that the
    resulting invariant is *inductive*, allowing the step
    subgoal to close.
    For example, consider the simple loop goal
    `x=1 |- [{?x>=1;^@; {x'=1}}*]x>=0`.
    x=1 is too strong as an invariant because as time
    elapses, x becomes larger than 1, so it is not true that
    x=1 |- [?x>=1;^@; {x'=1}]x=1.
    x>=0 is too weak as an invariant because Demon loses
    within the loop at the test when, e.g., x=0, so it is
    not true that x>=0 |- [?x>=1;^@; {x'=1}]x>=0.
    A suitable invariant is x>=1, which is weaker than the
    assumptions, stronger than the postcondition, and
    inductive in that if it holds initially, after running
    one iteration of the loop, it will continue to hold.

    Other advice for identifying an invariant:
    - A good heuristic is to start with the initial
      assumptions as the invariant, and then weaken them as
      needed, or start with the postcondition \varphi and
      strengthen it. Ask yourself: do the initial
      assumptions already look inductive? Does the
      postcondition already look inductive?
    - It is usually a good idea to retain all assumptions on
      signs of constants. These often end up being critical
      to inductiveness. You lose nothing by including
      assumptions in the invariant, unless they are
      non-inductive and stop holding after some time.
    - It is easy to get the invariant subtly wrong because
      of small inaccuracies or missing edge cases. But in
      the course of doing the proof, if you carefully
      analyze proof failures, you will discover the missing
      requirements, so do not hesitate to modify the
      invariant accordingly.
    - Finding the right invariant is the only way to do
      proofs about Angel loops in Demon modality games or
      Dually, Demon loops in Angel games 
      (i.e. `[\alpha*]\varphi` or `<\alpha*>\varphi`).
      The only way to prove the latter is to first
      convert to `[\alpha*]\varphi` form, e.g. using 
      unfold, and then apply the loop tactic.

    ### 10.5 On proving Angel choices in Angel games:

    This section applies dually also to Demon choices in
    Demon games.
    Consider an Angel choice in an Angel game, e.g.,
    `x=1 | x=-1 |- <{ {v:=1; ++ v:=-1;}^@; 
    {x'=v}}^@>(x>=1 | x<=-1)`.
    After applying `unfold`, we get a two subsequents:
    `x = 1 | x = -1 |- [{x'=1}](x >= 1  |  x <= -1),
    [{x'=-1}](x >= 1  |  x <= -1)`
    This reflects how we can choose which subsequent to prove
    based on which choice Angel would have strategically
    made.
    In this case, Angel would choose the first branch if x=1
    initially, and the second branch if x=-1 initially.
    Thus we could apply the following tactic to complete the
    proof, with the orL splitting on initial conditions and
    hideR choosing the right subsequent to then prove:
    ```
    unfold; orL(-1); <(
        hideR(2); dC("x>=1", 1); <(
            dW(1); QE,
            dIRule(1); <(
                QE,
                unfold; QE
            )
        ),
        hideR(1); dC("x<=-1", 1); <(
          dW(1); QE,
          dIRule(1); <(
              QE,
              unfold; QE
          )
        )
    )
    ```

    ### 10.6 Common Error Messages:

    #### Subgoal mismatch.
    Error messages of the form "<(e)(v) is only defined when
    len(e) = len(v), but 2!=0 subgoals (v)" indicates a
    branching combinator `<( ... )` has the wrong number of
    branches.
    In this example, a branching combinator for two branches
    was used, but the goal was actually already closed (0
    subgoals), and the fix is to remove the branching
    combinator since the subgoal is already proved.

    Sometimes, this error message is also reached by auto.
    This is a bug in the implementation of auto that is
    triggered on some formula shapes. If this happens, since
    you cannot control auto, just try to work around by using
    a different tactic or breaking down the proof into smaller
    steps.

    #### dW Bug.
    Sometimes, dW fails with an error message like 
    ```
        Not found: locator 'R=="a&b->c" of position tactic ...
        does not match anywhere in succedent of ... from ==>
        1: a -> c
    ```
    where typically a is the domain constraint of the ODE to
    which we are applying dW, c is the postcondition, and b
    consists of some formulas from the antecedent with only
    constants.

    This is a bug in KeYmaera X.
    A trick that often works around this bug is cutting in
    "1=1". Replace `dW(i); *subsequent tactic*` with
    `cut("1=1"); <(dW(i); *subsequent tactic*, QE("Z3") using
    "1=1")`.
    The syntax "using "1=1"" tells QE to ignore all antecedent
    and consequent formulas in this subgoal except "1=1".

    If the error persists, you can try first removing
    antecedent formulas with only constants that are
    appearing in the error message, and only then apply dW.
    Use hide(i) to hide the formula at position i.

    ---
    End of Bellerophon Tactic Guide
    ---

    ## Output Instructions

    ### Output Format

    In your response, first briefly summarize you reasoning.
    Then write a newline.
    Then, in a code block, write the new Bellerophon tactic
    you want KeYmaera X to try.

    ### Advice

    If you are not sure what to how to complete the proof, you
    can provide some initial steps.
    KeYmaera X will run these and you can review the open
    branches to complete the proof in later steps.
    Remember to use the print tactic at the end of branches to
    find out what their state is for future reference.

    Tactics can be sequentially combined using semicolons
    (e.g., `unfold; auto`. Note that there is no semicolon
    after the last tactic in a sequence).
    If the proof branches, e.g., after dC, then use the
    combinator `<(,)`, e.g., `dC; <( print("this is branch
    one."); print("this is a second print still in branch
    one."), print("this is branch two.") )`.

    Common mistakes to avoid:
    - When expanding on a previous tactic script, do not
      forget the initial `unfold` tactic because this can
      lead to subsequent tactic failing. You should retain
      the entire previous tactic and expand on it, unless
      there is an error requiring you to backtrack.
    - Do not modify the old tactic any more than strictly
      necessary. Tactics can be fragile and a single
      misplaced andL can break the proof.
    - Tactics should almost always begin with unfold to
      normalize the formula first.
    - Do not try more than one or two proof steps at once
      without checking if the proof state is like you expect.
      Do not try to solve more than one branch at a time
      because the complexity can quickly become unmanageable.
    - Do not forget to print the state of branches after QE or
      auto to confirm that they are indeed closed.
    - No unicode. Only ascii characters are allowed.

    You will next be shown 0 or more examples of how to answer
    such a prompt, which you can learn from, followed by the
    current question which you should answer.
- role: user
  content: |-
    The formula that you are trying to prove is:
    (t <= T & v = 0 & w = 0 & !(x = 0 & y = 0)) ->
    <{x' = v, y' = w, v' = om * w, w' = - om * v, t' = 1 &
    t <= T}^@>(!(x = 0 & y = 0) & (!x*w - y*v = 0 | (v = 0 &
    w = 0)))

    Now, please provide a Bellerophon tactic to try next.
    You should retain the entire previous tactic and expand
    on it, unless there is an error requiring you to
    backtrack.
    In your response, first briefly summarize your reasoning.
    Then write a newline.
    Then, in a code block, write the entire Bellerophon tactic
    you want KeYmaera X to try.
- role: assistant
  answer:
    mode: null
    content: |
      Intuitively we expect this formula to be valid because
      if initially v=w=0, then v and w remain 0, so x and y
      remain constant, so if we start not at the origin we
      stay not at the origin. In this way, it should in the
      end hold that !(x = 0 & y = 0) & (v = 0 & w = 0) which
      implies the postcondition.
      First lets unfold and differential-cut in the invariant
      v=0 & w=0.
      We can then print the state of the second branch to
      check that things look right so far.

      ```
      unfold;
      dC("v = 0 & w = 0", 1) <(
        label("Use cut"),
        label("Show cut");
        print("Show branch state.")
      )
      ```
- role: user
  content: |-
    The formula that you are trying to prove is:
    (t <= T & v = 0 & w = 0 & !(x = 0 & y = 0)) ->
    <{x' = v, y' = w, v' = om * w, w' = - om * v, t' = 1 &
    t <= T}^@>(!(x = 0 & y = 0) & (!x*w - y*v = 0 | (v = 0 &
    w = 0)))

    This is what you have been trying so far and the results.
    Pay careful attention to not repeat past mistakes and make
    progress based on what you have learned.

    We expect this formula to be valid because if initially
    v=w=0, then v and w remain 0, so x and y remain constant,
    so if we start not at the origin we stay not at the
    origin. In this way, it should in the end hold that
    !(x = 0 & y = 0) & (v = 0 & w = 0) which implies the
    postcondition.
    We have differential-cut in the invariant v=0 & w=0 and
    shown that is holds using the tactic below. We have
    printed the proof state of the "Show cut" branch and it
    was successfully closed.

    unfold;
    dC("w=0&v=0", 1); <(
      label("Use cut"); print("Use branch state."),
      label("Show cut");
      dC("v^2+w^2=0", 1); <(
        label("Use second cut"); dW(1) ; QE,
        label("Show second cut"); dIRule(1); <(
          unfold; QE("Z3");
          print("Show branch of dC, dI init. State after QE."),
          unfold; QE("Z3");
          print("Show branch of dC, dI inductiveness subgoal.
          State after QE.")
        )
      )
    )

    Both dIRule branches are proved.
    The proof state corresponding to the "Use cut" branch is:
    -1:  t<=T    LessEqual
    -2:  v=0     Equal
    -3:  w=0     Equal
    -4:  !(x=0&y=0) Not
     ==> 1:  [{x'=v,y'=w,v'=om*w,w'=-om*v,t'=1&t<=T&w=0&v=0}]
             (!(x=0&y=0)&(!x*w-y*v=0|v=0&w=0)) Box})
    So it remains to prove this goal.

    Now, please provide a Bellerophon tactic to try next.
    You should retain the entire previous tactic and expand
    on it, unless there is an error requiring you to
    backtrack.
    In your response, first briefly summarize your reasoning.
    Then write a newline.
    Then, in a code block, write the entire Bellerophon tactic
    you want KeYmaera X to try.
- role: assistant
  answer:
    mode: null
    content: |
      We want to show that after the ODE, !(x = 0 & y = 0) &
      (v = 0 & w = 0) and already have v=0&w=0 in the domain
      constraint. Intuitively, !(x = 0 & y = 0) because this
      is true initially neither x nor y changes. To cut in
      the fact that x and y do not change, lets dC in
      (x=x0 & y=y0) where x0 and y0 are the initial values of
      x and y, which we can introduce via discreteGhost
      (which should also automatically change the initial
      assumption !(x = 0 & y = 0) into !(x0 = 0 & y0 = 0)).
      Later, we can then show the "show" branch by dIRule
      because x and y do not change, and the "use" branch by
      dW followed by QE because x=x0, y=y0 and v=0 & w=0
      should be in the antecedents from the domain constraint
      and !(x0=0 & y0=0) should be in the antecedents because
      it has only free variables x0 and y0 which are not
      changed by the ODE.
      But first, lets print to make sure the subgoals look as
      expected.

      ```
      unfold;
      dC("w=0&v=0", 1); <(
          "Use":
          discreteGhost("x", "x0", 1);
          discreteGhost("y", "y0", 1);
          dC("x=x0&y=y0", 1); <(
              "Use-use":
              print("State of the use-use branch dW."),
              "Use-Show":
              print("State of use-show branch.")
          ),
          "Show":
          dC("v^2+w^2=0", 1); <(
              "Use":
              dW(1) ; QE,
              "Show":
               dIRule(1); <(
                unfold; QE("Z3"),
                unfold; QE("Z3")
          )
      )
      ```
- role: user
  content: |-
    The formula that you are trying to prove is:
    (((a >= 0 & a >= (-v)/T & end - (p + v*(T - t) + 1/2*a*
    (T - t)^2 + (v + a*(T - t))^2/(2*B)) >= 0) | (a < 0 &
    a >= (-B) & end - (p - v^2/(2*a)) >= 0) | (a < (-B) &
    end - (p + v^2/(2*B)) >= 0)) & A > 0 & B > 0 & T > 0 &
    v >= 0) -> <{p' = v, v' = a, t' = 1 & t <= T & v >= 0}^@>
    (end - (p + v^2/(2*B)) >= 0 & A > 0 & B > 0 & T > 0 &
    v >= 0)

    This is what you have been trying so far and the results.
    Pay careful attention to not repeat past mistakes and make
    progress based on what you have learned.

    Proposed Tactic:
    unfold;
    boxAnd(1); andR(1);
    print("State")
    .
    Outcome:   ===== State ==== ElidingProvable(Provable{
    ==> 1:  (a>=0&a>=(-v)/T&end-(p+v*(T-t)+1/2*a*(T-t)^2+
           (v+a*(T-t))^2/(2*B))>=0|a < 0&a>=-B&end-(p-v^2/
           (2*a))>=0|a < -B&end-(p+v^2/(2*B))>=0)&A>0&B>0&
           T>0&v>=0-><{{p'=v,v'=a,t'=1&t<=T&v>=0}}^@>
           (end-(p+v^2/(2*B))>=0&A>0&B>0&T>0&v>=0) Imply
      from
       -1:  a>=0&a>=(-v)/T&end-(p+v*(T-t)+1/2*a*(T-t)^2+
            (v+a*(T-t))^2/(2*B))>=0|a < 0&a>=-B&end-(p-v^2/
            (2*a))>=0|a < -B&end-(p+v^2/(2*B))>=0 Or
       -2:  A>0 Greater
       -3:  B>0 Greater
       -4:  T>0 Greater
       -5:  v>=0 GreaterEqual
    ==> 1:  [{p'=v,v'=a,t'=1&t<=T&v>=0}]end-(p+v^2/(2*B))>=0
            Box
      with
       -1:  a>=0&a>=(-v)/T&end-(p+v*(T-t)+1/2*a*(T-t)^2+
            (v+a*(T-t))^2/(2*B))>=0|a < 0&a>=-B&end-(p-v^2/
            (2*a))>=0|a < -B&end-(p+v^2/(2*B))>=0 Or
       -2:  A>0 Greater
       -3:  B>0 Greater
       -4:  T>0 Greater
       -5:  v>=0 GreaterEqual
    ==> 1:  [{p'=v,v'=a,t'=1&t<=T&v>=0}](A>0&B>0&T>0&v>=0)
            Box}) =====
    print done (no progress, 1s)
    Done .

    Now, please provide a Bellerophon tactic to try next.
    You should retain the entire previous tactic and expand
    on it, unless there is an error requiring you to
    backtrack.
    In your response, first briefly summarize your reasoning.
    Then write a newline.
    Then, in a code block, write the entire Bellerophon tactic
    you want KeYmaera X to try.
- role: assistant
  answer:
    mode: null
    content: |
      We see the proof has split into two goals, the first
      with consequent `[{p'=v,v'=a,t'=1&t<=T&v>=0}]end-(p+v^2/
      (2*B))>=0` and the second with consequent
      `[{p'=v,v'=a,t'=1&t<=T&v>=0}](A>0&B>0&T>0&v>=0)`. The
      second one should be easy to prove, so we will try that
      first with the auto tactic.
      We will print the state after trying auto to see if it
      was successful, or if more work is required. We should
      also print the state of the harder branch so that we
      can plan our next steps there.
      Our tactic will carefully retain the initial
      `unfold; boxAnd(1); andR(1)` to ensure we don't lose the
      progress we have already made.

      ```
      unfold; boxAnd(1); andR(1) <(
        label("harder");
          print("State in harder branch"),
        label("easier");
          auto; print("State after auto on easier branch")
      )
      ```
- role: user
  content: |-
    The formula that you are trying to prove is:
    (kA > 0 & kB > 0 & kC > 0 & kT >= 0 & T > 0 & A >= 0 &
    B >= 0 & C >= 0 & Temp >= 0 & (kB*A <= kA*B & kA*Temp +
    kT*A <= kA*Tmax | kA*B <= kB*A & kB*Temp + kT*B <= kB*
    Tmax)) -> < { { { isOn:=0; ++ isOn:=1; }; t:=0;
    { A' = - A*B*Temp*isOn*kA, B' = - A*B*Temp*isOn*kB,
    C' = A*B*Temp*isOn*kC, Temp' = A*B*Temp*isOn*kT,
    t' = 1 & t<=T & A>=0 & B>=0 & C>=0 & Temp>=0 }; }* }^@ >
    Temp<=Tmax

    Now, please provide a Bellerophon tactic to try next.
    You should retain the entire previous tactic and expand
    on it, unless there is an error requiring you to
    backtrack.
    In your response, first briefly summarize your reasoning.
    Then write a newline.
    Then, in a code block, write the entire Bellerophon tactic
    you want KeYmaera X to try.

\end{verbatim}

\noindent 3. Summarize summarizes the proof state in preparation for the next iteration of the verification loop.

\begin{verbatim}
- role: system
content: |
    You are given a step in an attempt to prove a
    differential game logic (dGL) formula in KeYmaera X,
    along with context about the previous state of the proof
    before the step, including the global plan and the
    things we have tried so far.
    Your task is to provide a new summary of the current
    state of the proof that now accounts for the latest step
    towards completing the proof and its outcome.

    Your proof summary will be used to then decide what the
    next step in the proof should be (if you decide to
    continue the proof) or how to change the formula (if you
    choose to revise the formula). Therefore it should
    remember all relevant information: previous mistakes and
    how you resolved them, location of sequents in subgoals
    that currently remain open, counterexamples discovered at
    the end of previous proof directions, etc. Expect that
    after many steps, the summary will be long. Be sure to
    include all relevant information.

    ## Input:
    1. The dGL formula that you are trying to prove.
    2. The summary of the proof state before this step.
    3. The proof tactic that you then ran this time to get
       one step closer to completing the proof.
    4. The outcome of running that tactic.

    Here, "step" means not a literal single step in the
    proof, but rather a new tactic that you applied that
    modifies the last one that was applied, either resolving
    errors or making progress towards completing the proof.

    ## More information:

    Here is more information about dGL:
    [This section is the same as in the GetTactic query.]

    Here is more information about KeYmaera X tactics:
    [This section is that same as in the Bellerophon Tactic
     guide in the GetTactic query.]

    ## Output Shape:

    You should return a detailed summary of the proof state.
    In the summary, you should include the following
    sections, many of which should carry over from the
    previous summary provided to you.

    1. The global proof plan so far. 
    Mostly carries over from previous summary unless the
    latest step surfaced an issue or new insight.
    2. (New) Previous Tactic (that is, the currently applied
    tactic in the latest step).
    Just copies over from input item 3. Do not make any
    changes or omissions.
    3. The current open subgoals with exact sequent locations
    if we know them.
    Copies over from input item 4.
    4. Any mistakes made so far and how we fixed them. This
    is important to avoid repeating the same mistakes in
    future steps. Guide the reader so they know what they
    must do to avoid repeating these mistakes.
    Partly carries over from previous summary. Might need to
    add new information if the latest step surfaced a new
    mistake or insight on how to fix a mistake.
    Remember to include all relevant context, such as the
    tactic leading up to the issue, the reason for the
    problem, and how to avoid it in the future.
    5. Previous proof directions that failed, if any, along
    with details such as the counterexamples discovered at
    the end of those directions.
    Mostly carries over from previous summary unless the
    latest step surfaced a new issue or insight.
    6. Any other relevant information about the proof state
    that will let a reader work on the proof from here
    assuming no other information besides the original
    formula to prove is provided to them.
    Partly carries over from previous summary. Might need to
    add new information if the latest step surfaced new
    relevant information. Remember that your summary and the
    original formula will be all the reader has to work from
    to proceed with the proof, so do not omit useful details.
    Some of these sections may be empty if the proof is still
    in early stages, but you should still include the
    section headers for completeness."
- role: user
content: |
    The formula that you are trying to prove is:
    (kA > 0 & kB > 0 & kC > 0 & kT >= 0 & T > 0 & A >= 0 &
    B >= 0 & C >= 0 & Temp >= 0 & (kB*A <= kA*B & kA*Temp +
    kT*A <= kA*Tmax | kA*B <= kB*A & kB*Temp + kT*B <= kB*
    Tmax)) -> < { { { isOn:=0; ++ isOn:=1; }; t:=0;
    { A' = - A*B*Temp*isOn*kA, B' = - A*B*Temp*isOn*kB,
    C' = A*B*Temp*isOn*kC, Temp' = A*B*Temp*isOn*kT,
    t' = 1 & t<=T & A>=0 & B>=0 & C>=0 & Temp>=0 }; }* }^@ >
    Temp<=Tmax

    Remember that nested duals alternate control back. For
    example, in the game `{(a ++ b)^@; (c++d);}^@`, the
    choice between `a` and `b` is controlled by Angel, while
    the choice between `c` and `d` is controlled by Demon.

    Here is an analysis of the formula, discussing what Angel
    vs. Demon controls and the game's control modes.
        1) Angel-controlled actions
    - None within the game body. The entire loop is dualized,
      so all otherwise-Angel constructs inside it are under
      Demon control.

    2) Demon-controlled actions
    - Looping: { … }*^@ — Demon chooses how many iterations
      to execute (including zero; any finite number).
    - Discrete choice each iteration: { isOn := 0; ++
      isOn := 1; } — Within the single ^@, Demon chooses
      whether to set isOn to 0 (OFF) or 1 (ON) for that
      iteration.
    - ODE duration each iteration: { A' = -A B Temp isOn kA,
      B' = -A B Temp isOn kB, C' = A B Temp isOn kC,
      Temp' = A B Temp isOn kT, t' = 1 & t <= T & A >= 0 &
      B >= 0 & C >= 0 & Temp >= 0 } — Demon controls how long
      to flow, subject to the evolution domain. He may stop
      immediately or at any time before hitting t = T or a
      state-constraint boundary (he cannot cross
      A,B,C,Temp < 0 or t > T).

        - Non-controlled deterministic assignments
    - t := 0 at the start of each iteration (resets the
      per-iteration clock; not a choice).
    - The chosen isOn := 0 or 1 is a deterministic assignment
      after Demon picks the branch.

    3) Control modes (qualitative behaviors)
    The per-iteration discrete choice creates two qualitative
    modes; Demon may switch modes between iterations by
    choosing isOn.

    - OFF mode (isOn = 0)
    - Dynamics: A' = B' = C' = Temp' = 0, t' = 1.
    - Behavior: State (A,B,C,Temp) is frozen; only the timer
      t increases up to T. This is an equilibrium for the
      chemistry/temperature.
    - Invariants: All polynomial combinations below remain
      constant (trivially, since derivatives are zero).

    - ON mode (isOn = 1)
    - Dynamics: A' = -kA A B Temp <= 0, B' = -kB A B Temp <= 0,
      C' = kC A B Temp >= 0, Temp' = kT A B Temp >= 0, with
      t' = 1 and domain t <= T, A,B,C,Temp >= 0.
    - Monotonicity: A(t), B(t) are nonincreasing; C(t), Temp(t)
      are nondecreasing. If any of A, B, or Temp is zero, all
      chemical rates are zero (equilibrium subset).
    - Polynomial invariants (valid in both modes because every
      derivative carries the common factor A B Temp isOn):
        - Stoichiometry: kB A - kA B = const (so the sign of
          kB A - kA B never changes; the “limiting reagent”
          relation stays fixed).
        - Thermal-reagent invariants:
        - kA Temp + kT A = const
        - kB Temp + kT B = const
        - Yield invariants (optional but useful): kC A + kA C =
          const and kC B + kB C = const.
    - Bounds from invariants (safety envelopes chosen by initial
      branch):
        - If initially kB A <= kA B and kA Temp + kT A <= kA Tmax,
          then for all future time Temp <= Tmax because A >= 0 and
          kA Temp + kT A stays constant.
        - If initially kA B <= kB A and kB Temp + kT B <= kB Tmax,
          then likewise Temp <= Tmax from the second invariant.
        - These two cases are disjoint branches selected by the
          initial stoichiometric ordering; the ordering is
          preserved by kB A - kA B = const.
    - Asymptotes/equilibria: Trajectories move along invariant
      manifolds until one reagent depletes (A → 0 or B → 0), then
      all rates drop to zero. Temperature approaches a finite
      limit bounded by the corresponding invariant, never
      exceeding Tmax under the given initial branch. If kT = 0,
      Temp is constant.

    4) Overall control pattern
    - Preconditions (outside the game) ensure positivity and pick
      one of two invariant safety envelopes: either
      kA Temp + kT A <= kA Tmax when kB A <= kA B, or
      kB Temp + kT B <= kB Tmax when kA B <= kB A. The
      stoichiometric ordering is invariant, so the chosen
      envelope remains applicable.
    - Iteration (repeats a Demon-chosen finite number of times):
    1) Demon picks isOn \in {0, 1} (OFF or ON for this iteration).
    2) Deterministic reset t := 0.
    3) Demon chooses how long to flow within the ODE domain
       (0 <= duration <= T).
        - OFF: no state change except t.
        - ON: A,B decay; C,Temp grow; all along the invariant
          manifolds keeping Temp within the selected bound.
    - Demon controls both switching (isOn) and dwell time (ODE
      duration) as well as the number of repetitions; Angel has
      no in-game choices. The qualitative modes are OFF
      (equilibrium) and ON (reaction with monotone decay/growth
      along invariant manifolds), with safety enforced by the
      polynomial invariants regardless of Demon's switching and
      timing.

    Carefully account for who controls which choice while
    constructing any counterexamples. The counterexample cannot,
    e.g., require Demon to make a choice that Angel controls.

    Then you performed the following proof attempt:
    We need an inductive invariant for the Demon loop. A
    natural invariant is the conserved linear bounds: either
    kA*Temp+kT*A<=kA*Tmax when kB*A<=kA*B, or
    kB*Temp+kT*B<=kB*Tmax when kA*B<=kB*A. Together with sign
    assumptions kA>0,kB>0,kT>=0 and A>=0,B>=0, this invariant
    implies Temp<=Tmax for the post branch. We set up the loop
    with this invariant, automatically discharge the init and
    post branches (splitting the disjunction in the post), and
    print the step branch after unfolding so we can handle ODE
    preservation next.

    ```
    unfold;
    loop("kA>0 & kB>0 & kT>=0 & A>=0 & B>=0 & ( (kB*A <= kA*B &
    kA*Temp + kT*A <= kA*Tmax) | (kA*B <= kB*A & kB*Temp + kT*B
    <= kB*Tmax) )", 1); <(
    /* init: assumptions imply invariant */
    unfold; auto; print("Init branch after auto."
        ),
    /* step: unfold body, we will handle preservation next */
    unfold; print("
        Step branch after unfolding body."),
    /* post: invariant implies Temp<=Tmax */
    andL(-1); andL(-1); andL(-1); andL(-1); andL(-1);
    orL(-1); <(
        andL(-1); QE("
        Z3"); print("Post branch case 1 closed (from kA*Temp +
        kT*A <= kA*Tmax)."),
        andL(-1); QE("Z3"); print("Post branch case 2 closed
        (from kB*Temp + kT*B <= kB*Tmax)."
        )
    )
    )
    ```

    The outcome was:
    Tactic andL applied at -1 on a non-matching expression in
    ElidingProvable(Provable{
    -1:  kA>0&kB>0&kT>=0&A>=0&B>=0&(kB*A<=kA*B&kA*Temp+kT*A<=
    kA*Tmax|kA*B<=kB*A&kB*Temp+kT*B<=kB*Tmax)
    And
    -2:  kA>0	Greater
    -3:  kB>0	Greater
    -4:  kC>0	Greater
    -5:  kT>=0
    GreaterEqual
    -6:  T>0	Greater
    ==> 1:  [{isOn:=0;++isOn:=1;}t:=0;{A'=-A*B*Temp*isOn*kA,
    B'=-A*B*Temp*isOn*kB,C'=A*B*Temp*isOn*kC,Temp'=A*B*Temp*
    isOn*kT,t'=1&t<=T&A>=0&B>=0&C>=0&Temp>=0}](kA>0&kB>0&kT
    >=0&A>=0&B>=0&%28kB*A<=kA*B&kA*Temp+kT*A<=kA*Tmax|kA*B
    <=kB*A&kB*Temp+kT*B<=kB*Tmax%29)
    Box
    from
    -1:  kA>0	Greater
    -2:  kB>0	Greater
    -3:  kC>0	Greater
    -4:  kT>=0	GreaterEqual
    -5:  T>0	Greater
    -6:  kA>0	Greater
    -7:  kB>0&kT>=0&A>=0&B>=0&(kB*A<=kA*B&kA*Temp+kT*A<=kA*
    Tmax|kA*B<=kB*A&kB*Temp+kT*B<=kB*Tmax)
    And
    ==> 1:  [{isOn:=0;++isOn:=1;}t:=0;{A'=-A*B*Temp*isOn*kA,
    B'=-A*B*Temp*isOn*kB,C'=A*B*Temp*isOn*kC,Temp'=A*B*Temp
    *isOn*kT,t'=1&t<=T&A>=0&B>=0&C>=0&Temp>=0}](kA>0&kB>0&
    kT>=0&A>=0&B>=0&%28kB*A<=kA*B&kA*Temp+kT*A<=kA*Tmax|kA*
    B<=kB*A&kB*Temp+kT*B<=kB*Tmax%29)
    Box})

    in <( (unfold;auto;print("Init branch after auto.")),
    (unfold;print("Step branch after unfolding body.")), 
    >>_<< )
    (Failed on branch (andL(-1);andL(-1);andL(-1);andL(-1);
    andL(-1);orL(-1);<( (andL(-1);QE("Z3");
    print("Post branch case 1 closed (from kA*Temp + kT*A <=
    kA*Tmax)."
    )), (andL(-1);QE("Z3");
    print("Post branch case 2 closed (from kB*Temp + kT*B <=
    kB*Tmax)."
    )) )))

    in <( (unfold;auto;print("Init branch after auto.")),
    (unfold;print("Step branch after unfolding body.")), 
    >>_<< )
    [Full trace elided]
    at edu.cmu.cs.ls.keymaerax.bellerophon.CoreLeftTactic.
    computeResult(BellerophonSyntax.scala:623)
    t... 40 more

    [launcher] Restarting KeYmaera X with sufficient stack
    space
    /usr/lib/jvm/java-17-openjdk-amd64/bin/java -Xss20M -da
    -jar /home/.../keymaerax.jar
    -launch -prove /home/.../keymaerax_conjectures/kA0.kyx
    -tactic unfold;
    loop("kA>0 & kB>0 & kT>=0 & A>=0 & B>=0 & ( (kB*A <= kA*B &
    kA*Temp + kT*A <= kA*Tmax) | (kA*B <= kB*A & kB*Temp + kT*B
    <= kB*Tmax) )", 1); <(
    /* init: assumptions imply invariant */
    unfold; auto; print("Init branch after auto."
    ),
    /* step: unfold body, we will handle preservation next */
    unfold; print("
    Step branch after unfolding body."),
    /* post: invariant implies Temp<=Tmax */
    andL(-1); andL(-1); andL(-1); andL(-1); andL(-1);
    orL(-1); <(
    andL(-1); QE("
    Z3"); print("Post branch case 1 closed (from kA*Temp + kT*A
    <= kA*Tmax)."),
    andL(-1); QE("Z3"); print("Post branch case 2 closed (from
    kB*Temp + kT*B <= kB*Tmax)."
    )
    )
    )
    -verbose -timeout 30
    KeYmaera X Prover 5.1.2
    Use option -help for usage and license information
    Proving entries from 1 files
    Proving /home/.../keymaerax_conjectures/kA0.kyx#Delphyne/kA0
    ...
    unfold... unfold done (transformed goal(s), 44ms)
    loop("kA>0&kB>0&kT>=0&A>=0&B>=0&(kB*A<=kA*B&kA*Temp+kT*A<=
    kA*Tmax|kA*B<=kB*A&kB*Temp+kT*B<=kB*Tmax)"
    , 1)... loop done (added 2 goal(s), 60ms)
    unfold... unfold done (no progress, 2ms)
    auto... auto done (proved, 16ms)
    print("Init branch after auto.")... INFO  ===== Init branch
    after auto. ==== ElidingProvable(Provable{   -1:  kA>0
    Greater
    -2:  kB>0	Greater
    -3:  kC>0	Greater
    -4:  kT>=0	GreaterEqual
    -5:  T>0	Greater
    -6:  A>=0	GreaterEqual
    -7:  B>=0	GreaterEqual
    -8:  C>=0	GreaterEqual
    -9:  Temp>=0	GreaterEqual
    -10:  kB*A<=kA*B&kA*Temp+kT*A<=kA*Tmax|kA*B<=kB*A&kB*Temp+kT*B
    <=kB*Tmax
    Or
    ==> 1:  kA>0&kB>0&kT>=0&A>=0&B>=0&(kB*A<=kA*B&kA*Temp+kT*A
    <=kA*Tmax|kA*B<=kB*A&kB*Temp+kT*B<=kB*Tmax)
    And proved}) =====
    print done (proved, 4ms)
    INFO  ===== Step branch after unfolding body. ====
    ElidingProvable(Provable{
    -1:  kA>0&kB>0&kT>=0&A>=0&B>=0&(kB*A<=kA*B&kA*Temp+kT*A
    <=kA*Tmax|kA*B<=kB*A&kB*Temp+kT*B<=kB*Tmax)
    And
    -2:  kA>0	Greater
    -3:  kB>0	Greater
    -4:  kC>0	Greater
    -5:  kT>=0
    GreaterEqual
    -6:  T>0	Greater
    ==> 1:  Temp<=Tmax	LessEqual
    from
    -1:  kA>0
    Greater
    -2:  kB>0	Greater
    -3:  kC>0	Greater
    -4:  kT>=0	GreaterEqual
    -5:  T>0	Greater
    -6:  kA>0	Greater
    -7:  kB>0	Greater
    -8:  kT>=0
    GreaterEqual
    -9:  A>=0	GreaterEqual
    -10:  B>=0	GreaterEqual
    -11:  kB*A<=kA*B&kA*Temp+kT*A<=kA*Tmax|kA*B<=kB*A&kB*
    Temp+kT*B<=kB*Tmax
    Or
    ==> 1:  Temp<=Tmax	LessEqual}) =====
    Done /home/.../keymaerax_conjectures/kA0.kyx#Delphyne/kA0
    (failed)
    FAILED Delphyne/kA0: tactic=user,tacticsize=-1,budget=30[s],
    duration=-1[ms],qe=-1[ms],rcf=-1,steps=-1


    Now please provide your summary.
\end{verbatim}

This ends the listing of the Verification Prompts.

\section{Synthesis Prompts}
\label{app:synthesis-prompts}

Similarly to \rref{app:verification-prompts}, this section lists the synthesis prompts.
Again prompts are listed in the order in which they are called during the synthesis process.

\noindent 1. AnalyzeGameLoop is the same as in \rref{app:verification-prompts}.

\noindent 2. PlanControlStrategy uses the analysis so far and the provided informal guideline to come upw ith a strategy that Angel should play.
\begin{verbatim}
- role: system
  content: |
    You are given a differential game logic (dGL) loop
    and a desired postcondition that Angel (the controller)
    aims to maintain.

    Your task is to propose a succinct plan describing the
    Angel's control strategy to guarantee the postcondition.
    You just need to find one control strategy that works,
    not necessarily the optimal or most general one. You may
    be provided guidelines about what kind of control
    solution to find. If so, the chose strategy should be in
    accordance with those guidelines.

    Background on dGL:
    [Identical to the dGL syntax guide of AnalyzeGameLoop
    followed
    by the Bellerophon Tactic guide of GetTactic.]

    You are given:
    - The full loop.
    - The postcondition that Angel aims to maintain.
    - A prior analysis of Angel vs. Demon controls and
      control modes.
    - Optionally, guidelines about what kind of control
      solution to find.

    Goal:
    Produce a strategy that maintains safety. If the control
    involves a Demon controlled loop, carefully account for
    infinite iterations.

    Keep the strategy simple and minimalist. Avoid
    unnecessarily long or involved control strategies.

    It is OK if your strategy requires special initial
    conditions or assumptions to work,  such as starting at
    specific equilibrium point or assuming the sign of some
    constants. Just be clear about what these assumptions
    are.

    Requirements for the plan:
    - State any initial conditions/assumptions the plan
      relies on, and what control mode it starts in. You
      must choose exactly one control mode to start in so as
      to make subsequent analysis easier. If the control
      mode requires specific initial conditions, this should
      be part of your assumptions.
    - After this, list Angel's strategy as a sequence of
      steps, corresponding to control modes.
        * Each step should start with an action word (e.g.,
          "Choose", "Set", "Switch to"), possibly with a
          conditional clause (e.g., "When X holds"). End
          with a temporal indicator (e.g., "repeatedly",
          "until Y", "forever").
        * Each step should focus on Angel's actions. Mention
          Demon only when necessary to motivate Angel's
          tactics, or when he forces a mode change.
        * Expect to transition between modes mainly by using
          Angel's controllable actions and choices.
        * Use clear English, no mathematical notation unless
          necessary for clarity. This will later be
          translated into formal constraints.
        * Produce a numbered list of Angel's decisions.
        * Be as concise as possible without superfluous
          words. Aim for no more than ~120 words.
        * Avoid hedging or verbose meta-commentary; write
          only Angel's action steps.

    The input includes the game and a prior analysis
    identifying Angel vs. Demon controls. Use that analysis
    to ensure that the plan accounts for any possible
    adversarial behavior.

    - Carefully consider the differential equation. There
      can be coupling between variables that affects your
      reasoning. 
    - Based on the dynamics, you may some times need to come
      up with upper/lower bounds of quantities and use that
      to derive sound preconditions. It's best to use
      simplistic, polynomial bounds so that they verify more
      easily.

    Output Structure:
    - Start with a one-line comment on what new region the
      strategy is trying to cover.
    - Then write section per the requirements listed above.
        * Write a description of initial
          conditions/assumptions, and mode.
        * Then proceed with the plan steps.

    You will be shown zero or more examples of how to
    answer, followed by a new question to answer.
- role: user
content: |-
    Full Game:
    {? A > 0 & B > 0 & kA > 0 & kB > 0 & kC > 0 & kT >= 0 &
    T > 0;^@}; {subgame_c: {{{{{isOn := 0;} ++ {isOn := 1;}}};
    {t := 0;}}; {subgame_j: {A' = - A * B * Temp * isOn * kA,
    B' = - A * B * Temp * isOn * kB, C' = A * B * Temp * isOn *
    kC, Temp' = A * B * Temp * isOn * kT, t' = 1 & t <= T &
    A >= 0 & B >= 0 & C >= 0 & Temp >= 0}
    }}*^@}

    Postcondition Angel aims to maintain after the loop:
    Temp <= Tmax

    Prior control analysis:
    1) Angel-controlled actions
    - None. The only potentially Angelic constructs (the
      choice isOn := 0 ++ isOn := 1 and the ODE duration) are
      inside a loop that is wrapped by a single dual ^@, so
      they are Demon-controlled.

    2) Demon-controlled actions
    - Initial test {? A > 0 & B > 0 & kA > 0 & kB > 0 & kC > 0 &
      kT >= 0 & T > 0}^@ (subsequent play assumes these hold;
      otherwise Demon would have lost).
    - Loop count {subgame_c: {...}*}^@: Demon chooses how many
      iterations (including zero).
    - Inside each iteration (because the whole loop is under
      one ^@):
    - Branch choice {{{isOn := 0;} ++ {isOn := 1;}}: Demon
      chooses whether the reaction is Off or On for that
      iteration's flow.
    - ODE duration in subgame_j: Demon controls how long to
      flow along
        A' = - A B Temp isOn kA,
        B' = - A B Temp isOn kB,
        C' =   A B Temp isOn kC,
        Temp' = A B Temp isOn kT,
        t' = 1
        subject to the evolution domain t <= T, A >= 0, B >= 0,
        C >= 0, Temp >= 0. He may stop anytime before hitting
        the domain boundary; he cannot exceed it.
    - Deterministic assignment t := 0 is not a choice (it
      always executes).

    3) Control modes (qualitative behaviors and invariants)
    - OFF mode (isOn = 0, or equivalently any boundary where
      A = 0 or B = 0 or Temp = 0 makes the vector field zero):
    - A' = B' = C' = Temp' = 0, t' = 1. Chemical state
      (A,B,C,Temp) is frozen; only the timer t increases until
      t = T.
    - Equilibrium for the chemical variables.
    - ON mode (isOn = 1, with A,B,Temp > 0):
    - Monotonicity: A' = -kA A B Temp <= 0, B' = -kB A B Temp
      <= 0, C' = kC A B Temp >= 0, Temp' = kT A B Temp >= 0
      (Temp' = 0 when kT = 0). Thus A,B decrease; C increases;
      Temp is nondecreasing.
    - Domain bounds: A,B,C,Temp remain nonnegative; 0 <= t <= T
      within each flow.
    - Polynomial invariants (valid in both ON and OFF, hence
      globally):
        - kB A - kA B = const.
        - C + (kC/kA) A = const and C + (kC/kB) B = const.
        - Temp + (kT/kA) A = const and Temp + (kT/kB) B = const
          (these degenerate correctly when kT = 0).
    - Equilibria within ON: if A = 0 or B = 0 or Temp = 0 is
      reached, the system becomes stationary (same as OFF for
      the chemical variables).
    - Asymptotic/terminal behavior: Trajectories move along the
      invariant line kB A - kA B = const toward the nonnegative
      axes; one reactant may be exhausted (A=0 or B=0) while the
      other remains positive. C accumulates according to the
      invariants; Temp can only increase (if kT>0) or remain
      constant (if kT=0).

    Special parameter submode:
    - Isothermal ON (kT = 0): Same as ON but Temp' = 0, so Temp
      is constant; other invariants and monotonicity still hold.

    4) Overall control pattern
    - Start with a Demon test ensuring A > 0, B > 0, kA,kB,kC
      > 0, kT >= 0, T > 0 (Angel may assume these).
    - Then repeat zero or more iterations (Demon loop):
    1) Demon chooses isOn \in {0,1}.
    2) Deterministically reset t := 0.
    3) Demon evolves the ODE for any duration up to the domain
       boundary (notably t <= T), stopping at will.
    - By sequencing iterations, Demon effectively applies a
      piecewise-constant ON/OFF control with segment durations
      at most T, steering the monotone reaction dynamics
      (consume A,B and produce C, increase Temp if kT>0) or
      freezing them when OFF.

    More information from previous proof attempts:
    Find the control solution for the situation where the
    reactant runs out before temperature can become too large.

    Provide the control strategy now.
\end{verbatim}

\noindent 3. The next query, which gets the LLM to guess a subvalue for a subgame, has three variants, one each for loops, nondeterministic assignments, and ODEs. We show the loop variant, GuessLoopInvControl, but the others are similar.

\begin{verbatim}
- role: system
  content: |
    Your task is to guess a loop invariant for a game loop
    given the loop body and the postcondition.

    A loop of the form {\alpha}*^@ executes the body \alpha
    zero or more times (and we must account for every
    possibility). A loop invariant I is a property that:
    1. Holds before the loop starts (precondition implies I)
    2. Is preserved by each iteration of the loop body (if I
       holds before \alpha, then I holds after \alpha)
    3. Combined with loop termination, implies the
       postcondition

    A loop of the form {\alpha}* executes the body \alpha
    zero or more times, and how many times is our choice.
    In this case, the invariant should be such if we run the
    loop enough (finite) times, the postcondition will be
    satisfied.

    While choosing the precondition/invariant, you need to
    consider the essential properties that remain true
    throughout all possible iterations of the loop body,
    considering both angel and demon choices.

    The loop invariant should be typically involves bounds,
    safety properties, or relationships between variables

    Return your answer as a logical formula in KeYmaera X
    syntax:
    - Use & for logical AND
    - Use | for logical OR  
    - Use ! for logical NOT
    - Use = for equality
    - Use <, <=, >, >= for comparisons
    - Use parentheses for grouping
    Use only variables you see in the problem. For example,
    if the problem has `x`, use `x`, not `x(0)`.

    Provide up to 4 lines describing your reasoning. Then
    after a blank line, write a code block with the loop
    invariant in KeYmaera X syntax.

    Some syntax instructions to keep in mind are:
    - Do not add comments to your code blocks, as this may
      cause parsing issues.

    Some information about dGL:
    "@" is the dual symbol that flips who controls the action
    between Angel and Demon.

    For example:
    In game "x := *", Angel gets to choose what value is
    assigned.
    The weakest precondition for such an assignment with
    postcondition P is the formula that holds *for some*
    value that could be assigned to x.

    In game "{x := *;}^@", any real number can be assigned,
    per Demon's decision.
    The weakest precondition for such an assignment with
    postcondition P is the formula that holds *for all*
    possible values that could be assigned to x.

    In game "{{ x:= *;}^@}^@" the control flips twice, so
    Angel gets to choose the assignment to x.

    - Carefully consider the differential equation. There
      can be coupling between variables that affects your
      reasoning. 
    - Based on the dynamics, you may some times need to come
      up with upper/lower bounds of quantities and use that
      to derive sound preconditions. It's best to use
      simplistic, polynomial bounds so that they verify more
      easily.
- role: user
  content: |
    Loop body:
    {isOn := 0; ++ isOn := 1;}^@; t := 0; {A' = - A * B *
    Temp * isOn
    * kA, B' = - A * B * Temp * isOn * kB, C' = A * B * Temp *
    isOn * kC, Temp' = A *
    B * Temp * isOn * kT, t' = 1 & t <= T & A >= 0 & B >= 0 &
    C >= 0 & Temp >= 0}^@

    Postcondition:
    Temp <= Tmax

    Overall game in which this loop appears (for context): 
    ? A > 0 & B > 0 & kA > 0 & kB > 0 & kC > 0 & kT >= 0 &
    T > 0;^@; {{isOn := 0; ++
    isOn := 1;}; t := 0; {A' = - A * B * Temp * isOn * kA,
    B' = - A * B * Temp * isOn
    * kB, C' = A * B * Temp * isOn * kC, Temp' = A * B * Temp
    * isOn * kT, t' = 1 & t
    <= T & A >= 0 & B >= 0 & C >= 0 & Temp >= 0}}*^@

    Here is an analysis of the game, the strategy that Angel
    is trying to play, and a log of things that you have
    already tried in the past.
    Carefully consider this information.

    Analysis of game:
    1) Angel-controlled actions
    - None. All controllable constructs in the body lie
      inside a single ^@ (the outer loop dual), so their
      control is flipped to Demon. There are no remaining
      Angel decisions.

    2) Demon-controlled actions
    - Initial test {? A > 0 & B > 0 & kA > 0 & kB > 0 & kC > 0 &
      kT >= 0 & T > 0}^@: Demon test. Angel may assume these
      hold initially; otherwise she already wins.
    - Loop {subgame_c; subgame_j}*^@: Demon loop. Demon
      decides how many iterations to execute (including
      zero).
    - Within each iteration:
    - Choice isOn := 0 ++ isOn := 1: Demon chooses whether
      the reaction is off or on during this iteration's flow
      (the ++ is inside the loop's ^@).
    - ODE duration in subgame_j: Demon controls how long to
      flow subject to the evolution domain (t <= T,
      A,B,C,Temp >= 0). He may stop immediately or any time
      before hitting the boundary. - Deterministic
      assignments (no player control once chosen to execute):
    - t := 0 each iteration (resets the local clock)
    - The chosen assignment to isOn is deterministic once the
      branch is selected.

    3) Control modes (qualitative behavior)
    - OFF mode (isOn = 0):
    - Dynamics: A' = B' = C' = Temp' = 0, t' = 1. State
      (A,B,C,Temp) is constant while time advances to at most
      T.
    - Equilibria: Entire state subspace is an equilibrium for
      the state variables.

    - ON mode (isOn = 1):
    - Dynamics: A' = -kA A B Temp <= 0, B' = -kB A B Temp <= 0,
      C' = kC A B Temp >= 0, Temp' = kT A B Temp >= 0 (with kT >=
      0). If any of A=0 or B=0 or Temp=0, then
      A'=B'=C'=Temp'=0 (equilibrium).     - Monotonicity (when
      A,B,Temp > 0): A and B strictly decrease; C and Temp
      strictly increase (Temp is constant if kT = 0).
    - Global linear invariants (valid in both ON and OFF,
      since OFF has zero derivatives):
        - kC A + kA C = const
        - kC B + kB C = const
        - Temp - (kT/kC) C = const
    - Bounds and asymptotes:
    
    - A(t), B(t) remain >= 0 and decrease to 0 at most.
    - From invariants, C is bounded above by
        C <= min(C0 + (kC/kA) A0, C0 + (kC/kB) B0).
    - Temp follows Temp = Temp0 + (kT/kC)(C - C0), hence
        Temp <= min(Temp0 + (kT/kA) A0, Temp0 + (kT/kB) B0).
    - Reaction effectively ceases upon reaching any
      equilibrium surface A=0 or B=0 or Temp=0; otherwise
      Demon may stop earlier at t = T. There are no
      oscillations; trajectories move toward these saturating
      bounds.
    - One-dimensional reduction:
        Using the invariants, C' becomes a cubic polynomial in
        C with roots at the limiting
        values where A or B hits zero (and a nonphysical root
        at negative C from Temp - (kT/kC)C
        if kT>0), confirming growth of C to a finite
        saturation.

    Special sub-modes:
    - Temperature-constant case (kT = 0): Temp' = 0; Temp is
      invariant. A,B decrease and C increases until A=0 or
      B=0 or Demon stops at t=T.

    4) Overall control pattern
    - Initial Demon test enforces positive parameters and
      T > 0.
    - The game proceeds through Demon-controlled iterations:
    1) Demon sets isOn \in {0,1}.
    2) Reset t := 0.
    3) Demon lets the ODE evolve for some duration up to T
       within the nonnegativity domain (or stops immediately).
    - Demon also decides how many such iterations occur.
    - Angel has no direct control; Demon's only lever is to
      toggle the reaction ON/OFF and pick
    durations/number of pulses. The state evolves along the
    linear invariant manifolds,
    with OFF holding the state and ON pushing it monotonically
    toward reagent depletion
    (A or B) with bounded increases in C and Temp.

    Angel control strategy:
    Covering the region where reactant depletion caps
    temperature growth before hitting Tmax.
    
    Initial conditions/assumptions and starting mode:
    - Assume Demon's test holds: A>0, B>0, kA,kB,kC>0, kT>=0,
      T>0.
    - Start in ON mode (isOn=1) for worst-case analysis of Temp
      growth.
    - Safety margin: Temp0 + min((kT/kA).A0, (kT/kB).B0) <= Tmax.
      (Special
        case: if kT=0, it suffices that Temp0 <= Tmax.)

    Angel's strategy:
    1) Do nothing;
    permit any Demon choices of ON/OFF, durations, and
    iterations, forever. The invariants
    cap Temp by Temp0 + min((kT/kA)A0, (kT/kB)B0), and reagent
    depletion (A or B→0) stops
    further increase, ensuring Temp <= Tmax throughout.

    Starting precondition computation.

    What is a suitable loop invariant for this loop?
\end{verbatim}

\noindent 4. GetNextActionCtrl is the query that decides what to do next: prove the current guess with the verification pipeline, or backtrack, and if so, to where?

\begin{verbatim}
- role: system
  content: |-
    You are trying to assign the preconditions of a
    differential game logic game.
    You have been proposing preconditions using backwards
    reasoning where you first proposed something for subgames
    closest to the end and then worked backwards recursively.
    Now you have reached a cross-roads and must choose a next
    action.
    You are given the original program, a log of what you
    were doing last, and the options of actions available to
    you.
    Reply with a brief explanation, followed by a newline,
    followed by a code block containing only the word for the
    action.

    ## Background information

    Here is some information about dGL.

    In differential game logic (dGL), there are two players:
    - **Angel**: The controllable player who makes choices we
      can control. Angel constructs include:
      - Angel choices: `a ++ b` (we choose between a and b)
      - Angel loops: `{a}*` (we choose how many iterations)
      - Angel tests: `?P` (we can proceed if P holds,
        otherwise we immediately lose)
      - Angel ODEs: `{x'=f(x) & Q}` (we control the duration)
      - Angel nondeterministic assignments: `x := *`
        (we choose the value)

    There is the dual operator `^@` that switches Angel and
    Demon roles in a game construct.
    Duals can be nested arbitrarily, so carefully check
    nesting levels to determine which player controls each
    action.
    Being within an odd number of `^@` operators indicates
    Demon control, while an even number (including zero)
    indicates Angel control.
    For example, in the game `{(a ++ b)^@; (c++d)}^@`, the
    choice between `a` and `b` is controlled by Angel, being
    within two `^@` operators, while the choice between `c`
    and `d` is controlled by Demon.
    Using this notation, we go over the Demon constructs
    below, indicating that they are Demon-controlled by
    wrapping them in a single `^@` operator.

    - **Demon**: The adversarial player who makes choices we
      cannot control. Demon has all the same constructs as
      Angel.
      - Demon choices: `{a ++ b}^@` (adversary chooses between
        a and b)
      - Demon loops: `{a*}^@` (adversary chooses how many
        iterations)
      - Demon tests: `{?P}^@` (adversary can proceed if P
        holds, and immediately loses otherwise. So Angel can
        *assume* in subsequent game play that P held at this
        point; otherwise she would already have won.)
      - Demon ODEs: `{x'=f(x) & Q}^@` (adversary controls the
        duration)
      - Demon nondeterministic assignments: `{x := *}^@`
        (adversary chooses the value)

    Additionally, the non-controlled operations are:
    - Assignment: `x := e` (deterministic assignment like in
      regular programs)
    - Sequential composition: `a; b` (first play a, then b)

    ## Notes
    - ODEs can sometimes introduce unmatched single quotes,
      so account for that when parsing.
    - In syntax such as { a++b; x:=* }*^@, the `^@` applies to
      the entire loop including its body. In this example,
      Demon controls the choice within the loop between a and
      b, as well as the assignment to x, because both are
      inside one `^@`.
    Some subgames are labeled with unique identifiers with
    the following syntax:
        <subgame_id>: <subgame>

    These subgame ids will help make your answers more
    specific.
    A *precondition* for a given game is a logical formula
    that, when it is true before the game starts, there
    exists a strategy for Angel to ensure that the
    *postcondition* holds after the game ends, regardless of
    how Demon plays.

    ## Input

    1. The current subgame for which you were trying to
       propose preconditions.
    2. overall_game: The overall game in which the current
       subgame appears.
    3. Analysis of the game and the control strategy Angel is
       trying to execute that the chosen preconditions should
       support.
    This is followed by a log of previous attempts at
    identifying preconditions for subgames and verifying
    whether they hold by stepping through the game and
    checking if the preconditions compose to ensure the
    postcondition is satisfied. Examine this history to
    figure out what just happened. Did you propose a bad ODE
    precondition? Did the precondition you proposed end up
    not being compatible with an earlier chosen loop
    invariant? Did you come up with a good invariant but you
    think it can be improved upon?
    4. The options you can choose for your next action.

    ## Possible Actions

    The preconditions that you proposed most recently didn't
    check out.
    Either it was not provably a precondition of the current
    subgame, or if part of a loop, your provided
    preconditions led to a precondition for the loop body
    that was not implied by the loop invariant.

    The error could lie in the most recent condition, or
    earlier in the process where you may have made a choice
    previously that backed you into a corner for this
    subgame, or maybe the condition is correct but too
    difficult for the automatic theorem prover to check. Now
    you must decide till what point to backtrack, or whether
    to try harder to prove the current subgame result using a
    tactic with KeYmaera X.

    You are given the original program, in which some points
    in the game are labeled as subgame_id.
    - `backtrack-to:id`: Backtrack to the subgame with the
      given id and propose a new precondition there. If you
      choose this option, in your reply, write
      `backtrack-to:id` where id is the id of the subgame you
      want to backtrack to. When backtracking, you can even
      choose the current subgame to retry. Loops are
      considered to occur "later" than their body, so if you
      backtrack to a subgame inside a loop, you won't be
      asked for a new loop invariant and the existing
      invariant will be used.
    - `try-proof`: Try to prove the current subgame using
      KeYmaera X with the precondition you proposed most
      recently. If you aren't sure why the most recent
      precondition was bad, this could be a good option
      because the manual proof attempt can give you feedback
      on what's wrong with the precondition.

    ## Output

    Write a few lines of explanation, followed by a newline,
    and then a code block containing the action you choose.
    For example, a valid code block is:

    ```
    try-proof
    ```

    I will now show you zero of more examples of how to
    respond, followed by the current case.
- role: user
  content: |-
    The current subgame is:
    subgame_p: {x' = a * x - b * x * y, y' = d * x * y - g *
    y}^@


    The overall game with labels is:
    {? x > 0 & y > 0 & a > 0 & b > 0 & d > 0 & g > 0 & x >=
    xmin & y >= ymin;^@}; {subgame_c: {{{{{{{subgame_j:
    xadd := *;^@
    }; {? xadd >= 0;^@}}; {subgame_l: yadd := *;^@
    }}; {? yadd >= 0;^@}}; {x := x + xadd;}}; {y := y +
    yadd;}}; {subgame_p: {x' = a * x - b * x * y, y' = d *
    x * y - g * y}
    }}*^@}

    Here is an analysis of the game followed by logs of
    precondition computation and checking so far:
    Analysis of game:
    1) Angel-controlled actions
    - ODE duration (subgame_p): In each iteration, Angel
      controls how long to flow along x' = a x - b x y, y' =
      d x y - g y (no evolution domain constraint, so any
      nonnegative duration, including 0).

    2) Demon-controlled actions
    - Initial test: {? x > 0 & y > 0 & a > 0 & b > 0 & d > 0 &
      g > 0 & x >= xmin & y >= ymin}^@. Angel may assume these
      hold initially; otherwise Demon would have already lost.
    - Looping: {...}*^@. Demon chooses how many iterations to
      perform (finite but unbounded, including zero).
    - Nondeterministic increments with constraints:
      - subgame_j: {xadd := *}^@ followed by {? xadd >= 0}^@:
        Demon chooses xadd and must satisfy xadd >= 0.
      - subgame_l: {yadd := *}^@ followed by {? yadd >= 0}^@:
        Demon chooses yadd and must satisfy yadd >= 0.
    - Deterministic updates x := x + xadd; y := y + yadd then
      apply each iteration.

    3) Control modes (qualitative dynamics)
    The hybrid iteration has two distinct phases per round:

    A. Discrete jump (Demon): x := x + xadd, y := y + yadd
    with xadd,yadd >= 0
    - Monotone, upward-only resets of both coordinates. Demon
      can make x,y arbitrarily large in a single iteration.
    - Effect on the continuous invariant (see below): H(x,y)
      = g ln x - d x + a ln y - b y has gradient
      - \delta H/\delta x = g/x - d, positive if x < g/d and negative if
        x > g/d.
      - \delta H/\delta y = a/y - b, positive if y < a/b and negative if y
        > a/b.
      So increasing a coordinate raises H when that coordinate
      is below its nullcline threshold, and lowers H when
      above. Demon's nonnegative jumps can thus move the state
      to higher or lower H-level sets depending on where (x,y)
      is relative to (g/d, a/b).

    B. Continuous Lotka–Volterra flow (Angel-controlled
    duration):
    - Nullclines:
      - x' = 0 when y = a/b (horizontal line).
      - y' = 0 when x = g/d (vertical line).
    - Unique interior equilibrium: E* = (g/d, a/b).
    - First integral (conserved along the ODE): H(x,y) = g ln
      x - d x + a ln y - b y = constant. Orbits are closed,
      encircling E*; no asymptotic convergence or divergence
      within the ODE phase.
    - Sign structure by regions (x>0,y>0 always preserved):
      - y < a/b, x > g/d: x' > 0, y' > 0 (both grow).
      - y > a/b, x > g/d: x' < 0, y' > 0.
      - y > a/b, x < g/d: x' < 0, y' < 0 (both decay).
      - y < a/b, x < g/d: x' > 0, y' < 0.
      Angel can stop the flow at any point on the current
      H-level set (including immediately with 0 time).
    - Simple bounds during any flow (using positivity and
      a,b,d,g>0):
      - x' = x(a - b y) <= a x \implies  x(t) <= x0 e^{a t}.
      - y' = y(d x - g) >= -g y \implies  y(t) >= y0 e^{-g t}.
      Positivity is invariant: x(t), y(t) stay > 0 for all
      flow times.

    Special submodes within the ODE:
    - Exact equilibrium: If (x,y) = (g/d, a/b), then x' = y' =
      0 (rest).
    - Oscillation: Otherwise, neutral cycles on H-level sets
      around E*; Angel chooses how far to traverse each cycle
      segment by choosing the duration.

    4) Overall control pattern
    - Start: Demon test establishes initial positivity and
      parameter bounds.
    - Iteration (repeats a Demon-chosen number of times):
      1) Demon picks xadd >= 0 (subgame_j) and yadd >= 0
         (subgame_l) and applies the discrete increases to x,y.
      2) Angel chooses a duration to flow along the
         Lotka–Volterra ODE (subgame_p), moving along the
         current closed orbit (conserved H).
    - Interaction:
      - Demon's lever: choose when to repeat and how large the
        nonnegative jumps are, thereby reselecting the H-level
        set before each flow.
      - Angel's lever: select the stopping point along the
        chosen H-level set each iteration (including skipping
        the flow by choosing zero duration).
      - Because Demon can pick unbounded xadd,yadd in any
        iteration, x and/or y can be made arbitrarily large;
        the ODE phase then oscillates around E* on the newly
        determined level set until Angel stops.

    Angel control strategy:
    Covering the “freeze the flow” region so Demon's
    nonnegative jumps alone preserve safety.

    Initial conditions/assumptions and start mode:
    - Assume the initial test holds: x>0, y>0, a,b,d,g>0, and
      x>=xmin, y>=ymin.
    - Start and remain in Immediate-stop ODE mode (Angel always
      ends the flow instantly).

    Angel's strategy:
    1) Choose zero ODE duration immediately whenever the
       Lotka–Volterra flow is offered, ending the continuous
       phase instantly, repeatedly.
    2) Repeat step 1 after every Demon discrete increment and
       update, forever (or until Demon stops the loop).

    Rationale: Discrete steps only increase x and y
    (xadd,yadd >= 0). By never flowing, x and y never
    decrease. Since the initial test ensures x>=xmin and
    y>=ymin, these inequalities remain true throughout all
    iterations.

    Starting precondition computation.
    Proposed the following loop invariant for loop {xadd := *;^@;
    ? xadd >= 0;^@; yadd := *;^@; ? yadd >= 0;^@; x := x + xadd;
    y := y + yadd; {x' = a * x - b * x * y, y' = d * x * y - g * y}}*^@.
    x >= xmin & y >= ymin
    Attempting to check invariant by precondition computation
    on loop body.
    Proposed precondition for subgame {x' = a * x - b * x * y,
    y' = d * x * y - g * y}^@ and postcondition x >= xmin &
    y >= ymin.
    Precondition or invariant: x >= xmin & y >= ymin.
    Check succeeded: False.
    Feedback: The formula check failed, indicating it is
    wrong, or too complex for the current level of
    automation. Please revise your guess or try a more
    detailed proof tactic.

    The options you can choose are:
    ['try-proof', 'backtrack-to:p', 'backtrack-to:c']
\end{verbatim}

\end{document}